\documentstyle[prl,eqsecnum,aps]{revtex}
\input{epsf}       

\begin{document}
\input psfig
\title{Studies of Gauge Boson Pair Production and Trilinear Couplings}

\author{                                                                        
S.~Abachi,$^{14}$                                                               
B.~Abbott,$^{28}$                                                               
M.~Abolins,$^{25}$                                                              
B.S.~Acharya,$^{43}$                                                            
I.~Adam,$^{12}$                                                                 
D.L.~Adams,$^{37}$                                                              
M.~Adams,$^{17}$                                                                
S.~Ahn,$^{14}$                                                                  
H.~Aihara,$^{22}$                                                               
G.A.~Alves,$^{10}$
% ...
\\                                                                              
\vskip 0.50cm                                                                   
\centerline{(D\O\ Collaboration)}                                               
\vskip 0.50cm                                                                   
}                                                                               
\address{                                                                       
\centerline{$^{1}$Universidad de los Andes, Bogot\'{a}, Colombia}               
\centerline{$^{2}$University of Arizona, Tucson, Arizona 85721}                 
\centerline{$^{3}$Boston University, Boston, Massachusetts 02215}               
\centerline{$^{4}$Brookhaven National Laboratory, Upton, New York 11973}        
\centerline{$^{5}$Brown University, Providence, Rhode Island 02912}             
\centerline{$^{6}$Universidad de Buenos Aires, Buenos Aires, Argentina}         
\centerline{$^{7}$University of California, Davis, California 95616}            
\centerline{$^{8}$University of California, Irvine, California 92697}           
\centerline{$^{9}$University of California, Riverside, California 92521}        
\centerline{$^{10}$LAFEX, Centro Brasileiro de Pesquisas F{\'\i}sicas,          
                  Rio de Janeiro, Brazil}                                       
\centerline{$^{11}$CINVESTAV, Mexico City, Mexico}                              
\centerline{$^{12}$Columbia University, New York, New York 10027}               
\centerline{$^{13}$Delhi University, Delhi, India 110007}                       
\centerline{$^{14}$Fermi National Accelerator Laboratory, Batavia,              
                   Illinois 60510}                                              
\centerline{$^{15}$Florida State University, Tallahassee, Florida 32306}        
\centerline{$^{16}$University of Hawaii, Honolulu, Hawaii 96822}                
\centerline{$^{17}$University of Illinois at Chicago, Chicago, Illinois 60607}  
\centerline{$^{18}$Indiana University, Bloomington, Indiana 47405}              
\centerline{$^{19}$Iowa State University, Ames, Iowa 50011}                     
\centerline{$^{20}$Korea University, Seoul, Korea}                              
\centerline{$^{21}$Kyungsung University, Pusan, Korea}                          
\centerline{$^{22}$Lawrence Berkeley National Laboratory and University of      
                   California, Berkeley, California 94720}                      
\centerline{$^{23}$University of Maryland, College Park, Maryland 20742}        
\centerline{$^{24}$University of Michigan, Ann Arbor, Michigan 48109}           
\centerline{$^{25}$Michigan State University, East Lansing, Michigan 48824}     
\centerline{$^{26}$Moscow State University, Moscow, Russia}                     
\centerline{$^{27}$University of Nebraska, Lincoln, Nebraska 68588}             
\centerline{$^{28}$New York University, New York, New York 10003}               
\centerline{$^{29}$Northeastern University, Boston, Massachusetts 02115}        
\centerline{$^{30}$Northern Illinois University, DeKalb, Illinois 60115}        
\centerline{$^{31}$Northwestern University, Evanston, Illinois 60208}           
\centerline{$^{32}$University of Notre Dame, Notre Dame, Indiana 46556}         
\centerline{$^{33}$University of Oklahoma, Norman, Oklahoma 73019}              
\centerline{$^{34}$University of Panjab, Chandigarh 16-00-14, India}            
\centerline{$^{35}$Institute for High Energy Physics, 142-284 Protvino, Russia} 
\centerline{$^{36}$Purdue University, West Lafayette, Indiana 47907}            
\centerline{$^{37}$Rice University, Houston, Texas 77005}                       
\centerline{$^{38}$Universidade Estadual do Rio de Janeiro, Brazil}             
\centerline{$^{39}$University of Rochester, Rochester, New York 14627}          
\centerline{$^{40}$CEA, DAPNIA/Service de Physique des Particules, CE-SACLAY,   
                   Gif-sur-Yvette, France}                                      
\centerline{$^{41}$Seoul National University, Seoul, Korea}                     
\centerline{$^{42}$State University of New York, Stony Brook, New York 11794}   
\centerline{$^{43}$Tata Institute of Fundamental Research,                      
                   Colaba, Mumbai 400005, India}                                
\centerline{$^{44}$University of Texas, Arlington, Texas 76019}                 
\centerline{$^{45}$Texas A\&M University, College Station, Texas 77843}         
}                                                                               

\maketitle

\begin{abstract}
The gauge boson pair production processes $W\gamma$, $WW$, $WZ$, and $Z\gamma$
were studied using $p \bar{p}$ collisions corresponding to an integrated 
luminosity of approximately $14~{\rm pb}^{-1}$ at a center-of-mass energy 
of $\sqrt{s} = 1.8$ TeV.  
Analysis of $W\gamma$ production  with subsequent $W$ boson decay to $\ell\nu$ 
$(\ell = e, \mu)$ is reported, including a fit to the $p_T$ spectrum of the 
photons which leads to limits on anomalous $WW\gamma$ couplings.  
A search for $WW$ production with subsequent decay to 
$\ell\bar{\ell}\nu\bar{\nu}$ $(\ell=e,\mu)$ is presented, leading to an upper
limit on the $WW$ production cross section and limits on anomalous $WW\gamma$
and $WWZ$ couplings.
A search for high $p_T$ $W$ bosons in $WW$ and $WZ$ production
is described, where one $W$ boson decays to an electron and a neutrino and the
second $W$ boson or the $Z$ boson decays to two jets.
A maximum likelihood fit to the $p_T$ spectrum of $W$ bosons
resulted in limits on anomalous $WW\gamma$ and $WWZ$ couplings.   
A combined fit to the three data sets which provided the tightest limits on
anomalous $WW\gamma$ and $WWZ$ couplings is also described.
Limits on anomalous $ZZ\gamma$ and $Z\gamma\gamma$ couplings are presented
from an analysis of the photon $E_T$ spectrum in $Z\gamma$ events 
in the decay channels $(ee, \mu\mu, {\rm and}\; \nu \nu)$ of the 
$Z$ boson. 
\end{abstract}

\vskip 2.0 cm
\normalsize 

\section{Introduction}
\label{sec-intro}
Interactions between gauge bosons, the $W$ boson, $Z$ boson and 
photon, are a consequence of the non-Abelian gauge symmetry of 
the Standard Model (SM).  The gauge boson self-interactions 
are described by the trilinear gauge boson vertices and contribute to gauge 
boson pair production in $p \bar{p}$ collisions.
The cross sections of these processes are relatively small within the SM.
The inclusion of non-SM (anomalous) couplings at 
the trilinear gauge boson vertices
enhances the production cross sections of gauge boson pairs,
especially at large values of the gauge boson transverse momentum $p_T$, and 
at large values of the invariant mass of the gauge boson pair system.
Observation of anomalous gauge boson pair production
would indicate physics beyond the SM.
Feynman diagrams for gauge boson pair production are shown
in Figs.~\ref{fig-vpair}(a)--(c), where $V_{0}, V_{1}$ and $V_{2}$ are the
$W$ boson, the $Z$ boson, or the photon.
Figures~\ref{fig-vpair}(a) and (b) are described by well-known couplings
between the gauge bosons and quarks.
Figure~\ref{fig-vpair}(c) shows the trilinear coupling diagram.
Numerous phenomenological studies\cite{numerous-studies} of
the characteristics of gauge boson self-interactions have been
performed in anticipation of hadron and $e^{+}e^{-}$ collider experiments
where direct measurements of the coupling parameters are possible by studying
gauge boson pair production processes. 

This paper describes studies of gauge boson pair production 
and the corresponding trilinear gauge boson coupling parameters using data
from $\sqrt{s}=1.8$
 TeV $p{\bar p}$ collisions taken with the D{\O} detector during 
the 1992-1993 Tevatron collider run at the Fermi National Accelerator 
Laboratory.  
Four processes were studied:
$W\gamma$ production, where the $W$ boson decayed into $e\nu$ or $\mu\nu$;
$W$ boson pair production, where both of the $W$ bosons decayed into
$e\nu$ or $\mu\nu$; $WW$ and $WZ$ production, where one $W$ boson decayed
into $e\nu$ and the second boson decayed hadronically; and
$Z\gamma$ production, where the $Z$ boson decayed into $e^+e^-$, 
$\mu^+\mu^-$, or $\nu\bar{\nu}$. 

This paper presents the details of analyses whose results have already been 
published~\cite{wgdzero,WWdilD0,lpchenprl,d0zg1,d0zg2}.  In addition, 
it presents limits on anomalous trilinear couplings from the combined 
$W\gamma$, $WW$, and $WZ$ analyses. 

\subsection{$WW\gamma$ and $WWZ$ Trilinear Gauge Boson Couplings}
A formalism has been developed\cite{lagrangian} to describe the $WW\gamma$ and 
$WWZ$ interactions for models beyond the SM using an effective Lagrangian.
The $WW\gamma$ and $WWZ$ vertices that satisfy Lorentz invariance 
and conservation of C and P can be described by a Lagrangian 
with two overall coupling strengths $g_{WW\gamma}=-e$ and $g_{WWZ} =
-e \cot{\theta}_w$ and 
six coupling parameters $g_{1}^{V}, \kappa_{V}$ and $\lambda_{V}$, where
$V = \gamma , Z$.
In the SM, $\Delta g_1^V = g_1^V - 1 =0, \Delta\kappa_V = \kappa_V - 1 = 0$,
and $\lambda_V = 0$. 
The anomalous couplings are parameterized as form factors with a 
scale, $\Lambda$, in order to avoid unitarity violation of the gauge
boson pair cross section at asymptotically high energies: e.g., 
$\lambda_\gamma ({\hat s}) = \lambda_\gamma /(1 + {\hat s}/\Lambda^2 )^2$.  
The $WW\gamma$ and $WWZ$ coupling parameters, in the static limit, 
are related to the magnetic dipole moments ($\mu_W$) 
and electric quadrupole moments ($Q_W^e$) of the $W$ boson:
$\mu_W=\frac{e}{2M_W}(1+\kappa+\lambda)$ and 
$Q_W^e=-\frac{e}{M_W^2}(\kappa-\lambda)$\cite{KIM},
where $e$ and $M_W$ are the charge and the mass of the $W$ boson. 
A more detailed discussion of the effective Lagrangian for the $WW\gamma$ and 
$WWZ$ interactions can be found in Appendix 1.

\begin{figure}
\vspace{-1.0in}
\centerline{\epsfbox{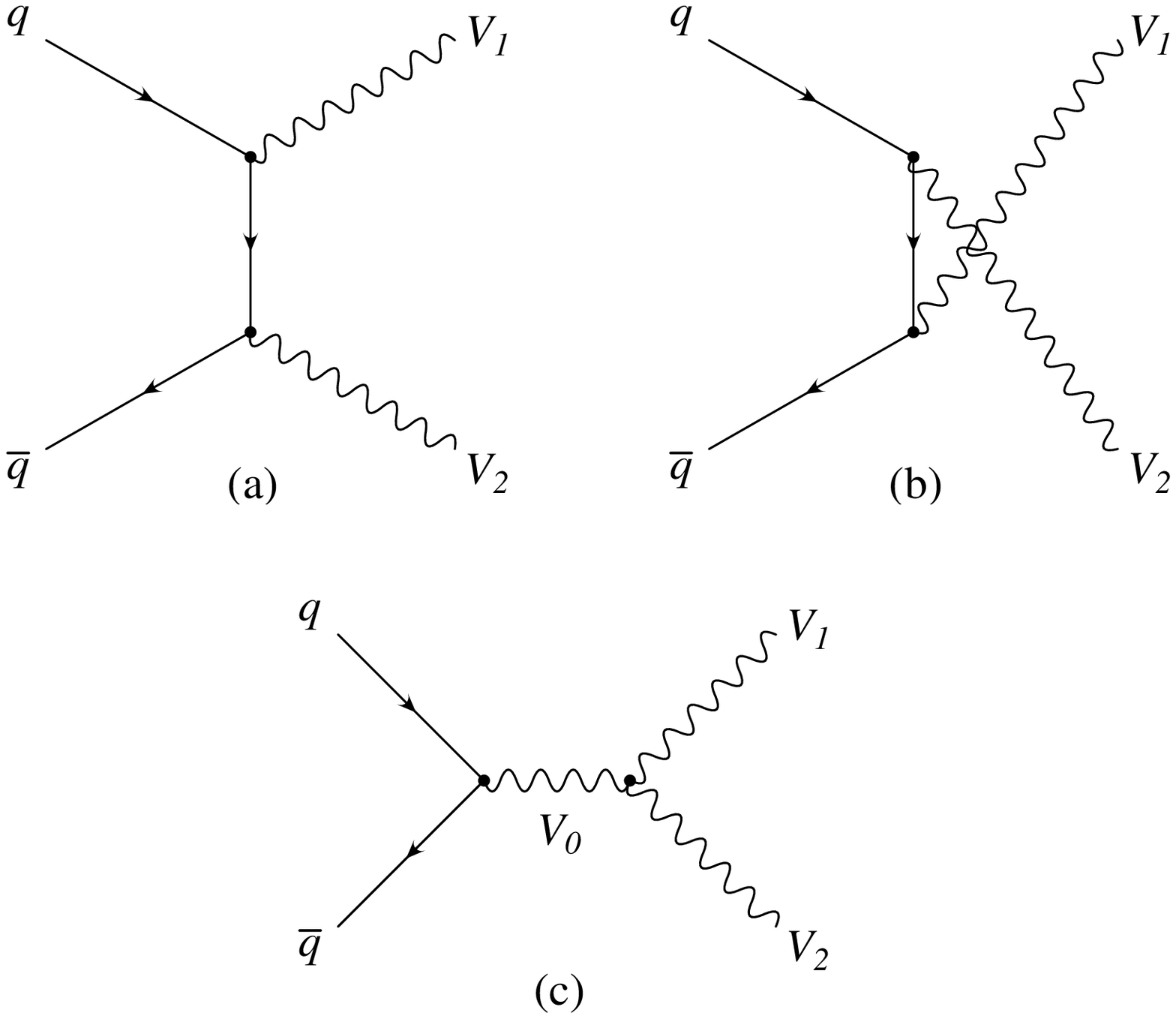}}
\vspace{1.0in}
\caption{Leading-order Feynman diagrams for vector boson pair production.
The assignment of $V_0$, $V_1$, and $V_2$ depends on the final state:
$W\gamma$, $WW$, $WZ$ or $Z\gamma$. }
\label{fig-vpair}
\end{figure}

\begin{figure}
\vspace{-0.5in}
\centerline{\epsfbox{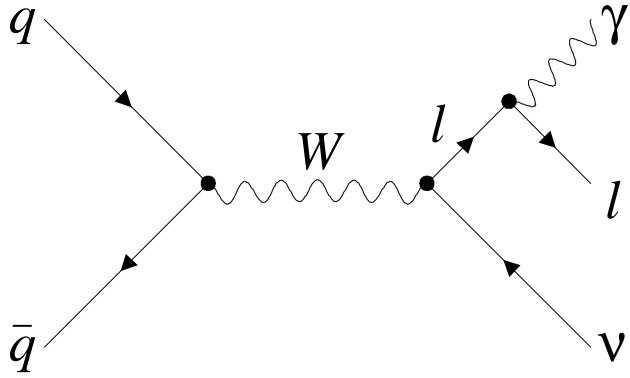}}
\caption{The leading-order radiative Feynman diagram for $W\gamma$ 
production where the photon is the result of bremsstrahlung from a final 
state lepton.}
\label{fig-wgrad}
\end{figure}

\begin{figure}
\vspace{-0.5in}
\centerline{\epsffile{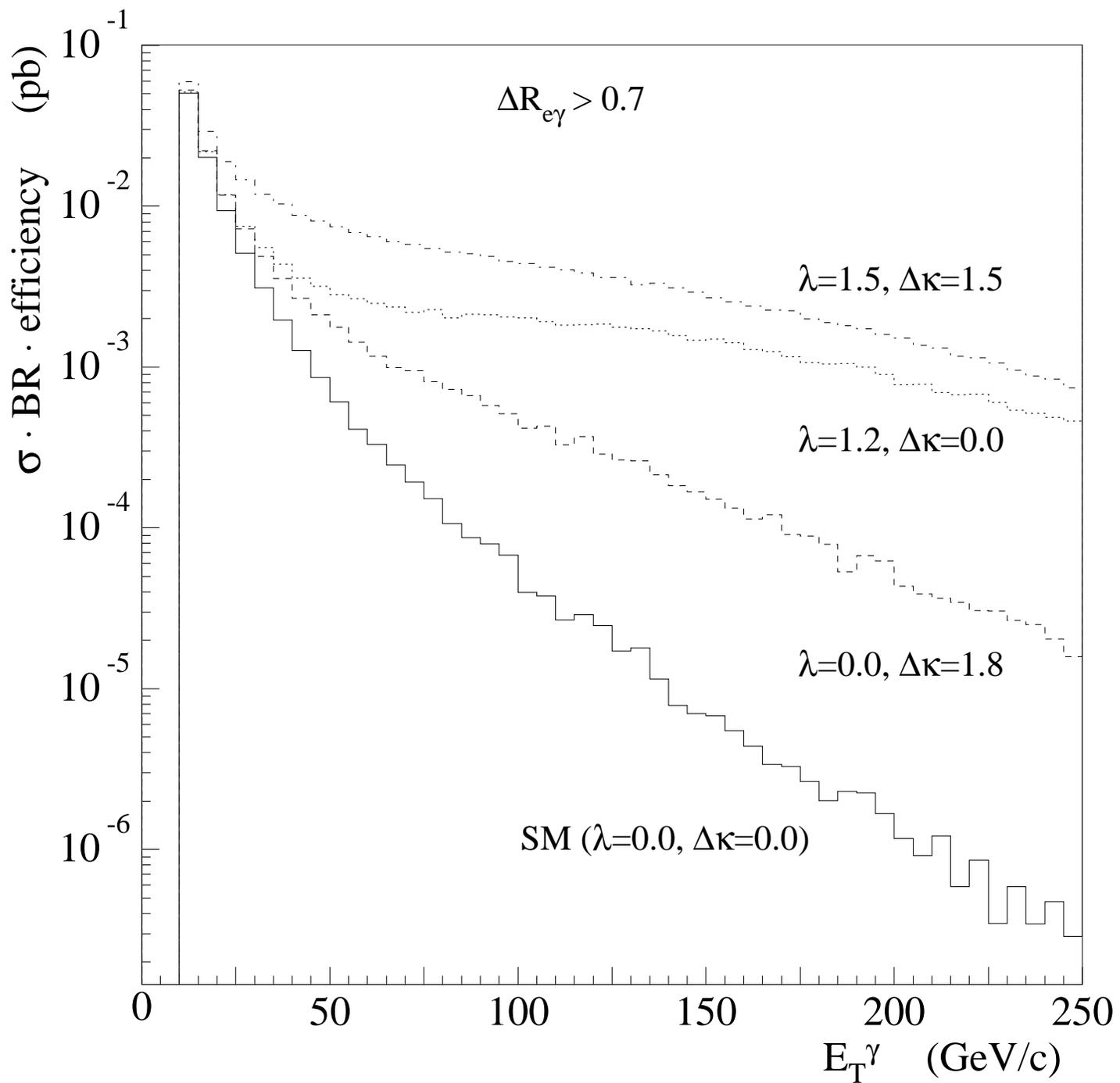}}
\caption{The predicted $E_T$ spectrum of photons in the $W\gamma$ production 
for SM and anomalous $WW\gamma$ couplings.  Radiative diagrams are included. 
The requirement that $\Delta R_{{\rm e}\gamma}>0.7$ has been 
made; otherwise the cross section diverges at low $E_T^{\gamma}$.}
\label{fig-wgpt}
\end{figure}

The $W\gamma$ production process has the highest 
cross section among the gauge boson pair production processes at the Tevatron.
Feynman diagrams for the $q{\bar q}'\rightarrow W\gamma$ process are
obtained by substituting $V_{0} =  V_{1} = W$ and $V_{2} = \gamma$
in Figs.~\ref{fig-vpair}(a)--(c). A delicate cancellation 
takes place between the amplitudes~\cite{citewg}   
that correspond to the $u$ and $t$-channel quark
exchange, shown in Figs.~\ref{fig-vpair}(a) and (b), and $s$-channel 
production with a $W$ boson as the mediating particle, shown in 
Fig.~\ref{fig-vpair}(c).
A $W$ boson is identified by its leptonic decay products: a high
$p_T$ charged lepton $\ell \; (\ell = e,\mu)$; and large missing transverse
energy (\hbox{$\rlap{\kern0.25em/}E_T$}) due to the undetected neutrino. 
Single $W$ boson production, followed by 
radiation of a photon from the charged lepton from the $W$ boson decay,
also contributes to the $\ell\nu\gamma$ final state. This process 
is shown in Fig.~\ref{fig-wgrad}.
The photon from the radiative decay is preferentially emitted along the 
direction of the charged lepton; the process can be suppressed by 
imposing a minimum 
separation requirement, $\Delta R_{\ell\gamma}$, between the charged 
lepton and the photon  where $\Delta R_{\ell\gamma}$ is the distance in 
pseudorapidity and azimuth. For $\sqrt{s} =1.8$ TeV $p{\bar p}$ collisions,
the predicted cross section times branching fraction for $W\rightarrow e\nu$
or $\mu \nu$ for $W\gamma$ final states with photon transverse energy 
$(E_{T}^{\gamma}) > 10$ GeV and $\Delta R_{\ell\gamma} > 0.7$ is 12.5 
pb. Figure~\ref{fig-wgpt} shows the $E_T$ spectrum of photons for the SM
and non-SM production processes predicted by the leading order 
theory\cite{BaurZep}. 
The distributions for non-SM $WW\gamma$ couplings exhibit a large increase
in the cross section at high $E_T^{\gamma}$. 
The $W\gamma$ process is sensitive only to $WW\gamma$ couplings, not to
$WWZ$ couplings.  
It is more sensitive to $\lambda_\gamma$ than to $\Delta\kappa_\gamma$, 
since the amplitudes related to $\lambda_\gamma$ and $\Delta\kappa_\gamma$
are proportional to ${\hat s}$ and $\sqrt{{\hat s}}$, respectively.

Limits on the $WW\gamma$ trilinear couplings from studies of $W\gamma$
production have been reported\cite{wgua2,wgcdf,wgdzero} previously by the
UA2, CDF, and D\O \  Collaborations.
In this paper, the results from the D\O \ 
Collaboration, with the most restrictive $WW\gamma$ limits, are presented 
in more detail than in the recent publication \cite{wgdzero}.  

\begin{figure}
\centerline{\epsffile{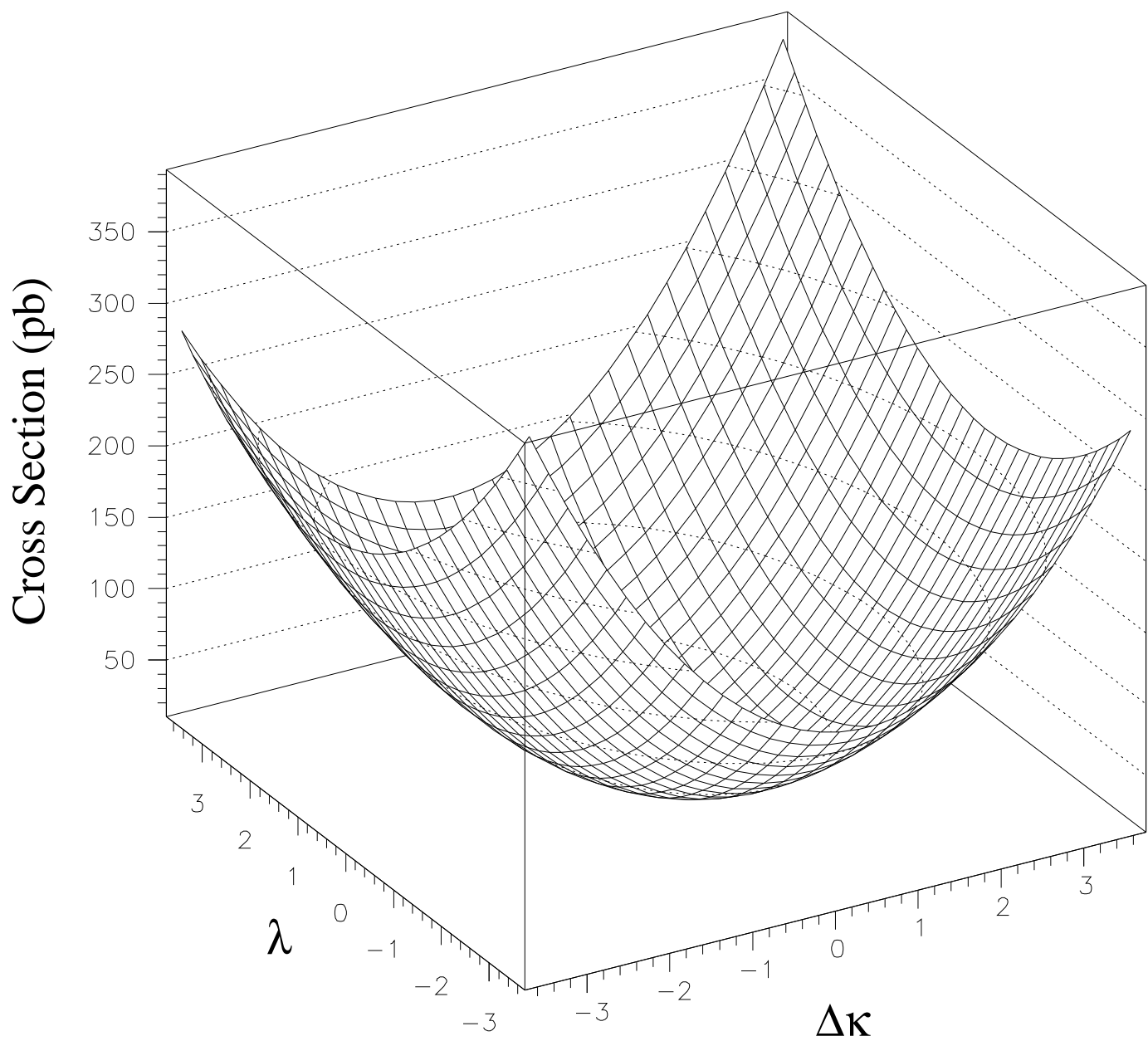}}
\caption{The predicted cross section for $WW$ production as a function of 
anomalous coupling parameters $\lambda$ and $\Delta\kappa$, assuming the 
$WWZ$ and $WW\gamma$ couplings are equal, with $\Lambda = 1000$ GeV. }
\label{fig-wwxs}
\end{figure}

The process $p\bar{p}\rightarrow WW+X$ is predicted to have a cross section 
of $9.5$ pb\cite{ohnemus} at next-to-leading order at $\sqrt{s}=1.8$ TeV.  
The Feynman diagrams for the $WW$ production processes
are obtained by substituting $V_{0} = \gamma$ or $Z$, $V_{1} = W^{+}$
and $V_{2} = W^{-}$ in Figs.~\ref{fig-vpair}(a)--(c).
Destructive interference, similar to that occuring in $W\gamma$ production,
occurs between the $u$ and $t$-channel amplitudes 
and the $s$-channel amplitude\cite{delicate-cancellation} 
with a photon or a $Z$ boson as the mediating particle.
The former processes contain the well-known couplings between the $W$ bosons
and quarks and the latter the $WW\gamma$ and $WWZ$ trilinear couplings.
$W$ boson pair production is sensitive to both of the $WW\gamma$ and $WWZ$ 
couplings.  It is approximately a factor of two more sensitive to the $WWZ$ 
couplings, due to the higher value of the overall coupling 
$g_{WWZ} = -e \, {\rm cot}\theta_w$, 
than to the $WW\gamma$ couplings with $g_{WW\gamma} = -e$ and is 
therefore complementary to $W\gamma$ production.
The predicted\cite{HWZ}  cross section for $WW$ production,
as a function of anomalous coupling 
parameters $\lambda \equiv \lambda_\gamma = \lambda_Z$ and
$\Delta\kappa \equiv \Delta\kappa_\gamma = \Delta\kappa_Z$, where the 
$WW\gamma$ and $WWZ$ couplings are assumed to be equal and $\Lambda =
1000$ GeV, is shown in Fig.~\ref{fig-wwxs}.

The details of the recently published analysis\cite{WWdilD0}, in which 
an upper limit on the $WW$ cross section was obtained from 
the observed number of signal events in the dilepton decay modes, are 
presented. The limit on the cross section was translated into the limits on 
anomalous coupling parameters.  

\begin{figure}[b]
\epsfysize = 5.0in
\epsfxsize = 7.0in
\centerline{\epsffile{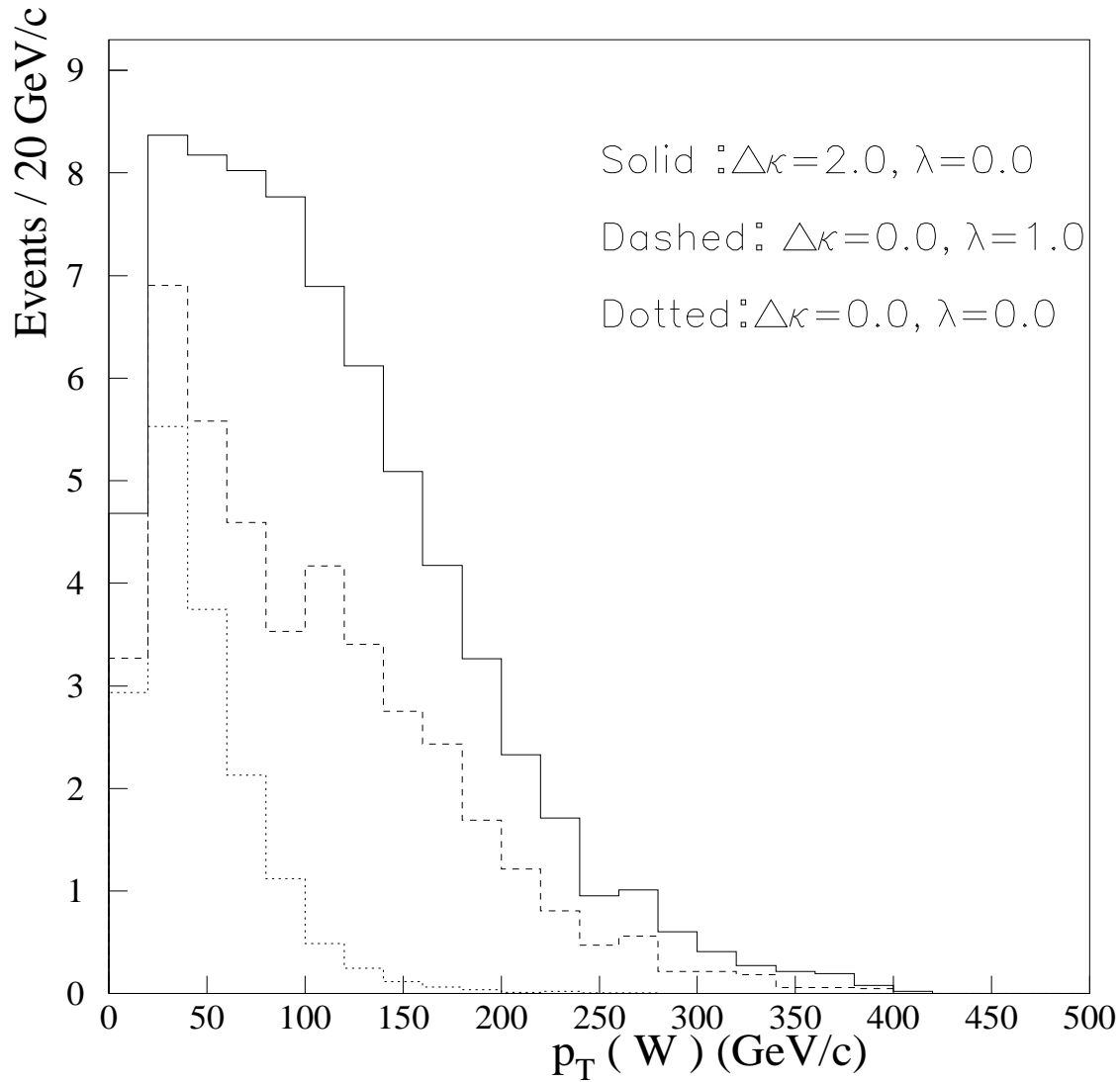}}
\vspace{2.0in}
\caption{The $p_T$ spectrum of $W$ bosons in $WW/WZ$ production
corresponding to approximately $10$ pb$^{-1}$ of collisions. The 
theoretical assumptions for the anomalous coupling spectra are that 
$\lambda_{\gamma}=\lambda_{Z}$ and $\Delta \kappa_{\gamma}=\Delta \kappa_Z$, 
with $\Lambda=1000$ GeV. }
\label{fig-wzpt}
\end{figure}

For $p\bar{p}$ collisions at $\sqrt{s}=1.8$ TeV, the cross section 
predicted\cite{ohnemusWZ} at next-to-leading order for $WZ$ production 
is $2.5$ pb. The Feynman diagrams for $WZ$ production are obtained
by substituting $V_{0} = V_{1} = W$ and $V_{2} = Z$ in
Figs.~\ref{fig-vpair}(a)--(c). While $WW$ production is sensitive to both
the $WW\gamma$ and $WWZ$ couplings, $WZ$ production depends only on
the $WWZ$ couplings.   
The $WW$ and $WZ$ decay channels in which one $W$ boson decays into an 
electron and a neutrino and the second $W$ boson or the $Z$ boson decays 
hadronically were studied in order to obtain an upper limit on the cross 
section and to restrict possible anomalies in the coupling parameters.
In these processes, the $W$ and $Z$ bosons that decay hadronically to produce 
two jets in the detector cannot be differentiated due to the limitations of 
jet energy resolution.   
Figure~\ref{fig-wzpt} shows the $p_T$ spectrum for $W$ bosons in 
$WW$ and $WZ$ production from the leading-order theoretical 
prediction\cite{HWZ}. This paper describes in detail an analysis summarized in 
Ref.~\cite{lpchenprl}, in which the $E_T$ spectrum 
of the $W$ bosons, produced with two or more jets which could have come from 
a hadronic $W$ or $Z$ boson decay, was compared to the expected SM signal plus
background to set limits on anomalous $WW\gamma$ and $WWZ$ couplings.
The CDF Collaboration has studied the $l\nu$ jet-jet decay mode and
reported\cite{CDFWZ} limits on anomalous $WW\gamma$ and $WWZ$ couplings. 

A new result on the anomalous couplings from a combined fit is 
presented. Since the $W\gamma$, $WW$ to dileptons, and $WW/WZ$ to electron
plus jets analyses measured the same coupling parameters, a combined fit
to all three data sets was performed, yielding improved
limits on the anomalous couplings compared to the individual analyses.

\subsection{$Z\gamma\gamma$ and $ZZ\gamma$ Trilinear Gauge Boson Couplings}
\begin{figure}
\vspace{-0.5in}
\centerline{\epsfbox{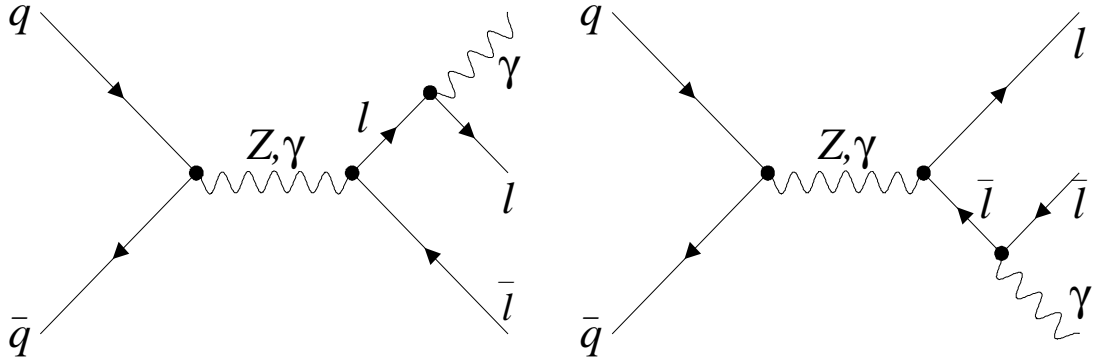}}
\caption{The leading-order radiative Feynman diagrams for $Z\gamma$ 
production where the photon is the result of bremsstrahlung from a final 
state lepton. This decay mode only applies to final states involving
charged leptons. }
\label{fig-zgrad}
\end{figure}

The interactions of pairs of neutral gauge bosons, the $Z$ boson and the
photon, can be studied through the $Z\gamma$ production process.
The Feynman diagrams for the $q{\bar q}' \rightarrow Z\gamma$ processes
are obtained by substituting $V_{0} = \gamma$ or $Z$, $V_{1} = Z$ and
$V_{2} = \gamma$ in Figs.~\ref{fig-vpair}(a)--(c).  There are no $ZZ\gamma$ 
and $Z\gamma\gamma$ couplings of the type shown in Fig.~\ref{fig-vpair}(c) 
in the SM; thus, there is no destructive interference of the $u$ and
$t$-channel amplitudes, such as occurs in $W\gamma$, $WW$, and $WZ$ production. 
A $Z$ boson is identified by its leptonic decay products, a pair of
high $p_T$ leptons $(\ell = e$ or $\mu$), or by the imbalance of momentum in
the event due to not detecting the neutrino pair.
The Drell-Yan production of a $Z$ boson or virtual photon,
followed by radiation of 
a photon off the charged lepton from the $Z$ boson or virtual
photon decay products, also contributes to the charged lepton final states,
as shown in Fig.~\ref{fig-zgrad}.
As with the final state radiation from $W$ boson decay products, 
the photon from the 
$Z$ boson decay products is preferentially emitted along the charged lepton
direction; the process can be suppressed by imposing a cut on the 
separation between the charged lepton and the photon.
The most general Lorentz and gauge invariant $ZV\gamma$ vertex is described
by eight coupling parameters $h_{i}^{V} (i = 1,...,4)$\cite{gl20}.
The anomalous couplings are parameterized as form factors
$h_{i}^{V} = h_{i0}^{V}/(1 + {\hat s}/\Lambda^2 )^n$, where ${\hat s}$ is
the square of the invariant mass of the $Z\gamma$ system, 
$n = 3$ for $h_{1,3}^{V}$, and $n = 4$ for $h_{2,4}^{V}$. This is discussed 
in more detail in Appendix 1.
Figure~\ref{fig-zgpt} shows the $E_T$ spectra of photons predicted for the SM
and the non-SM model production processes.
The distributions for the non-SM $ZZ\gamma$ and $Z\gamma\gamma$
couplings exhibit a large increase of the cross section at high $E_T$.
Limits on the anomalous coupling parameters are obtained from a maximum
likelihood fit to the $E_T$ spectrum of the photons, 
as in $W\gamma$ production.
Previously, CDF has published limits on the $ZZ\gamma$ and 
$Z\gamma \gamma$ anomalous couplings\cite{cdfzg} using the $ee$ and $\mu\mu$ 
final states.  Recent results from the D\O \ experiment
are presented here in more detail than in the previous\cite{d0zg1,d0zg2} 
summaries and include the $ee$, $\mu\mu$, and neutrino final states.

\subsection{Outline of Paper}
The paper is organized as follows. In Section~\ref{sec-pid}
the D\O \ detector and the techniques used to identify particles from the 
collisions are discussed.
Section~\ref{sec-mc} is a summary of various Monte Carlo
modelling tools used in these analyses.  Section~\ref{sec-ds} discusses the 
1992-1993 collider run and data samples.  
Section~\ref{sec-wg} describes a measurement of the $WW\gamma$
coupling parameters using $W\gamma$ events where the $W$ boson
decays into a high $p_T$ electron or muon and a neutrino.
In Section~\ref{sec-wwdilep} the results of a search of $WW\rightarrow 
(\ell\bar{\nu}) (\bar{\ell '}\nu ')$ process are presented.
This is followed by a description and the 
results of the analysis for $WW$ and $WZ$ 
production with subsequent decay to $e\nu$ and at least two jets in Section~
\ref{sec-wwev}.
Section~\ref{sec-cl} describes the combined limits on
the anomalous couplings from all of the $W\gamma$ and $WW/WZ$
analyses.
Section~\ref{sec-zg} presents a measurement of $ZZ\gamma$ and $Z\gamma\gamma$
coupling parameters using
$Z\gamma$ production events where the $Z$ boson decays into $ee$, $\mu\mu$ or
$\nu\bar{\nu}$.
The results are summarized and reviewed in Section~\ref{sec-conclusions}. 
Appendix 1 contains a detailed review of trilinear gauge boson couplings 
and the effective Lagrangian. Finally, Appendix 2 contains a discussion 
of the binned likelihood fitting procedure used in determining the anomalous 
couplings limits. 

\section{Detector}
\label{sec-pid}

\begin{figure}
\centerline{\epsffile{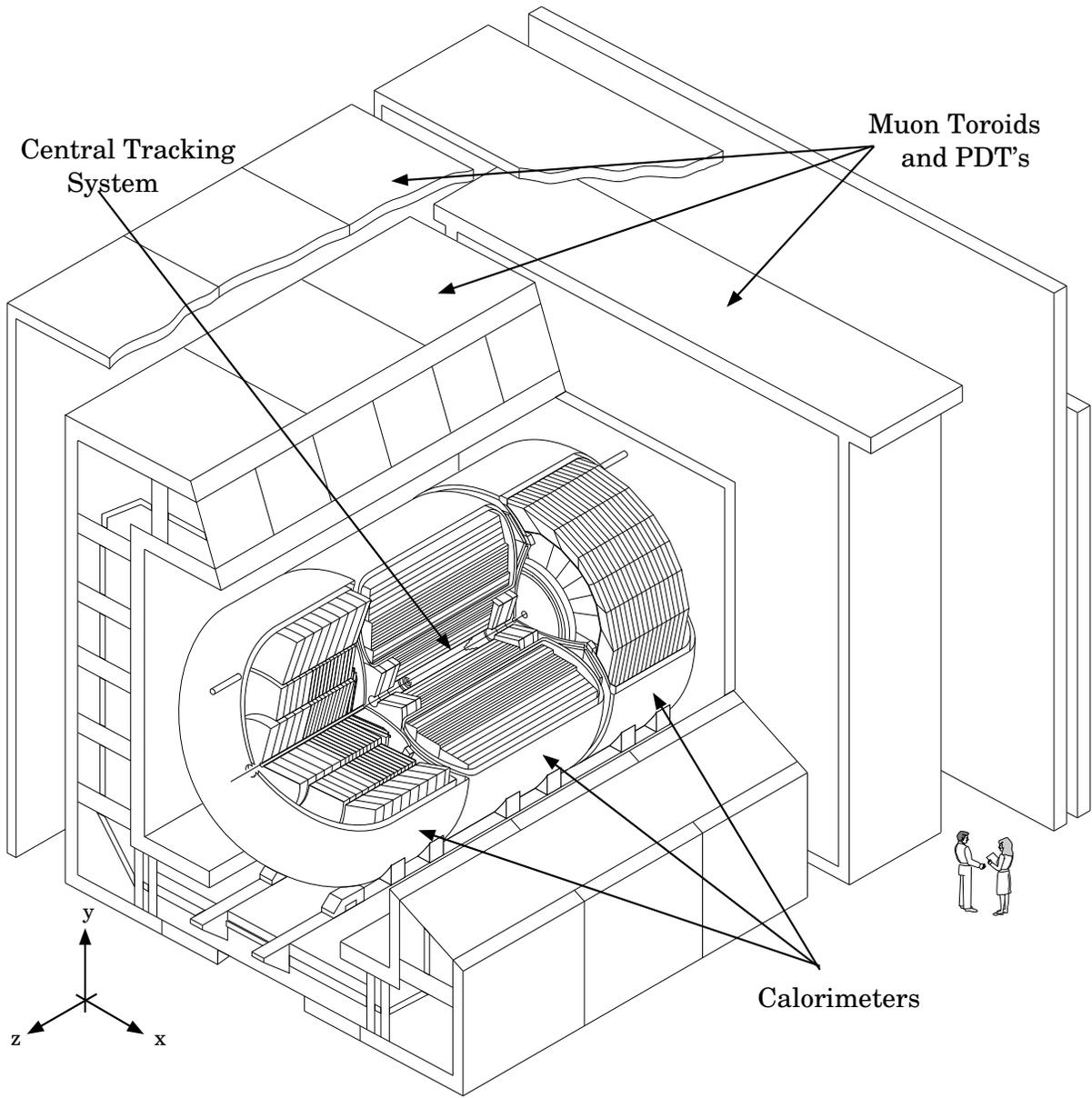}}
\caption{ Perspective view of the D\O\ detector. Also shown are the movable 
support platform, the Tevatron beampipe centered within the detector and the 
Main Ring beampipe which penetrates
the muon system and calorimeter above the detector center.}
\label{fig-D0}
\end{figure}

 The major components of the D{\O} detector\cite{d0nim} were 
a nonmagnetic central tracking detector system, a hermetic liquid--argon
uranium calorimeter and a muon spectrometer system with a toroidal magnetic 
field. 
A perspective view of the 
detector is shown in Fig.~\ref{fig-D0}, depicting the three major systems.

\begin{figure}
\centerline{\epsffile{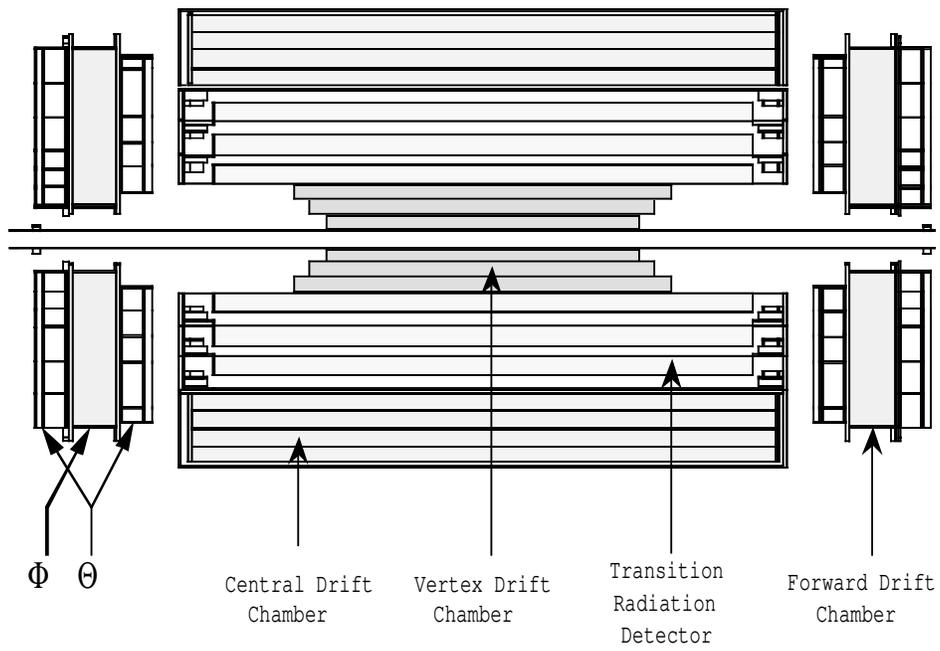}}
\vspace{1.5in}
\caption{The D\O\ central tracking detector system.}
\label{fig-tracking}
\end{figure}

The central tracking detectors (CD), shown in
Fig.~\ref{fig-tracking}, included the 
Vertex Drift Chamber (VTX), the Transition Radiation Detector (TRD),
the Central Drift Chamber (CDC) and two Forward Drift Chambers (FDC's).
The VTX, TRD and CDC were arranged in three concentric layers around the
beamline, from the beampipe out to the central calorimeter. 
The wires in the FDC's were oriented perpendicular to the beamline.
The entire CD was contained in the cylindrical volume
$(r = 76 \; {\rm cm}, z = \pm 135 \; {\rm cm})$ bounded by the calorimeter
cryostats.
The CD measured the trajectory of charged particles
with a resolution of 2.5 mrad
in $\phi$ and 28 mrad in $\theta$, where $\phi$ and $\theta$ are the azimuthal
and polar angles of the track, respectively, and covered  
the region $|\eta| \leq 3.2$ in pseudorapidity, where
$\eta = -{\rm ln (\rm tan}\frac{\theta}{2})$.
From these measurements, the position of the interaction vertex along the 
beam direction  was determined with a resolution of 8 mm. 
The presence of a CD track or hits pointing towards a shower 
was the key element for distinguishing electrons from photons. 
The CDC and FDC's also provided ionization energy loss measurement for 
separating single electrons from closely-spaced photon conversion pairs
where the photon converted before it reached the tracking detector.  

\begin{figure}
\epsfysize = 6.0in
\epsfxsize = 6.0in
\centerline{\epsffile{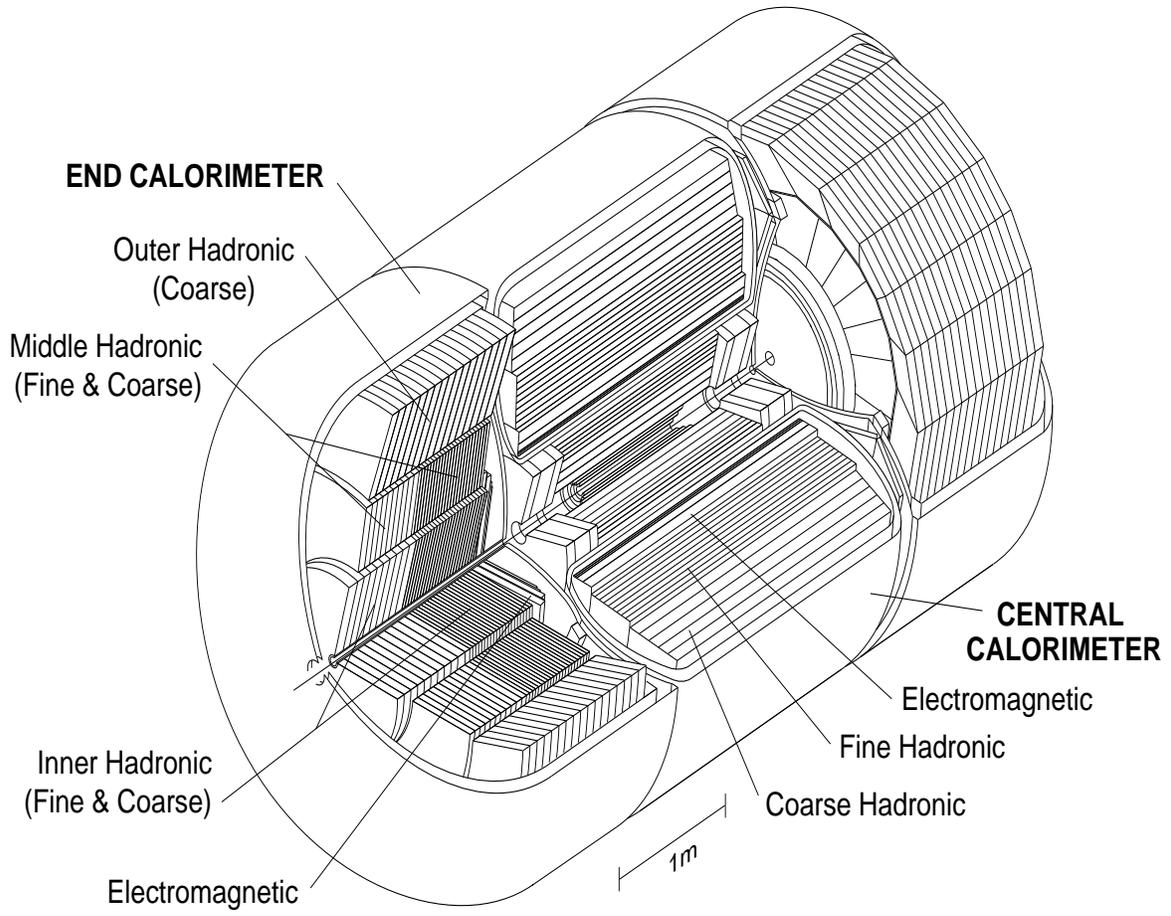}}
\vspace{0.5in}
\caption{The D\O\ calorimeters.}
\label{fig-calorimeter}
\end{figure}

The calorimeter 
was divided into three parts, each enclosed in a steel cryostat to contain the
liquid argon: the central calorimeter (CC) and two end 
calorimeters (EC's) as shown in Fig.~\ref{fig-calorimeter}. Each consisted of
an inner electromagnetic (EM) section, a finely-segmented hadronic 
section (FH) and a coarsely-segmented hadronic section (CH). 
The scintillator-based intercryostat detectors (ICD's),
which improved the energy resolution for jets that straddled the central and 
end calorimeters, were inserted into the space between the 
cryostats.  The calorimeter covered the pseudorapidity range $|\eta| \leq 4.2$.
The boundaries between the CC and EC's were chosen to be approximately 
perpendicular to the beam direction and to match the transition between the 
CDC and FDC's. 

 Each EM section was divided into four longitudinal layers forming a 
thickness of 21 radiation lengths. The hadronic sections were divided into 
four (CC) or five (EC) layers and were 7--10 nuclear interaction lengths 
thick.  The calorimeter was transversely segmented into projective towers with 
$\Delta \eta \times \Delta \phi = 0.1 \times 0.1$, where $\phi$ is the
azimuthal angle. 
The third layer of the EM calorimeters, where the maximum energy deposition
from EM showers was expected to occur, was segmented more finely into
cells with $\Delta \eta \times \Delta \phi = 0.05 \times 0.05$.
The azimuthal position resolution for electrons with energy above 50 GeV 
was approximately 2.5 mm.

The calorimeter provided the energy measurement for electrons, photons, 
charged hadrons, and jets.
The energy resolution of the D\O\ calorimeter was measured in a test
beam for electrons and pions.
The energy resolution for electrons and photons was 
$\sigma_{E}/E = 15\%/\sqrt{E({\rm GeV})}\oplus 0.4\%$.  
For charged pions and jets the resolutions were
approximately $50\%/\sqrt{E({\rm GeV})}$ and $80\%/\sqrt{E({\rm GeV})}$,
respectively \cite{d0nim,NIMTB}.
The transverse energy of a neutrino was inferred from the undetected
transverse energy, \hbox{$\rlap{\kern0.25em/}E_T$},
which is the negative of the vector sum of all the
transverse energies in the calorimeter cells.
Using a minimum bias data sample, the resolution for each
component of the missing transverse energy, 
\hbox{$\rlap{\kern0.25em/}E_x$} and \hbox{$\rlap{\kern0.25em/}E_y$}, 
was measured to be $1.1\;{\rm GeV} + 0.02(\sum E_T)$,
where $\sum E_T$ is the scalar sum of transverse energies
in all calorimeter cells.
For one analysis (see Section~\ref{sec-zgvvg}) the 
\hbox{$\rlap{\kern0.25em/}E_T$} was calculated from the negative of the vector 
sum of the transverse energies in towers with $E_T>200 (400)$ MeV in the 
EM (FH1) calorimeters.  The hadronic calorimeter scale was determined by 
comparing the transverse energy of the recoil against that of the electron 
pair in $Z\rightarrow ee + X$ events. The resolution of the $x$ and $y$ 
components of the \hbox{$\rlap{\kern0.25em/}E_T$} was 
$\sigma = \sqrt{(4.88 \, {\rm GeV})^2 + (1.34 \cdot P_T^{\rm recoil})^2}$.

The muon spectrometer system, shown in Fig.~\ref{fig-D0},  
consisted of solid-iron toroidal magnets and
sets of proportional drift tubes (PDT's). It provided identification of muons 
and determination of their trajectories and momenta.  Since muons from decays
of $W$ and $Z$ bosons are primarily in the central region, the analyses 
presented here used only the wide angle muon spectrometer (WAMUS) which
subtended the region $|\eta| \leq 2.5$. The WAMUS system consisted of three 
layers: the ``A-layer" with four planes of PDT's,
located between the calorimeter and the toroid magnets; and 
the ``B-" and ``C-layers" each with three planes of PDT's,
located outside the toroid magnets.  The toroids 
were magnetized to $\pm 1.9$ T.  The wires in the drift tubes and 
the magnetic field 
lines in the toroids were oriented transversely to the beam direction. 
The muon system mounted on the central (forward) muon toroid, covering 
approximately $|\eta|<1$ $(1<|\eta|<1.7)$ is referred to as the ``CF (EF)" 
region.

The total material in the calorimeter and iron toroids varied from 
13 to 19 interaction lengths, making background from hadronic punchthrough
negligible.  Because of the small tracking volume, the background to prompt 
muons from in-flight decays of $\pi$ and ${\rm{K}}$ mesons was also 
negligible.

The muon momentum $p$ was determined from its deflection angle in the magnetic
field of the toroid.  The momentum resolution was limited by multiple 
scattering in the calorimeter and toroid, knowledge of the magnetic field 
integral, and the accuracy of the deflection angle measurement.  The momentum 
resolution, determined from $J/\psi\rightarrow \mu\mu$ and 
$Z\rightarrow \mu\mu $ events,
was $ \sigma (1/p)= 0.18 (p-2)/p^2 \oplus 0.008  $  ($p$ in ${\rm  GeV/c}$),
where $\oplus$ indicates addition in quadrature.

 \subsection{Trigger}
A multi--level, multi--detector trigger system was used for selecting 
interesting events and recording them to tape.
A coincidence between hits in two hodoscopes 
of scintillation counters (level 0), centered around the beampipe,
was required in order to register 
the presence of an inelastic collision.
These counters also served as the 
luminosity monitor for the experiment.  
The level 1 and level 1.5 triggers were 
programmable hardware triggers which made decisions based on combinations of 
detector-specific algorithms.  The level 2 trigger was a farm of 48 Vax 
4000/60 and 4000/90 computers which filtered the events based on 
reconstruction of the information available from the front-end electronics.
Losses from the Main Ring beam, usually involved in the production
of antiprotons, 
caused backgrounds in the muon system and calorimeter. In the analyses 
presented here, triggers which occurred at the times when a Main 
Ring proton bunch passed through the detector were not used.
Similarly, triggers which occurred during the first 0.4 seconds of the 
2.4 second antiproton production cycle 
were vetoed. These ``Main Ring vetoes"
accounted for approximately 25\% trigger deadtime. 

At each level of the trigger, the D\O \ trigger system gathered the 
results from each of the detector-specific triggers and filters. In this
way, trigger decisions could be made from combinations of different 
detector-specific results.  Table~\ref{tab-trigs} is a compilation of the
triggers used in the various analyses presented in this paper.

\subsubsection{Calorimeter Trigger}
The level 1 triggers for electromagnetic showers were based on analog sums of 
transverse energy in calorimeter towers with  
$\Delta \eta \times \Delta \phi = 0.2 \times 0.2$
and with two longitudinal sections, EM and FH. 
The level 1 EM trigger required transverse energy in the EM section of a 
trigger tower to be above preselected thresholds.  
The level 2 EM algorithm identified electrons and photons by forming clusters,
around level 1 trigger towers, of transverse energy read out from the four 
layers of the EM calorimeter and the first layer of the FH calorimeter. 
The clusters were of size $\Delta \eta \times \Delta \phi = 0.3 \times 0.3$,
centered on the highest $E_T$ tower in the cluster.
The longitudinal and transverse profile of the cluster had to satisfy
the following requirements which were designed to discriminate electrons
and photons from hadronic showers.
The fraction of the cluster energy in the EM
section had to exceed a value which depended on the energy and
location of the cluster in the calorimeter.
The transverse shape classification was based on the energy deposition 
pattern in the third EM layer.  The difference of the energy depositions in two 
regions, of size $\Delta \eta \times \Delta \phi = 0.25 \times 0.25$ and 
$0.15 \times 0.15$ and centered around the cell with highest $E_T$, had to be
within a range which depended on the total cluster energy.

Another calorimeter based trigger was the ``Missing $E_T$" trigger.
At level 2 the \hbox{$\rlap{\kern0.25em/}E_T$} was formed from the 
negative of the vector
sum of the $E_T$ deposited in the calorimeter and ICD cells, corrected 
for the vertex position.

\subsubsection{Muon Trigger}
The muon level 1 and level 1.5 triggers required coincidences of hits in the 
PDT's consistent with a muon originating from the collision region.  The 
level 1 algorithm combined coincidences of hits in PDT cells into 60 cm-wide 
hodoscopic elements. 
If a combination of hodoscopic elements matched 
a preprogrammed pattern of a muon track, the event was accepted.  
In the central region, three layers of PDT's, each with at least 
two hit planes, were required except in regions where detector 
services and support limited the coverage of the one of the layers.  
In the forward region,
defined approximately as $1.0 \leq |\eta| \leq 2.5$, three layers of PDT's 
were required, with at least three hit planes in the A-layer and two hit planes 
in both the  B- and C-layers. The trigger required a minimum $p_T$ of 
$3$ GeV/c and became fully efficient at about 6 GeV/c.
The level 1 trigger efficiency was
$(79 \pm 3) \%$ for the central region and $(36 \pm 12) \%$ for the 
forward region.  
At level 1.5, the hodoscopic elements had half-cell resolution, 
providing a sharper $p_T$ turn-on.  A three-layer requirement made at this
trigger level reduced the acceptance of the central muon system by 
approximately 15\%.

At level 2, the first stage of the muon reconstruction algorithm,
which consisted of the pattern recognition and initial track fit,  was 
performed.
To minimize processing time, the search for muon candidates
in the forward region was restricted to the sectors which had a level 1 
trigger.
A valid level 2 trigger was a three-dimensional muon track with
hits in at least two planes of two PDT's.
The level 2 muon trigger program calculated several quantities that
provided information on the quality of the muon track including:
the goodness of track fit in the PDT drift view and along the PDT wire, 
the projections of the track to the interaction point in both views and
the number of hits used to fit the track. 
A track quality variable was defined as the number of these quantities
that failed the standard criteria.  In addition,
in the forward region, muon candidates formed with less than six hits
on the track were discarded, since they were likely to be random background
hits in coincidence. 
In the central region, cosmic ray muons were identified if there was 
evidence of a single muon penetrating the entire detector;  muon candidates 
with a track within $20 \deg$ in $\phi$ and $5\deg$ in $\theta$ or hits 
within 60 cm (roughly $5 \deg$) of the projection of the muon track into the 
opposite side PDT's were rejected.
The muon was accepted by level 2 if the $p_T$ was above the desired threshold 
and if the track quality variable was zero (``tight" standards),
or one (``loose" muon standards). 
The muon level 2 trigger efficiency was determined to be $(95 \pm 3) \%$ 
excluding effects of the chamber efficiencies and geometrical acceptance
for the ``loose" muon requirements of the MU-ELE trigger (see 
Table~\ref{tab-trigs}).
\vbox{
\begin{table}
\begin{tabular}{|c|c|c|c|c|}
Trigger Name  & Level 1 & Level 1.5 & Level 2 & Analyses \\ \hline \hline
\small{MU-MAX}
              &1 $\mu$, $|\eta| \leq 1.7 $
                        &1 $\mu$, $|\eta| \leq 1.7$
                                    &1 $\mu$ (tight), $p_T\geq 15$ GeV
                                              &$WW\rightarrow \mu\mu$ \\ \hline
\small{MU-ELE}&1 EM tower, $E_T \geq 7$ GeV 
                        & -         & 1 e or $\gamma$, $E_T\geq 7$ GeV  
                                              &$WW\rightarrow e\mu$ \\
              &1 $\mu$, $|\eta|\leq 1.7$
                        &           & 1 $\mu$ (loose), $p_T \geq 5$ GeV/c
                                              &$W\gamma \rightarrow \mu \nu 
                                                 \gamma$            \\
              &         &           &         &$Z\gamma \rightarrow \mu \mu
                                                 \gamma$            \\ \hline
\small{ELE-HIGH}
              &1 EM tower, $E_T \geq 14$ GeV
                        & -         &1 e or $\gamma$, $E_T\geq 20$ GeV
                                              &$WW/WZ \rightarrow$ \\ 
              &         &           &         &$ e\nu$ jet jet \\
              &         &           &         &$Z\gamma\rightarrow\nu\nu
                                                   \gamma$     \\  \hline 
\small{ELE-MAX}
              &1 EM tower, $E_T \geq 10$ GeV
                        & -         &1 e or $\gamma$, $E_T\geq 20$ GeV 
                                              &$W\gamma \rightarrow e \nu
                                              \gamma$                     \\
              &         &           & \hbox{$\rlap{\kern0.25em/}E_T$} $\geq
                                      20$ GeV &                \\ \hline
\small{ELE-2-HIGH}
              &2 EM towers, $E_T \geq 7$ GeV
                        & -         &2 e and/or $\gamma$, $E_T\geq 10$ GeV
                                              &$WW\rightarrow ee$ \\ \hline
\small{ELE-2-MAX}
              &2 EM towers, $E_T \geq 7$ GeV
                        & -         &2 e and/or $\gamma$, $E_T\geq 20$ GeV
                                              & $Z\gamma \rightarrow ee 
                                                 \gamma$            \\ 
\end{tabular}
\caption{Triggers used in the analyses presented in this paper.
The \small{ELE-2-MAX} trigger
was a subset of the \small{ELE-2-HIGH} trigger which included shower shape cuts
on the EM candidates. }
\label{tab-trigs}
\end{table} }

\subsection{Muon Identification}  
Muons were identified as tracks in the muon PDT's
in association with tracks in the CD and energy deposits in the calorimeter.
The momentum of the muon was computed from the deflection of the 
track in the magnetized toroid.  The track fit used a least-squares 
calculation which considered seven parameters: four describing the position
and angle of the track before the calorimeter, two describing
the effects due to multiple Coulomb scattering, and the inverse of the 
muon momentum, $1/p$. This seven parameter fit was applied to 16 data points:
vertex position measurements in the $x$ and $y$ directions, the angles
and positions of the track segments before and after the calorimeter and 
outside of the toroid magnet, and two angles representing the multiple 
scattering of the muon in the calorimeter.  The fit determined the charge of
the muon and which CD track, if any, matched the muon.  
The muon momentum was then 
corrected for the energy lost in the calorimeter using a 
{\small GEANT}-based~\cite{GEANT} detector model.

In the following, the quantities used to describe the muon
tracks are presented.  The definitions for muons differ slightly
among the various analyses because of the nature and magnitude of the 
backgrounds. 
Table~\ref{tab-muonid} 
lists the five different definitions of muons in the analyses described in
this paper.

\subsubsection{Muon Track Quality}
\begin{figure}
\vspace{-2.25in}
\epsffile{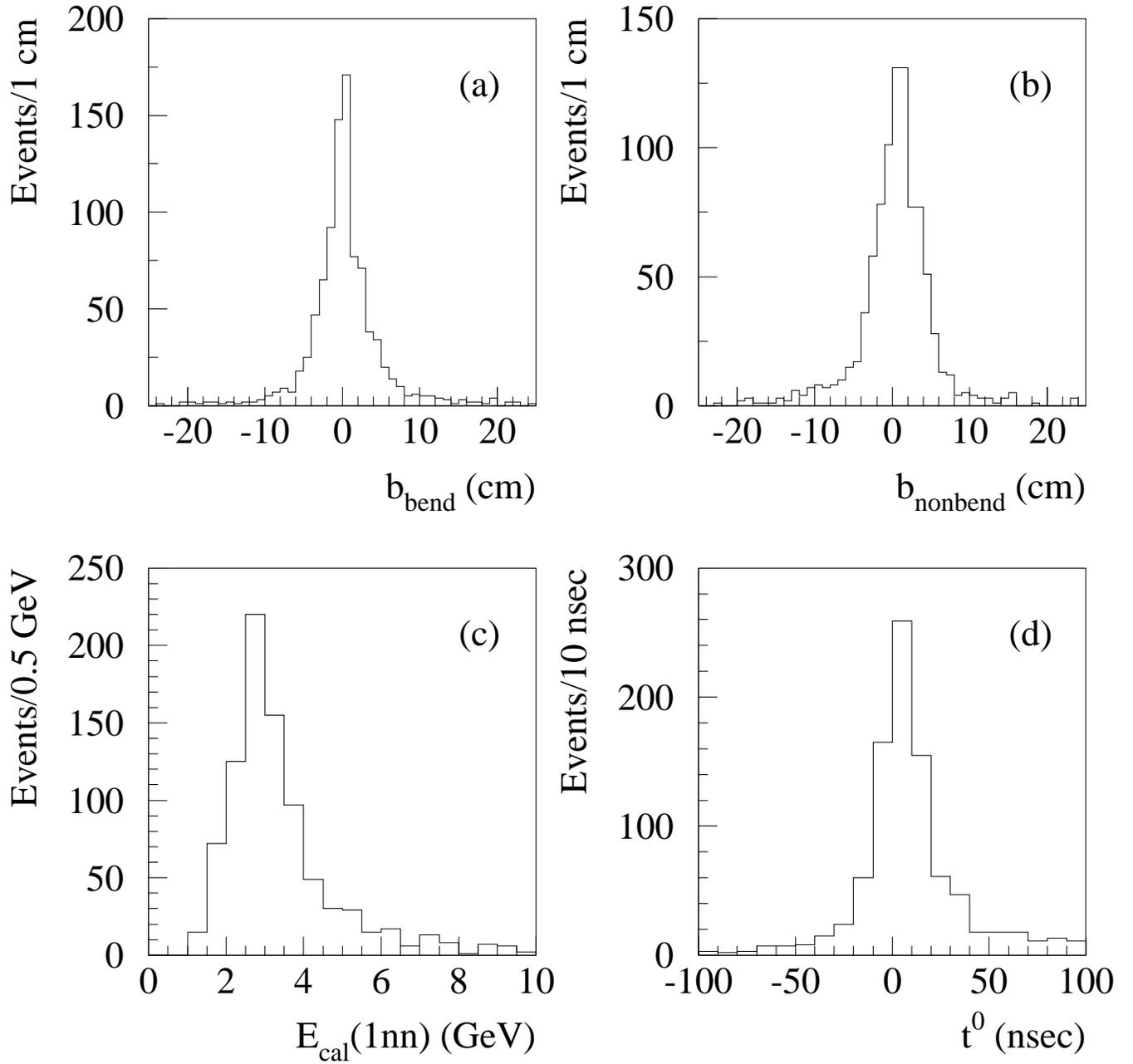}
\caption{Muon selection variables including (a) the bend view impact parameter,
(b) the nonbend view impact parameter, (c) the energy in the 
 calorimeter tower plus the nearest neighboring towers 
around the muon, and (d) the $t_0$
 resulting from the track fit with the muon time-of-origin as a parameter. }
\label{fig-muid}
\end{figure}

The muon reconstruction algorithm defined a muon track quality, similar
to that used in the level 2 trigger, which contained information 
about the number of hits on the track from each layer of muon PDT's, the
track impact parameters, and the goodness of the track fit.  
If the track did not satisfy criteria on more than one of the above 
quantities, the muon candidate was rejected.  
Figures~\ref{fig-muid}(a) and (b) show the impact parameters
in the track bend view ($r-z$ plane), $b_{\rm bend}$, and in the 
track nonbend view ($x-y$ plane), $b_{\rm nonbend}$,
for muons which satisfied all of the other selection criteria.  
The 3D impact parameter was sometimes
used in lieu of the combination of the $r-z$ and $x-y$ selection criteria.

\subsubsection{Fiducial Requirements}
Muons which passed through the region between the CF and EF toroid
magnets near $|\eta| \approx 0.9$ may have traversed a smaller amount 
of magnetized iron and thus have a reduced momentum resolution.
To reject these poorly measured muons, all of the muon identification
definitions in this paper except ``Loose II" required the minimum magnetic
field integral along the muon track, $\int B dl$, to be at least $ 2.0$
Tesla-meters.

The ``Tight II", ``Tight III", and ``Loose II" definitions required that the 
muon have hits in the A-layer, between the calorimeter and the toroid magnet.  
Making this requirement reduced the fake-muon background in the forward region
but also reduced the acceptance by limiting the pseudorapidity coverage to 
approximately $|\eta|<1.7$.  A cut on $|\eta| < 1.7$ was used to restrict 
the muon tracks to those totally contained within the WAMUS spectrometer.

\subsubsection{Central Detector Track-Match} 
Muon candidates were required to have a confirming track in the CD
within a range in both the polar and azimuthal angles.
This reduced the backgrounds from cosmic ray muons and from combinations
of random hits.

\subsubsection{Calorimeter Confirmation}
Muons deposited energy in the calorimeter as they passed through it. 
It was required that at least $ 1$ GeV of energy was deposited in 
the tower which contained the projection of the muon track through the 
calorimeter and the nearest neighboring towers. 
Figure~\ref{fig-muid}(c) shows the energy deposition in these towers around
muons which passed the other selection requirements.

\subsubsection{Cosmic Ray Identification}
The muon track was refitted with the timing of the muon track with respect
to the $p\bar{p}$ interaction as a floating parameter, $t_0$.
This allowed identification of cosmic ray muons. 
Figure~\ref{fig-muid}(d) shows the value of $t_0$ which resulted in the best 
track-fit for muons which passed all other selection requirements.
The ``Tight I" and ``Tight II" definitions required $t_0 < 100$ ns.

\subsubsection{Muon Isolation}
Muons from the decay of pions, kaons, and heavy quarks were reduced by
requiring that the muon be isolated from other jet activity.  This was done
in three ways.  One isolation variable (2NN) was defined by summing the 
energy deposited in the calorimeter cells hit by the muon and two nearest 
neighbors, subtracting the energy expected to have been deposited by the muon,
and dividing the difference by the uncertainty.  This was required to 
be less than five standard deviations.  Another isolation variable (Halo) 
was defined as the difference between the energy deposited in a cone of size
$\Delta  R = 0.6$ and the energy deposited in a cone of size $\Delta R = 0.2$,
where $\Delta R = \sqrt{ \Delta \eta ^2 + \Delta \phi ^2}$,
around the muon in the calorimeter.  This was required to be less than 
$8$ GeV. The third isolation criterion ($\Delta R_{\mu-{\rm{jet}}}$ )
was that muons were spatially separated 
from the axis of any jet with $E_T \geq 10$ GeV by at least 
$\Delta  R = 0.5$. 
\vbox{
\begin{table}
\begin{tabular} {|c|c|c|c|c|c|}
Selection            & Tight I  & Tight II & Tight III & Loose I & Loose II
                                                            \\ \hline \hline
Analysis             & $WW\rightarrow \mu\mu\nu\bar{\nu}$
                                &$W\gamma \rightarrow \mu\nu \gamma$
                                           &$Z\gamma \rightarrow \mu\mu \gamma$
                                                       &$WW\rightarrow e\mu
                                                        \nu \bar{\nu}$
                                                                 & 
                                $Z\gamma \rightarrow \mu\mu \gamma$ \\ \hline
Muon Quality         &$\surd$   & $\surd$  & $\surd$   & $\surd$ & $\surd$
                                                                     \\ \hline
Back-to-Back         &          &          &           &         &   \\
   Muons Removed     &$\surd$   & $\surd$  & $\surd$   & $\surd$ & $\surd$
                                                                     \\ \hline
Minimum              &          &          &           &         &   \\ 
Field Integral (T-m) & 2.0      &  2.0     & 1.9       & 2.0     & - \\ \hline 
3-Layers Required    & -        & -        & $\surd$   &  -      & - \\ \hline
A-Layer Required     & -        & $\surd$  & $\surd$   &  -      & $\surd$
                                                                     \\ \hline
Isolation            & 2NN Cut  & $\Delta R_{\mu-{\rm{jet}}}$
                                           & $\Delta R_{\mu-{\rm{jet}}}$
                                                       & $\Delta R 
                                              _{\mu-{\rm{jet}}}$ &
                                         $\Delta  R_{\mu-{\rm{jet}}}$  \\ 
Requirement          & Halo Cut &$>0.5$    &$>0.5$     &$>0.5$   &$>0.5$
                                                                   \\ \hline
Impact Parameter     & $|3D| \leq 22$ cm
                                & $|RZ| \leq22$ cm  
                                           & $|RZ| \leq22$ cm
                                                       &$|3D|\leq$ 22cm 
                                                                 &$|RZ|\leq$
                                                                 22cm  \\
                     &          & $|XY|\leq$ 15 cm  
                                           &          &          &   \\ \hline
t$_0$                & $\leq 100$ ns
                                & $\leq 100$ ns
                                           & -        & -        & - \\ \hline
CD Match             & $\delta \phi \leq 0.45$
                                &$\delta \phi \leq 0.25$
                                           &$\delta \phi \leq 0.25$
                                                       &$\delta \phi \leq 0.45$
                                                                 &
                                                     $\delta \phi \leq 0.25$   
                                                                     \\
                     &$\delta \theta \leq 0.45$ 
                                &$\delta \theta \leq 0.30$
                                           &$\delta \theta \leq 0.30$
                                                       &$\delta \theta \leq 
                                                              0.45$ 
                                                                  &
                                                      $\delta \theta \leq 0.30$
                                                                     \\ \hline
Cal Confirm          &$\surd$   &$\surd$   &$\surd$    & $\surd$  & $\surd$ \\
\end{tabular}
\caption{Summary of the  various muon identification definitions used in 
the analyses presented in this paper.  }
\label{tab-muonid} 
\end{table}}

\subsection{Electron and Photon Identification}
\begin{figure}
\vspace{-2.5in}
\centerline{\epsffile{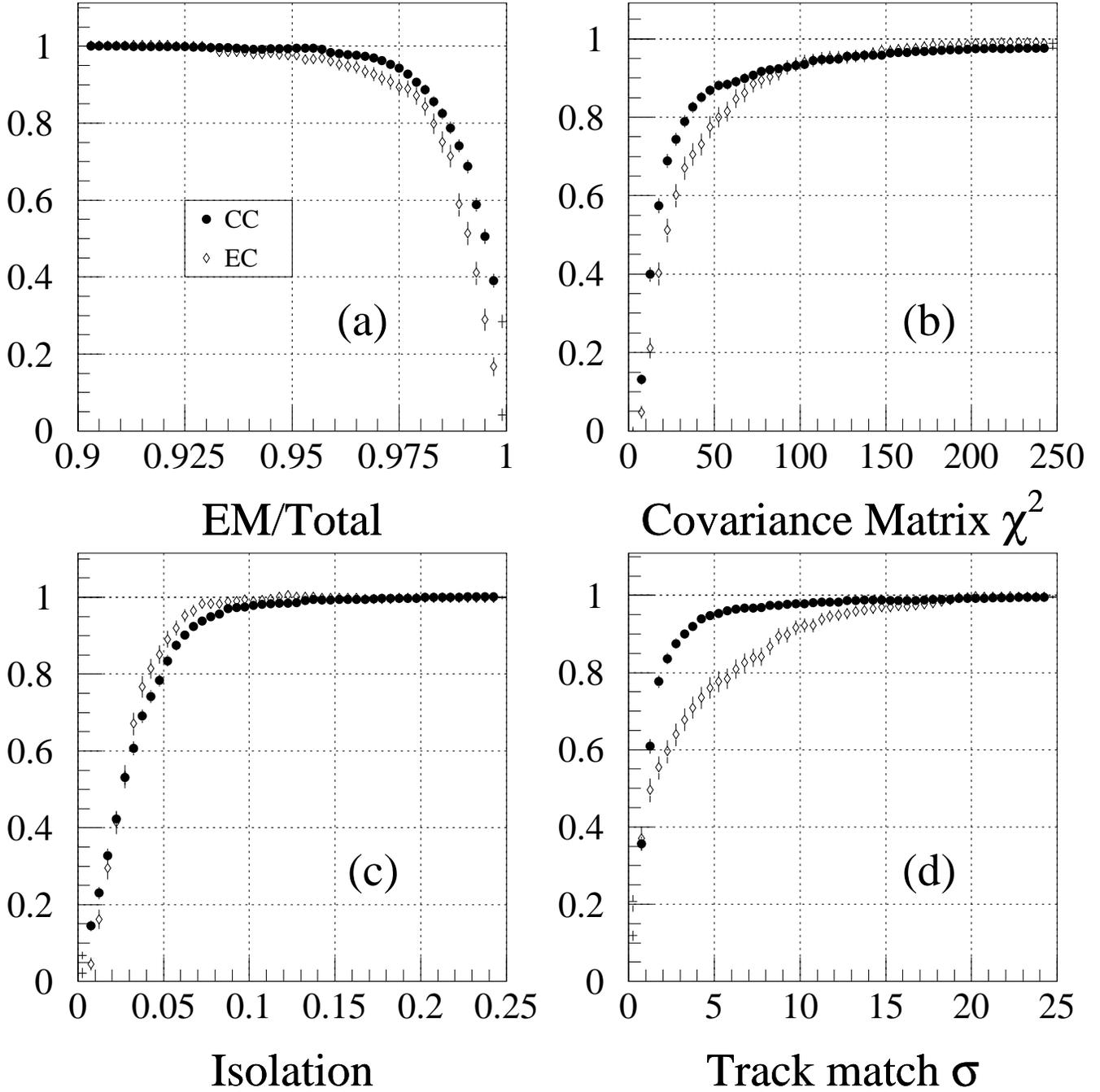}}
\caption{Electron selection efficiencies for 
 (a) the electron selection efficiency as a function of 
the fraction of the energy deposited in the EM calorimeter,
(b) the H-matrix $\chi^2$, (c) the isolation variable $f_{iso}$, and 
(d) the track-match-significance, $TMS$. The solid circles are for CC
electrons and the open diamonds are for EC electrons. }
\label{fig-emvars}
\end{figure}

Electrons and photons were identified by the properties of the shower 
in the calorimeter and the presence, or lack thereof, of a matching track
in the CD.
Using a nearest-neighbor algorithm, 
clusters were formed  from adjacent EM 
towers containing significant energy deposition.  The clusters for which
the energy in the EM and first FH section of the calorimeter 
divided by the sum of the energies in the EM and all hadronic sections
(EMF) was greater 
than or equal to $0.9$ were flagged as possible electrons or photons.  
Figure~\ref{fig-emvars}(a) shows the fraction 
of electrons from Z boson decays for which the EMF is above the value given on
the abscissa.  More detailed analysis of the calorimeter and tracking
chamber information was then used to refine the sample as is described below. 
A summary of electron definitions is presented in 
Table~\ref{tab-emid}. The photon definitions are summarized in 
Table~\ref{tab-phoid}.

\subsubsection{Fiducial Coverage of EM Calorimeter}
\label{sec-emfid}
All of the analyses presented in this paper made identical selection
on the fiducial coverage of the EM calorimeter.  It was required that 
the electron or photon have pseudorapidity  within the range $\pm 2.5$.
Furthermore, the EM calorimeter had a gap in the coverage at the 
transition between the CC and the EC.  The four longitudinal layer coverage 
of the CC ended at pseudorapidity of 1.1 and the four longitudinal layer 
coverage of the EC returned at pseudorapidity of 1.5.  Therefore, all 
analyses of photons  and electrons in this paper made a fiducial selection 
which removed this transition region. 
The ``Tight" photon identification criteria went one step further,
requiring that photons in the CC have pseudorapidity within the range
$\pm 1.0$. The samples used in the discussion of photon and electron
identification presented below have these fiducial selections already applied.
  
\subsubsection{Covariance Matrix $\chi^2$}
The electron or photon shower shape was characterized by the fraction
of the cluster energy deposited into each layer of the calorimeter.  
These fractions were correlated, depending on the depth of the 
start of the shower and on the energy of the incident particle.  
In order to reject background using the shower shape, including these
correlations, a comparison was made 
between the candidate and a reference sample of Monte Carlo electrons with 
energies ranging from 10 to 150 GeV.  This comparison (H-matrix $\chi^2$) 
was carried out in 41 observables: the fractional energies in layers 
1, 2, and 4 of the EM calorimeter; the fractional energy in each cell of 
a $6\times 6$ array of cells in layer 3 centered on the most energetic tower 
in the EM cluster; the logarithm of the total 
energy of the electron cluster, taking into account the depth 
dependence on the incident energy; and the position of the event vertex 
along the beam direction, taking into account the dependence of the shower 
shape on the incident angle.  A separate reference shower shape 
was available as a function of $\eta$, assuming $\phi$ symmetry.  
Figure~\ref{fig-emvars}(b)  shows the fraction of electrons from Z boson
decays for which the value 
of the H-matrix $\chi^2$ is less than the value given on
the abscissa. Requiring that the H-matrix $\chi^2 < 100 (200)$ in the CC (EC)
gave an efficiency of $94.9\pm0.8\%$ ($100.0^{+0.0}_{-1.0}\%$)
for electrons with $E_T > 25$ GeV.

\begin{figure}
\centerline{\epsffile{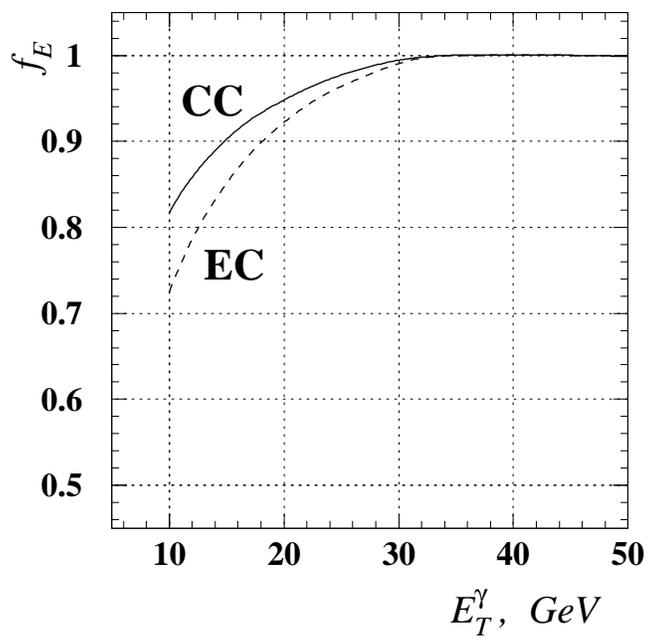}}
\vspace{-3.0in}
\caption{The efficiency of the ``Loose" photon H-matrix $\chi^2$ selection 
criteria as a function of $E_T^{\gamma}$. }
\label{fig-HMeffy}
\end{figure}

The efficiency for the H-matrix 
selection decreased if the $E_T$ fell below 25 GeV. 
The efficiency as a function of photon $E_T$ was measured in a test beam for
both the CC and EC.  This dependence was a dominant source of systematic 
uncertainty in the efficiency for low $p_T$  photons. 
 Figure~\ref{fig-HMeffy} shows the 
efficiency versus $p_T$ for the H-matrix selection criteria for low $p_T$ 
``Loose" photons.  

\subsubsection{Cluster Isolation}
The EM clusters were required to be isolated from other particles in the 
event in order to reduce the background from hadronic jets with high
EM content.  The isolation variable was
\begin{equation}
{\rm{f_{iso}}} = \frac{E(0.4)-EM(0.2)}{EM(0.2)},
\end{equation}
where $E(0.4)$ was the energy deposited in all the calorimeter cells in a 
cone of radius $ R = 0.4$ around the electron or photon 
and $EM(0.2)$ was the 
energy deposited in the EM calorimeter in a cone of radius $ R=0.2$. 
For EM objects with 
$E_T<20$ GeV, there was deterioration of the efficiency of the $f_{iso}$
selection criteria.  This was modeled with a turn-on curve in a way 
similar to the H-matrix efficiency described above.
Figure~\ref{fig-emvars}(c) shows the fraction of electrons from Z boson
decays passing an $f_{iso}$ selection criterion.
Requiring ${\rm{f_{iso}}}<0.10$ was $97.6\pm0.6\%$ ($98.5\pm1.4\%$) efficient
for CC (EC) electron candidates.  

\subsubsection{Electromagnetic Fraction}
The ``Tight" photon identification criteria included the requirement 
that the energy deposited in the four EM layers be at least $96\%$ of 
total energy in the calorimeter in a cone around the shower maximum. 
This was in addition to the EMF requirement discussed above.

\subsubsection{Electron Track Match}
\label{sec-tms}
Electrons and photons were distinguished from each other by the presence of a 
track consistent with the passage of a charged particle in the CD which
pointed to the EM cluster in the calorimeter.
An electron had such a track; a photon did not. The efficiency for 
track-finding was $86.7\pm1.4\%$ in the CDC and $86.1\pm1.8\%$ in the FDC's.
By demanding a good spatial match between the 
cluster and the track, backgrounds due to accidental overlaps of charged 
particles with photons in EM jets were reduced.  
In the calorimeter, the 
shower centroid, $\vec{x}_c$, was determined  from the weighted mean of the 
coordinates $\vec{x}_i$  of all cells containing the shower,
$\vec{x}_c = \sum_i w_i \vec{x}_i/ \sum_i w_i$.
The weights were defined as $w_i = {\rm max}(0,w_0+ln{\frac{E_i}{E}})$,
where $E_i$ was the energy in the $i$th cell, $E$ the energy of the cluster,
and $w_0$ a parameter chosen to optimize the position resolution. The
logarithmic weighting was motivated by the exponential lateral profile of an 
electromagnetic shower.  The azimuthal position matching resolution in 
the CC and EC was measured to be $\approx 2.5$ mm. 
The CD track was extrapolated to the shower centroid and the significance of 
the track match, $TMS$, was formed between the position of the track and the 
centroid. For the CC this quantity was 
\begin{equation}
TMS_{CC} = \sqrt{(\frac{\Delta \phi}{\delta_{\Delta\phi}})^2 +
                (\frac{\Delta z }{\delta_{\Delta z}})^2 },
\end{equation}
where $\Delta \phi$ was the azimuthal mismatch, $\Delta z$ the mismatch 
along the beam direction, and $\delta_x$ was the resolution for the 
observable $x$. For the EC, $\Delta z$ was replaced by $\Delta r$, the 
mismatch transverse to the beam.  
Figure~\ref{fig-emvars}(d) shows the fraction of electrons from $Z$ boson
decays for which the track match significance variable is less than the
criterion on the abscissa. Requiring 
$TMS \leq 10$ was $98.0\pm0.6\%$ ($91.5\pm1.8\%$) efficient for CC (EC) 
electron  candidates. 

\subsubsection{Electron Track Ionization}
\begin{figure}
\centerline{\epsffile{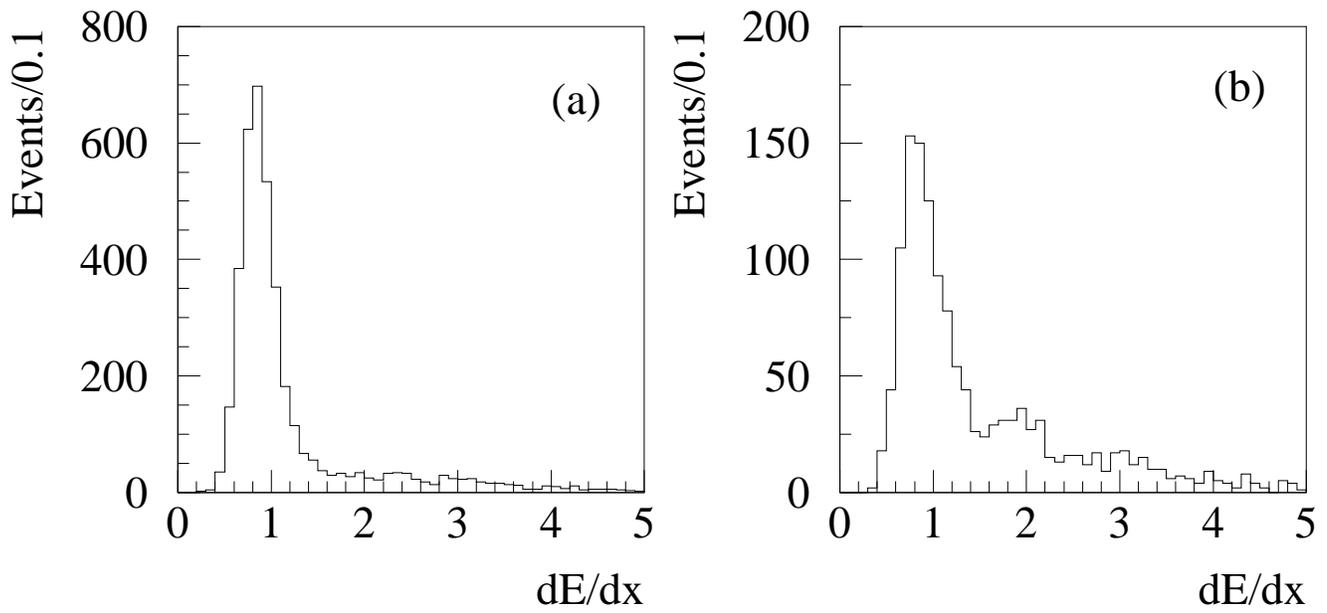}}
\caption{(a) the $dE/dx$ measured in the CDC for electrons from Z boson 
decays. (b) the $dE/dx$ measured in the CDC for EM clusters in an inclusive 
jet sample. }
\label{fig-dedx}
\end{figure}

The tracks from $e^+e^-$ pairs produced in photon 
conversions due to interactions with 
material in the tracking chambers were often reconstructed as a single track. 
For such pairs, the ionization in the tracking chambers was expected to be
twice that of a single charged particle.  The distribution of ionization
per unit length ($dE/dx$) for electrons from $Z\rightarrow ee$ decays and 
from EM clusters in an inclusive jet sample are shown in 
Figs.~\ref{fig-dedx}(a) and \ref{fig-dedx}(b). 
Most electrons had $dE/dx \approx 1$.
The ionization in the inclusive jet sample shows a two-peaked structure.
The lower peak, at $dE/dx \approx 1$ was due to single charged particles.
The higher peak came from unresolved $e^+e^-$ pairs.  This background was
rejected by removing electron candidates with $dE/dx \sim 2$. 
The veto requirement for CC(EC) was
$1.6 \leq dE/dx \leq 3.0 \; (1.6 \leq dE/dx \leq 2.6)$ and was 
$94.4 \pm 1.1 \; (75.2 \pm 3.7)$\% efficient for electrons from $Z$ bosons.

\subsubsection{Loss of Photons due to Track Overlaps}
Some photons were mislabeled as electrons because of spatial overlap of the 
photon with a random track. The inefficiency introduced 
was estimated by looking for a track or tracks in a cone randomly oriented
in $\phi$ but at the same $\eta$ location as the electrons in $Z$ boson 
decays to $ee$.  The assumption is that the probability of finding such a 
track or tracks at a given $\eta$ is the same in $Z$ boson production and 
double vector boson production.  
The probability for a random track overlap was found to be $6\pm1\%$ 
and $15\pm1\%$ for the CC and EC, respectively, the latter being higher due 
to the higher density of tracks in the forward direction. 

\subsubsection{Photon Conversions}
\begin{figure}
\epsffile{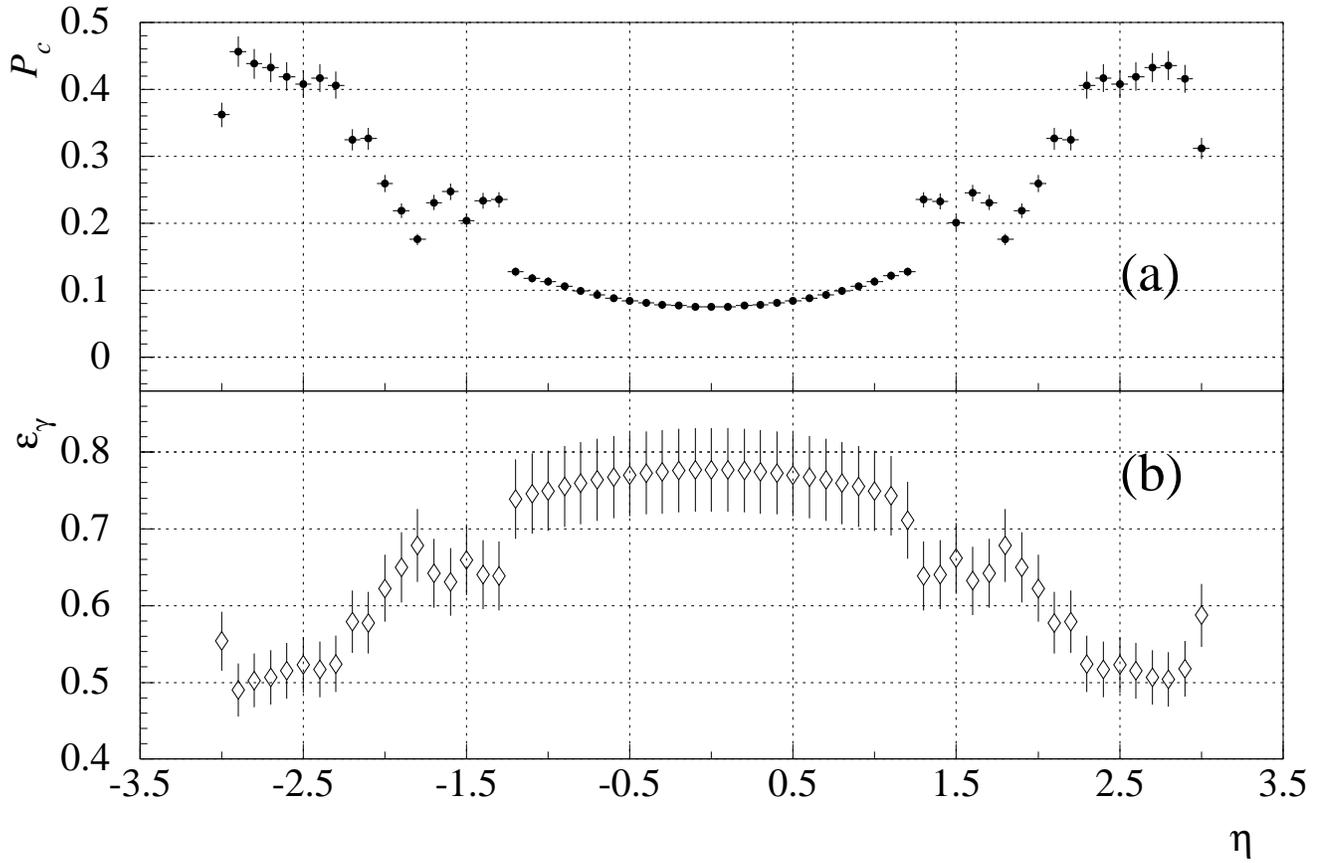}
\caption{(a) $\eta$-dependent probability for photon conversion ($P_c$) in the
material in front of the CDC and FDC ; (b) $\eta$-dependent efficiency of the 
photon identification ($\epsilon_{\gamma}$) for high $p_T$ photons
for the $Z\gamma$ ``Loose" selection criteria.  
The uncertainty shown includes the statistical uncertainty plus a common 
systematic uncertainty  of 5\%.}
\label{fig-phoconv}
\end{figure}

Some photons were lost when they converted to ${e^+e^-}$ pairs in material
in front of the CDC or FDC. The conversion probability was calculated using
the {\small{GEANT}} simulation of the D\O\ detector.  This probability 
depended on the pseudorapidity of the photon and is shown in 
Fig.~\ref{fig-phoconv}. Averaged over the CC (EC) it amounted to a 
10\% (26\%) loss of photons. There is a  systematic uncertainty of 5\%.

\subsubsection{Photon-Vertex Projection}
\label{sec-emvtx}
\begin{figure}
\epsffile{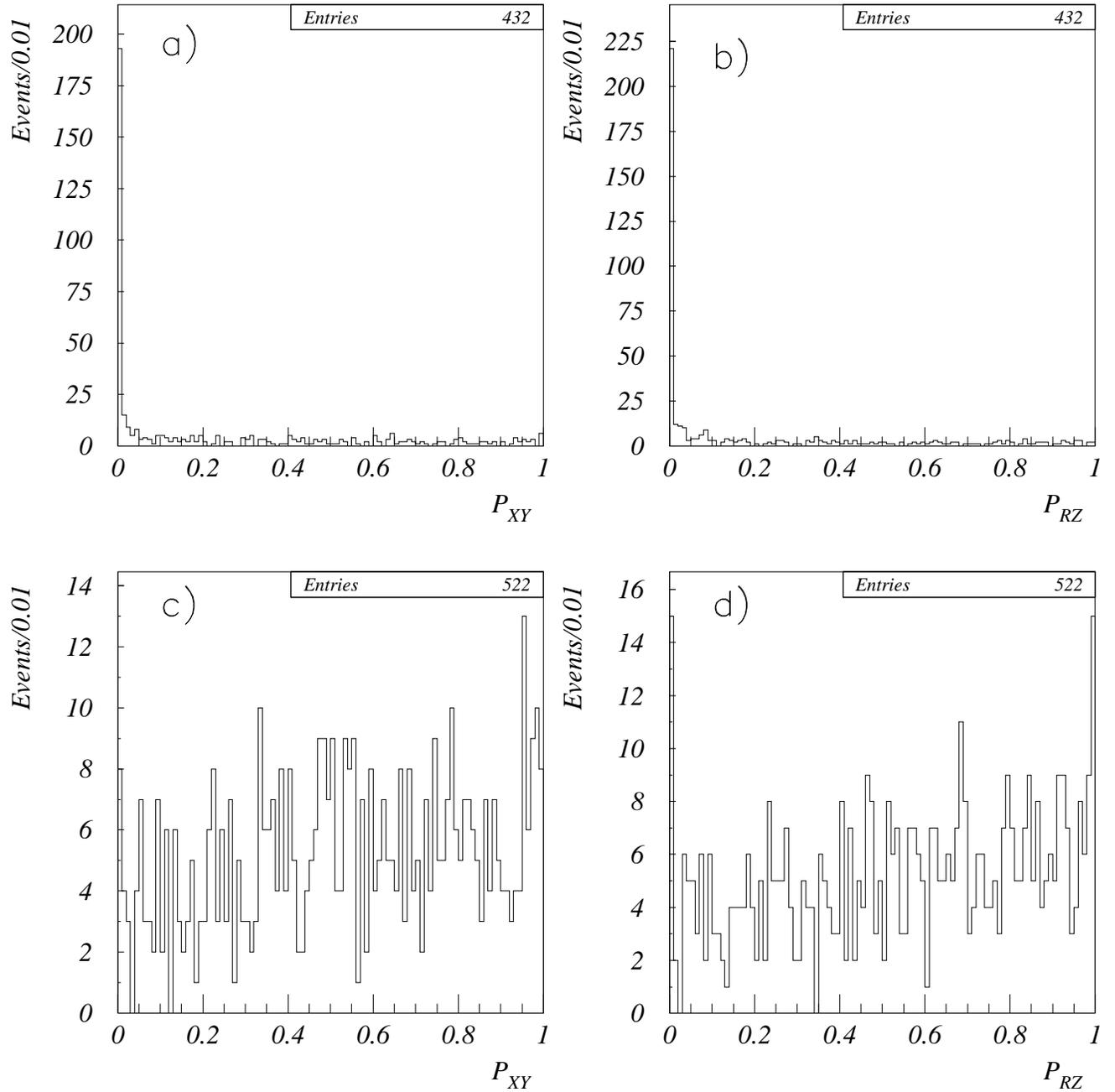}
\caption{Probability distributions $P_{XY}$ and $P_{RZ}$. 
(a) and (b) are from photons resulting from  cosmic ray bremsstrahlung. 
(c) and (d) are from  electrons from $Z$ boson decays.  }
\label{fig-EMVTX}
\end{figure}

An algorithm, {\small EMVTX}, was developed to reduce the  background from
cosmic ray or beam-related muon bremsstrahlung which produced photons 
inconsistent with having originated at the event vertex.  
The energy-weighted centers of 
the cluster in each of the four layers of the EM calorimeter plus the vertex 
position, and their uncertainties, were used to compute two dimensional fits.
The resulting $\chi^2$ was then converted into a probability for the 
photon to have originated at the vertex. It was required that the probability 
of the $RZ$ and $XY$ projections, $P_{RZ}$ and $P_{XY}$, each exceed $1\%$.  
Comparison of the $P_{RZ}$ and $P_{XY}$ distributions for electrons from 
$Z$ bosons and from photons resulting from cosmic ray bremsstrahlung are 
shown in Fig.~\ref{fig-EMVTX}.  In case there were multiple vertices
in the event, the one with the highest $P_{RZ}$ was selected as the 
vertex for the interaction. This vertex was then used in computing 
the missing transverse
energy and $E_T^{\gamma}$.  The vertex resolution provided by this 
algorithm was approximately 17 (11) cm in the $RZ$ $(XY)$ planes. 

\subsubsection{Hits Along Photon Roads}
The backgrounds to photons from 
electrons and high-EM content jets with unreconstructed tracks were  
reduced by looking for hits in narrow roads
between the vertex and the EM cluster in the calorimeter.  In particular, 
a background to 
$Z\gamma\rightarrow \nu\nu\gamma$ was $W\rightarrow e\nu$ events where the 
electron was misidentified as a photon. The tracking algorithm could have been 
confused by extra hits or have missed the track because the wrong vertex was 
selected (in the case of multiple 
vertices) or because the vertex was reconstructed poorly and the track pointed 
in a different direction.  An algorithm was developed, called 
{\small HITSINFO},  to identify this background.  
Roads were defined between the cluster and each reconstructed vertex as well 
as the vertex position obtained by the {\small EMVTX} algorithm. The road 
size depended on the tracking chamber.  The following road sizes were used:
\begin{eqnarray*}
 \Delta\theta_{VTX} = 0.005, & \Delta\phi_{VTX}=0.012;  \\
 \Delta\theta_{CDC} = 0.050, & \Delta\phi_{CDC}=0.0075; \\
 \Delta\theta_{FDC} = 0.005, & \Delta\phi_{FDC}=0.015; 
\end{eqnarray*} 
where the angles, $\Delta\theta$ and $\Delta\phi$, were the half-opening 
angles of the roads in the $RZ$ and $XY$ planes. 

The roads were examined for tracks and hits. The photon candidate was
required to have no tracks from any vertex.  Further requirements
were made on the fraction of available wires hit, the number of reconstructed 
track segments, and on the number of hits, depending on the tracking 
sub-detector.  The  selection criteria were optimized using the
$Z \to ee$ sample with  one of the electrons  being misidentified as a
photon due to  tracking  chamber  inefficiency. The  efficiency 
was  calculated  using a  sample of ``emulated" photons obtained by 
rotation  of the positions of the  electron energy clusters by 
90 degrees in $\phi$ and then applying the selection criteria.
\vbox{
\begin{table}
\begin{tabular}{|c|c|c|c|c|c|}
 Selection        
           & Tight I
                    & Tight II
                             & Tight III
                                      & Tight IV
                                           & Loose I         \\ \hline \hline
 Analysis  & $WW\rightarrow  ee \; {\rm and}\; e\mu$
                    & $Z\gamma \rightarrow ee\gamma$
                             &$W\gamma \rightarrow e\nu \gamma$
                                      & $WW/WZ \rightarrow e\nu$ jet jet
                                           & $Z\gamma \rightarrow ee\gamma$ 
                                                             \\ \hline
 EM Fraction
           & $>0.90$&$>0.90$ &$>0.90$ &$>0.90$  
                                           &$>0.90$              \\ \hline
 Track Match   
           & $\surd$& $\surd$&$\surd$ & $\surd$
                                           & -                   \\ \hline
 CC (EC) $\chi^2$ 
           &$ < 100 (100)$     
                    & $< 100 (200)$
                             & $< 100 (200)$
                                      & $< 100 (100)$
                                           & $< 100 (200)$       \\ \hline
 Isolation &$< 0.1$ & $< 0.1$&$<0.15$ & $< 0.1$
                                           & $< 0.1 $            \\ \hline
 $TMS$ 
           &$<10\sigma$
                    & $<10 \sigma$
                             & $<10 \sigma$
                                      & $<5 \sigma$
                                           & -                   \\ \hline
 $dE/dx$   &$\surd$ & -      & -      & -  & -              \\ \hline \hline
CC Efficiency 
           &$72.9\pm2.3\%$ 
                    &$78.1\pm2.3\%$ 
                             &$79\pm2\%$  
                                      &$76.7\pm1.6\%$            
                                           &$90.2\pm1.3\%$  \\ \hline
EC Efficiency
           &$51.0\pm3.6\%$
                    &$70.8\pm3.4\%$ 
                             &$78\pm3\%$
                                      &$62.0\pm3.1\%$
                                            &$97.1\pm2.9\%$       \\ 
\end{tabular}
\caption{Summary of the various electron identification
definitions used in the analyses presented in this paper. }
\label{tab-emid} 
\end{table}}
\vbox{
\begin{table} 
\begin{tabular}{|c|c|c|}
Selection   &  Tight    & Loose                      \\ \hline \hline
Analysis    & $Z\gamma \rightarrow \nu \nu \gamma$
                        &  $W\gamma $                \\ 
            &           & $Z\gamma \rightarrow ee \gamma \;
                         {\rm and} \; \mu\mu \gamma$ \\ \hline 
EM Fraction 
            & $>0.96$   & $>0.90$                    \\ \hline
CC (EC) $\chi^2$
            & $< 100 (100)$ 
                        & $< 100 (200)$              \\ \hline
Isolation   & $<0.1$    & $<0.1$                     \\ \hline
Matching Track 
            & Veto
                        & Veto                       \\ \hline 
{\small EMVTX} 
            & $\surd$   & -                          \\ \hline
{\small HITSINFO} 
            & $\surd$   & -                          \\ \hline \hline
CC Efficiency 
            & $57\pm 2\%$
                        & $74\pm 7\%$                \\ \hline 
EC Efficiency 
            & $56\pm 4\%$
                        & $58\pm 5\%$                \\ 
\end{tabular}
\caption{Summary of the various photon identification 
 definitions used in the analyses presented in this
paper. The ``Loose" efficiencies do not include the $p_T$ dependent 
effects important at low $p_T$. 
Similarly, the ``Tight" efficiency applies to only the high $p_T$ 
photons within the fiducial region. }
\label{tab-phoid}
\end{table}}

\subsection{Jet Reconstruction}
Jets were reconstructed using cone algorithms with cone sizes, $\Delta 
{\cal R}$, 
of $0.3$, $0.5$, and $0.7$ for the analyses presented in this paper.  
The algorithm was as follows.  First, a jet candidate was identified by 
forming preclusters of size $\Delta \eta \times \Delta \phi = 0.3 \times 0.3$, 
centered on the highest $E_T$ tower in the cluster, from a list of jet towers
with $E_T\geq 1.0$ GeV ordered by $E_T$.  Next, the jet direction was 
determined by an iterative process.  A cone of size $\Delta  {\cal R}$ 
was placed around a new $E_T$ weighted jet center of towers and
the process was repeated until the jet direction became stable.
If two jets shared energy, they were combined or split, based on the
fraction of energy shared relative to the $E_T$ of lower $E_T$ jet.
If the shared energy was greater than 50\% of the lower $E_T$ jet,
the jets were merged.

The jet energy was corrected for a number of effects.  These included
energy contributed to the jet from the underlying event, energy from the
jet which escaped the jet cone, energy lost due to the zero-suppression,
as well as the overall jet energy scale.  A cone of radius 
$\Delta {\cal R} = 0.7$ 
was selected by all analyses presented here, except for the $WW/WZ\rightarrow 
e\nu$ jet jet analysis, where a cone of radius $\Delta {\cal R} = 0.3$ 
was used, and the $Z\gamma\rightarrow \nu\bar{\nu}\gamma$ analysis, where a 
   cone of radius $\Delta {\cal R} = 0.5$ was used. 
The small cone size was advantageous for detecting two closely-spaced jets 
expected from high-$p_T$ $W$ boson decays.  The larger cone size had smaller
$E_T$ corrections.

\subsection{Missing Transverse Energy}
The missing transverse energy in the calorimeter
\hbox{$\rlap{\kern0.25em/}E_T^{{\rm cal}}$} was defined as 
\begin{equation}
\rlap{\kern0.25em/}E_T^{{\rm cal}} = \sqrt{ \rlap{\kern0.25em/}
E_{Tx}^{{\rm cal^2}} + \rlap{\kern0.25em/}E_{Ty}^{{\rm cal^2}} },
\end{equation}
where 
\begin{equation}
\rlap{\kern0.25em/}E_{Tx}^{{\rm cal}}  = 
-\sum_i{E_i\sin{\theta_i}\cos{\phi_i} } - \sum_j{\Delta E_x^i},
\end{equation}\begin{equation}
\rlap{\kern0.25em/}E_{Ty}^{{\rm cal}}  = 
-\sum_i{E_i\sin{\theta_i}\sin{\phi_i} } - \sum_j{\Delta E_y^i},
\end{equation}
The first sum is over all the cells in the calorimeter and ICD. The second 
sum is over all the corrections in $E_T$ applied to all the electrons and jets 
in the event.  
In order to obtain the best resolution, the corrections
$\Delta E_T^i$ were those from reconstructing the event with a jet of 
cone size $\Delta {\cal R}=0.7$.

The sources of $\rlap{\kern0.25em/}E_{T}$ included neutrinos,
which escaped undetected, and the energy imbalance due to the 
resolution of the calorimeter and muon system. 
The missing transverse energy was corrected if there were muons
in the event.  The transverse momenta of the muons were removed from the 
\hbox{$\rlap{\kern0.25em/}E_T^{{\rm cal}}$} to form the total missing 
$E_T$, whose components were:
\begin{equation}
\rlap{\kern0.25em/}E_{Tx} = 
\rlap{\kern0.25em/}E_{Tx}^{{\rm cal}} - \sum_i{p_x^{\mu_i}},
\end{equation}
\begin{equation}
\rlap{\kern0.25em/}E_{Ty} =
\rlap{\kern0.25em/}E_{Ty}^{{\rm cal}} - \sum_i{p_y^{\mu_i}}.
\end{equation}
In that which follows in this paper, analyses not involving muons did not 
distinguish
between \hbox{$\rlap{\kern0.25em/}E_T^{{\rm cal}}$} and 
\hbox{$\rlap{\kern0.25em/}E_T$}; the ``cal" superscript is then ignored. 

\section{Monte Carlo Simulation}
\label{sec-mc}
In order to determine effects due to experimental limitations such as detector
acceptances, resolutions, and efficiencies on the expected signal and
background, and to provide a cross-check for many of the quantities 
measured with the data, simulations of the detector and trigger were
developed. Various levels of sophistication were used, depending on 
the detail required. 

\subsection{Detector simulation programs}
\label{sec-detsimG}
The most detailed model of the detector was the {\small GEANT}~\cite{GEANT} 
simulation.  The D{\O} implementation of {\small GEANT}, {\small D{\O}GEANT},
included details of the geometry of individual detectors, instrumental 
efficiencies and resolutions, and particle responses. 
The performance of the {\small D{\O}GEANT} program was confirmed by
comparing the simulation results with the data taken from test 
beams\cite{NIMTB}, cosmic ray muons and ${\bar p}p$ collisions.  
It was typically used to predict and cross-check the effect of variations
in the particle identification requirements on the efficiency for leptons
and jets.  It was also used to predict and cross-check the 
effect of changing the kinematic requirements on the number and 
characteristics of some of the signals and backgrounds.  

In a typical application, an event generator such as 
{\small PYTHIA}\cite{PYTHIA} or
{\small ISAJET}\cite{ISAJET}
was used to create a list of particles produced in the 
collision.  The simulation converted this into a Monte Carlo event
with the same format as the digitized information from the real collision.
This Monte Carlo event was then reconstructed in the same way as the data. 
As it was a very detailed detector simulation, it consumed relatively large 
quantities of computer resources.  This limited its application to problems 
of manageable scale. 

In order to speed-up the {\small GEANT} simulation, the calorimeter
response for electrons, photons, and hadrons could be modeled using a 
database of particle showers called the Shower Library. 
The Shower Library~\cite{raja} was created by storing the energy 
deposition in each calorimeter cell for each shower that was generated using 
{\small GEANT} in the full shower mode. Each shower was stored in a list 
together with its particle identity, momentum, pseudorapidity, azimuth,
and collision vertex origin.  When using the Shower Library to simulate 
the response of a particle in a Monte Carlo event, a shower of the 
appropriate type was selected randomly from the library and added
to the event. 
This method was useful, for example, in determining the efficiency for 
dijets in the $WW/WZ\rightarrow e\nu$ jet jet analysis presented in 
Section~\ref{sec-wwev}, where the advantage of speed made it possible to 
create a parameterization of the efficiency.  

An even faster simulation, {\small D\O FAST}, with correspondingly less detail, 
used simplified geometrical structures of the D{\O} detector and 
parameterizations of the detector response 
including energy (momentum) resolutions, particle identification efficiencies,
and trigger turn-on curves obtained from the data and described in the previous
section.  Careful comparisons were performed between {\small D{\O}FAST} 
and {\small D{\O}GEANT} for the processes with Standard Model
couplings to ensure that nothing important was lost in using the former.
This simulation was used, for example, to model the acceptance 
for the grid points in the anomalous coupling parameter space, where 
dozens of grid points were used, each with 10,000 to 100,000 MC events.
A similar fast Monte Carlo was used to estimate the background
from $Z$ boson decays for the $WW\rightarrow$ dileptons analyses.  

\subsection{Trigger Simulation}
\label{sec-trigsim}
In order to optimize and cross-check the efficiency of the triggers described 
in  Table~\ref{tab-trigs}, and to provide a method for finding the efficiency
of combinations of separate triggers, a detailed simulation of the trigger 
algorithm was made.  The list of available triggers, particularly those used 
for monitoring the higher-$p_T$ triggers, changed from time-to-time
over the course of the run as the luminosity increased.  Occasionally the 
algorithms were improved as our understanding of the detector improved.
The trigger simulation, {\small TRIGSIM}, was used to pre--test the changes 
in the trigger.  The output from the {\small GEANT} simulation was
processed by the simulator using level 1 (L1.0) and level 1.5 (L1.5) hardware
and level 2 (L2.0) software simulations.  The L2.0 simulation 
used software identical to that used in the
L2 computers.  The results were then compared to the arrays of available 
triggers and the events were marked as passed or failed.  The simulator 
was cross-checked against the actual trigger using real data as input events.  

\section{Data Samples}
\label{sec-ds}
During the 1992--1993 $p\bar{p}$ collider run, the Fermilab
Tevatron, operating at a center of mass energy of $\sqrt{s}=1.8$ TeV,
delivered a total integrated luminosity of $\int L dt = 21.8$ pb$^{-1}$. 
Typical instantaneous luminosities of $4\times10^{30}$ cm$^{-2}$sec$^{-1}$
were attained.  D\O \ collected $14.4$ pb$^{-1}$ to tape. The difference 
between delivered and collected luminosity was dominated by the dead time
incurred due to operation of the Main Ring accelerator.  
A small part of the data was lost due to
operational difficulties and hardware problems (bad runs) at the time of data 
collection.  

The luminosity was calculated by measuring the rate for $p\bar{p}$ 
nondiffractive inelastic collisions using the level 0 scintillation counter
hodoscopes. The normalization for the luminosity measurement and the 
$5.4\%$ systematic uncertainty came from the $p\bar{p}$ inelastic cross 
section and the uncertainty in the acceptance of the 
counters\cite{L0unc}.
The final integrated luminosity varied from trigger to trigger for a number 
of reasons.  The muon L1.5 triggers started operating approximately
six weeks after the muon L1.0 and calorimeter triggers. 
The muon triggers tended to be prescaled 
at high luminosities because they had higher L1.0 and L1.5 trigger rates
than the calorimeter triggers.  
Finally the analyses which used only EM objects could use luminosity
collected while the muon system had hardware problems whereas the muon 
system, which relied on the calorimeter as part of muon identification, 
could not use luminosity collected when the calorimeter had a problem. 
The luminosity for a given trigger may have varied slightly from analysis
to analysis depending on the bad run list used. 
Table~\ref{tab-triglum} shows the total integrated luminosity, after bad run 
removal, for each trigger used in the analyses presented in this paper. 
\vbox{
\begin{table}
\begin{tabular}{|cc|}
Trigger            & $\int Ldt$ (pb$^{-1})$        \\ \hline \hline
\small{MU-MAX}     & $12.2\pm0.7$                  \\ 
\small{MU-ELE}     & $13.8\pm0.7$                  \\ 
\small{ELE-HIGH}   & $13.7\pm0.7$ $(13.1\pm0.7)$   \\ 
\small{ELE-MAX}    & $13.8\pm0.7$                  \\ 
\small{ELE-2-HIGH} & $14.3\pm0.8$                  \\
\small{ELE-2-MAX}  & $14.3\pm0.8$                  \\
\end{tabular}
\caption{Integrated luminosity for each trigger after accounting for 
the effects of the Main Ring and for bad runs due to hardware problems.  
The ELE-HIGH trigger has separate luminosities depending on whether the 
calorimeter or calorimeter plus muon system were checked.}
\label{tab-triglum}
\end{table}}

\section{$W\gamma$ Analysis}
\label{sec-wg}
A measurement of the $WW\gamma$ couplings
using $p\bar{p}\rightarrow \ell\nu\gamma+X$ $(\ell=e,\mu)$ events is
presented in this section.
These events contained the $W\gamma$ production processes, 
$p\bar{p}\rightarrow W\gamma $, followed by $W\rightarrow \ell\nu$ or
the final state radiation process 
$W\rightarrow \ell \nu \rightarrow \ell\nu\gamma$,
as shown in Figs.~\ref{fig-vpair} and \ref{fig-wgrad}. Anomalous 
coupling parameters would enhance the $W\gamma$ production cross section,
leading to an excess of events with
high transverse energy photons, well separated from the
charged lepton.  Figure~\ref{fig-wgpt} shows the photon spectrum 
from SM and anomalous couplings predicted by theory\cite{BaurZep}. 
The procedure of the analysis was to obtain a candidate sample, estimate the 
background contribution as a function of photon $E_T$, and compare the
background-subtracted candidate photon spectrum with that expected 
from various anomalous $WW\gamma$ couplings.
In the following, the electron and muon channels are referred to as
$W(e\nu)\gamma$ and $W(\mu\nu)\gamma$, respectively.

\subsection{Event selection}
The $W(e\nu)\gamma$ candidates were obtained by searching for
events with an isolated high $E_T$ electron, large missing 
transverse energy, and an isolated photon.
The data sample was taken with a L1.0 trigger that required at 
least one EM tower with $E_T > 10$ GeV and the {\small ELE-MAX} trigger 
at level 2, that required an isolated EM cluster with $E_T \ge 20$ GeV and 
\hbox{$\rlap{\kern0.25em/}E_T \ge 20 $} GeV, as described in 
Table~\ref{tab-trigs}. The data sample corresponded to an integrated 
luminosity of $13.8 \pm 0.7 \,{\rm pb}^{-1}$.  The electron was required to
pass the ``Tight III"  requirements of Table~\ref{tab-emid}  and the photon 
to pass the ``Loose" requirements as described in Table~\ref{tab-phoid}.  
The electron and photon were
required to be within the fiducial region of the calorimeter, as discussed 
in Section~\ref{sec-emfid}, and at least $0.01$ radians away from the 
azimuthal boundaries of the 32 EM modules in the CC.
Kinematic selection was made requiring $E^e_T>25 $ GeV, 
\hbox{$\rlap{\kern0.25em/}E_T$} $> 25$ GeV, and $M_T > 40$ 
GeV/c$^2$, where $M_T$ is the transverse mass of the electron and 
\hbox{$\rlap{\kern0.25em/}E_T$} vector defined as 
\begin{equation}
M_T=\left[ 2E^e_T \rlap{\kern0.25em/}E_T (1-\cos{\phi^{e\nu}}) \right] ^{1/2},
\end{equation}
and $\phi^{e\nu}$ is the angle between the electron $E_T$ and the 
\hbox{$\rlap{\kern0.25em/}E_T$}.

The $W(\mu\nu)\gamma$ candidates were obtained by searching 
for events with an isolated high $p_T$ muon and an isolated photon
in the data sample taken with the {\small MU-ELE} trigger 
described in Table~\ref{tab-trigs}.
The sample corresponded to an integrated luminosity of
$13.8 \pm 0.7 \, {\rm pb}^{-1}$. 
A muon track satisfying the ``Tight II" definition of Table~\ref{tab-muonid}
was required. 
Kinematic selection was made requiring $p_T^{\mu}> 15$ GeV/c and 
\hbox{$\rlap{\kern0.25em/}E_T > 15$} GeV.  To reduce background from 
$Z\gamma$ events, 
where the \hbox{$\rlap{\kern0.25em/}E_T$} resulted from muon
$p_T$ mismeasurement, events were rejected if they contained an additional muon 
track with $p_T^{\mu} > 8$ GeV/c.   

The requirements on photons were the same for both the electron and muon 
samples. The photon  was required to  have $E_T^{\gamma} \ge 10$ GeV. 
The separation between the photon and charged lepton, 
$\Delta{\cal{R}}_{\ell\gamma}$, was required to be $\geq 0.7$.
This requirement suppressed the contribution of the final state
radiation process, and minimized the probability for a photon
cluster to merge with a nearby calorimeter cluster associated with
an electron or a muon.
The above selection criteria yielded 11 $W(e\nu)\gamma$ candidates and
12 $W(\mu\nu)\gamma$ candidates.

\subsection{Efficiencies}
The trigger and offline lepton selection efficiencies,
shown in Table~\ref{tab-wgeff}, were primarily determined using
$Z\rightarrow \ell\ell$ events, requiring only one of the leptons 
to pass the trigger and selection criteria. Thus the second lepton provided 
an unbiased sample to measure efficiencies.
The efficiency for the \hbox{$\rlap{\kern0.25em/}E_T$} requirement of
the ELE-MAX trigger was calculated using the events which passed the
ELE-HIGH trigger, which had no \hbox{$\rlap{\kern0.25em/}E_T$} requirement.
The detection efficiency of the photons with $E_T>25$ GeV was determined
using electrons from $Z$ decays.
For photons with lower $E_T$ there was a decrease in detection efficiency
due to  the cluster shape requirement, determined
using test beam electrons, and the isolation requirement, which was
determined by measuring the energy in a cone of radius
${\cal R}=0.4$ rotated randomly in azimuth in the inclusive $W(e\nu)$ sample.
Combining this $E_T$--dependent efficiency with the probability
of losing a photon due to $e^+e^-$ pair conversion,
$0.10~(0.26)$ in the CC~(EC),  and due to
an overlap with a random track in the event, with probability 
$0.065~(0.155)$, 
the overall photon selection efficiency was estimated to be
$0.43\pm 0.04$ ($0.38\pm 0.03$) at $E_T^\gamma=10$ GeV which increased
to $0.74\pm 0.07$ ($0.58\pm 0.05$) for $E_T^\gamma> 25$ GeV.

The kinematic and geometrical acceptance was calculated
as a function of coupling parameters
using the Monte Carlo program of Baur and Zeppenfeld~\cite{BaurZep}, 
in which the $W\gamma$ production
and radiative decay processes were generated to leading order, and
higher order QCD effects were approximated by a K-factor of 1.34.
The ${\rm MRSD-^{\prime}}$ structure functions~\cite{MRSD} were used and
the $p_T$ distribution of the $W\gamma$ system was simulated using the observed
$p_T$ spectrum of the $W$ in the inclusive $W(e\nu)$ sample.
Table~\ref{tab-wgeff} lists the acceptances for the SM production of
$W(\ell\nu)\gamma$.
\vbox{
\begin{table}
\begin{tabular}{|ccccc|}  
     &\multicolumn{2}{c}{$W(e\nu)\gamma$} 
                          & \multicolumn{2}{c|}{$W(\mu\nu)\gamma$} \\ 
     &\multicolumn{2}{c}{$E_T(e)>25$ GeV} 
                          & \multicolumn{2}{c|}{$p_T(\mu)>$15 GeV/c} \\ 
     & $|\eta|< 1.1$ 
             & $1.5< |\eta|< 2.5$
                          & $|\eta|< 1.0$     & $1.0<|\eta|< 1.7$ \\ 
                                                                \hline \hline
$\epsilon_{trig}$  
     &$0.98\pm 0.02$ 
             & $0.98\pm 0.02$  
                          & $0.74\pm 0.06$    & $0.35\pm 0.14$    \\
$\epsilon_\ell$ 
     & $0.79\pm 0.02$ 
             & $0.78\pm 0.03$ 
                          & $0.54\pm 0.04$    & $0.22\pm 0.07$    \\
$\epsilon_{Acc}^{SM}$ &\multicolumn{2}{c}{$0.11 \pm 0.01$} 
                          & \multicolumn{2}{c|}{$0.29 \pm 0.02$}   \\
\end{tabular}
\caption{Summary of trigger ($\epsilon_{trig}$) and 
  lepton selection ($\epsilon_\ell$) efficiencies and geometrical
  acceptances ($\epsilon_{Acc}^{SM}$) for the SM $W\gamma$ production events.}
\label{tab-wgeff}
\end{table}}

\subsection{Backgrounds}
\label{sec-wgbacksec}
The background estimate, summarized in Table~\ref{tab-wgback},
included contributions from:
$Z\gamma$, where the $Z$ decays to
$\ell\ell$, and one of the leptons was undetected or was
mismeasured by the detector and contributed to \mbox{$\not\!\!E_T;$}
$W\gamma$ with $W\rightarrow \tau\nu$ followed by
$\tau\rightarrow \ell\nu\bar{\nu}$; and
$W+{\rm jet(s)}$, where a jet was misidentified as a photon.
The backgrounds due to $Z\gamma$ were estimated using
the $Z\gamma$ event generator of Baur and Berger~\cite{gl20}
followed by a full detector simulation using the {\small GEANT}
program~\cite{GEANT}.
It should be noted that $\sigma(Z(\ell\ell)\gamma)/\sigma(W(\ell\nu)\gamma)$
is about 0.5 (rather than 0.1 which is the ratio of cross sections
of $Z\rightarrow \ell\ell$ and $W\rightarrow \ell\nu$),
since the $W(\ell\nu)\gamma$ process is suppressed by 
interference between the production diagrams and since the $Z$ boson has 
twice as many leptons from which a photon can be radiated.
The background due to $W\gamma\rightarrow (\tau\nu)\gamma$ was estimated
from the ratio of the detection efficiencies of
$W\rightarrow \tau\nu \rightarrow e\nu{\bar \nu}$ and
$W\rightarrow e\nu$ processes.
The ratio was found to be $0.019 \pm 0.002$, using the 
{\small ISAJET}~\cite{ISAJET}
event generator followed by the {\small GEANT} detector simulation.

\begin{figure}
\vspace{-3.0in}
\centerline{\epsffile{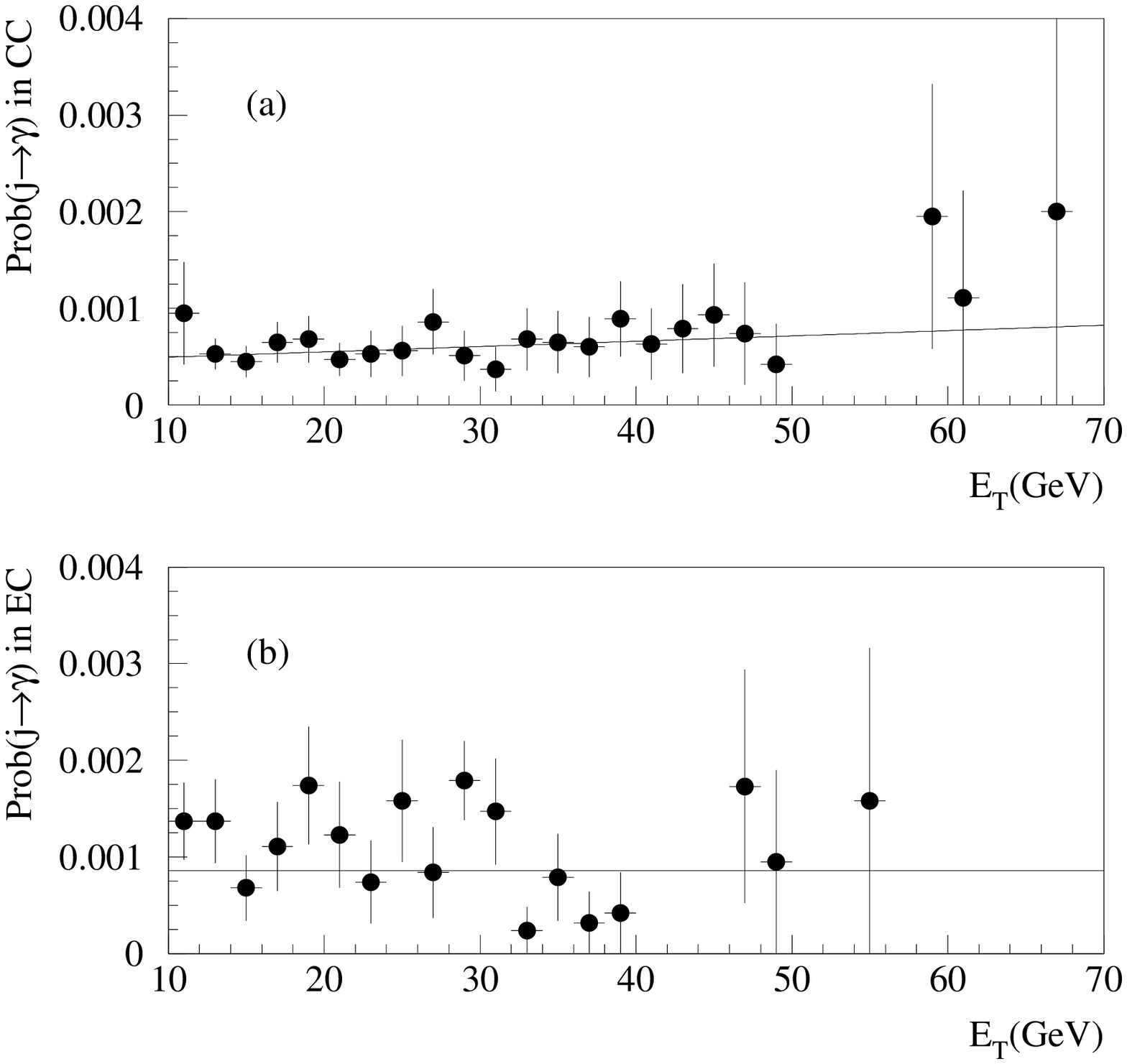}}
\vspace{-0.5in}
\caption{The probability of a jet to be misidentified as a photon as
a function of $E_T$ (before the removal of the contribution from 
direct photons) for the ``Loose" photon selection criteria 
in the CC (a) and EC (b). }
\label{fig-fake}
\end{figure}

The $W+{\rm jets}$ background was estimated
using the  probability, ${\cal P}(j\rightarrow ``\gamma ")$,
for a jet to be misidentified as a photon.
The probability was determined as a function of $E_T$ of the jet
by measuring the fraction of jets in a sample of multi-jet events
that passed the photon identification requirements.
Of course, some of the ``fake rate" was due to real photons in the jet sample.
The fraction of direct photon events in the multi-jet sample was estimated 
using the differences in the transverse and longitudinal shower shapes of
multiple photons from meson decays and single photons\cite{DG}. In the 
$E_T$ range 10 to 50 GeV, $25\%\pm 25\%$ of the ``fake" photons in the 
background sample were attributed to direct photons.  
This fraction was subtracted from ${\cal P}(j\rightarrow ``\gamma ")$.
The misidentification probability was found to be ${\cal P}(j\rightarrow 
``\gamma ") \sim 4\times 10^{-4}$
($\sim 6\times 10^{-4}$)  in the CC (EC)
in the $E_T$ region between  10 and 40 GeV.
The measured probability, before direct photon subtraction, 
for a jet to mimic a photon is shown in Fig.~\ref{fig-fake}.

The total numbers of $W+{\rm jets}$ background
events were estimated to be $1.7\pm 0.9$ and $1.3\pm 0.7$
for $W(e\nu)\gamma$ and
$W(\mu\nu)\gamma$, respectively,
by applying ${\cal P}(j\rightarrow ``\gamma ")$
to the observed $E_T$ spectrum of jets in the
inclusive $W(\ell\nu)$ sample.
The uncertainty on the background estimates was dominated by
the uncertainty on ${\cal P}(j\rightarrow ``\gamma ")$
due to the direct photon subtraction.
Bias in the $W+{\rm jets}$ background estimate
due to a possible difference in jet fragmentation (e.g.~the number of
$\pi^0$'s in a jet) between jets in the $W$ sample and those in the 
multi-jet sample was investigated by parameterizing
${\cal P}(j\rightarrow ``\gamma ")$ as a function of
the EM energy fraction of the jet.
No statistically significant difference was found between
the background estimates with and without the parametrization.
The estimated $W+{\rm jets}$ background also included
the background from  $\ell$+jets, where $\ell$ was a jet misidentified 
as an electron, a cosmic ray muon or a fake muon track, 
since it was derived from the observed inclusive $W\rightarrow \ell\nu$ event
sample.

Other backgrounds considered and found to be negligible included those from
single photon events where a jet was misidentified as an electron, 
and $ee+X$ events where 
an electron was misidentified as a photon due to tracking inefficiency.
\vbox{
\begin{table}
\begin{tabular}{|lcc|}
 & $W(e\nu)\gamma$ & $W(\mu\nu)\gamma$ \\ \hline \hline
Source: & & \\
$~~~W+{\rm jets}$ & $1.7\pm 0.9$   & $1.3\pm 0.7$ \\
$~~~Z\gamma$      & $0.11\pm 0.02$ & $2.7\pm 0.8$ \\
$~~~W(\tau\nu)\gamma$ & $0.17\pm 0.02$ & $0.4\pm 0.1$ \\ \hline
Total Background        & $2.0\pm 0.9$ & $4.4\pm 1.1$ \\ \hline \hline
Data & 11 & 12 \\
\end{tabular}
\caption{Summary of $W(e\nu)\gamma$ and $W(\mu\nu)\gamma$ data
and backgrounds.}
\label{tab-wgback}
\end{table}}

\subsection{Cross section and limits on the coupling parameters}
After subtraction of the estimated backgrounds
from the observed number of events,
the number of signal events was found to be
$$N^{W(e\nu)\gamma}=9.0^{+4.2}_{-3.1}\pm 0.9,
~~~N^{W(\mu\nu)\gamma}
=7.6^{+4.4}_{-3.2}\pm 1.1,$$
where the first uncertainty is statistical, calculated following
the prescription for Poisson processes with background given in 
Ref.~\cite{PDG}, and the second is systematic.

\begin{figure}
\centerline{\epsffile{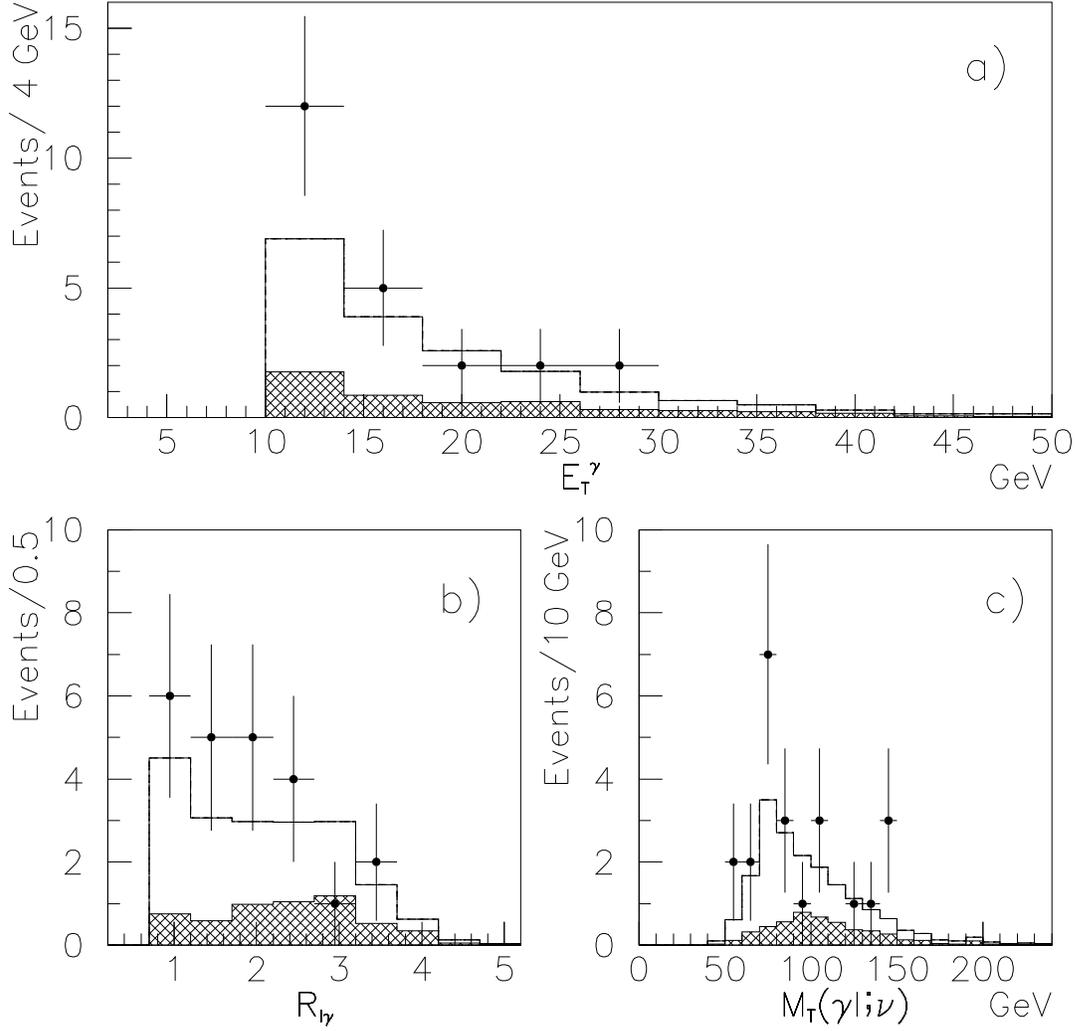}}
\caption{Distribution of (a)~$E_T^\gamma$,
(b)~$\Delta R_{\ell\gamma}$ and (c)~$M_T(\gamma\ell;\nu)$
for the $W(e\nu)\gamma$ + $W(\mu\nu)\gamma$ combined sample.
The points are data. The shaded areas represent the estimated
background, and the solid histograms are the expected signal from the Standard
Model  plus the estimated background.}
\label{fig-ET}
\end{figure}

Using the acceptance for  SM couplings of  $0.11\pm 0.01$ for $W(e\nu)\gamma$
and $0.29\pm 0.02$ for $W(\mu\nu)\gamma$ and the efficiencies quoted above,
the $W\gamma$ cross section (for photons with $E_T^\gamma> 10$
GeV and $\Delta{\cal R}_{\ell\gamma}>0.7$) was calculated
from a combined $e+\mu$ sample:
$$\sigma(W\gamma)=138^{+51}_{-38}({\rm stat})\pm 21({\rm syst})\ \ {\rm pb},$$
where the systematic uncertainty includes
the uncertainties in the $e/\mu/\gamma$
efficiencies,
the choice of the structure functions and
the $Q^2$ scale at which the structure functions are evaluated,
the $p_T$ distribution of the $W\gamma$ system, and
the integrated luminosity calculation.
The systematic uncertainties from the sources
other than trigger and lepton selection efficiencies and geometrical
acceptances are listed in Table~\ref{tab-wgsysunc}.
The measured cross section agrees with the SM prediction
of $\sigma_{W\gamma}^{SM}=112\pm 10\ {\rm pb}$ within the uncertainty.
Figure~\ref{fig-ET} shows
the data and the SM prediction plus the background
in the distributions of $E_T^\gamma,$  $\Delta {\cal R}_\ell\gamma,$
and the three-body ``cluster" transverse mass defined by
$M_T(\gamma\ell;\nu)=[((m_{\gamma\ell}^2+|{\bf E_T^\gamma}
+{\bf E_T^\ell}|^2)^{\frac{1}{2}}+{\not\!\!E_T})^2-|{\bf E_T^\gamma}+
{\bf E_T^\ell}+{\bf \not\!\!E_T}|^2]^{\frac{1}{2}}.$
Final state radiation events and background events composed most of the 
expected signal with $M_T(\gamma\ell;\nu)\leq M_W$. 
Of the 23 observed events,  11 events had $M_T(\gamma\ell;\nu)\leq M_W$.
\vbox{
\begin{table}
\begin{tabular}{|cc|}
                           & Uncertainty    \\ \hline \hline
Luminosity                 & 5.4\%          \\ 
Structure Function choice  & 6.0\%          \\ 
$P_T^{W\gamma}$            & 3.9\%          \\ 
Conversion Probability     & 5.0\%          \\ 
Random track overlap       & 1.0\%          \\ 
Photon selection efficiency& 7.0\%          \\ \hline \hline
Total                      & 12.5\%         \\
\end{tabular}
\caption{The values of systematic uncertainties in the $W\gamma$ cross
section and coupling limit measurements, other than those of trigger,
lepton selection and acceptance.}
\label{tab-wgsysunc}
\end{table}}

\begin{figure}
\epsffile{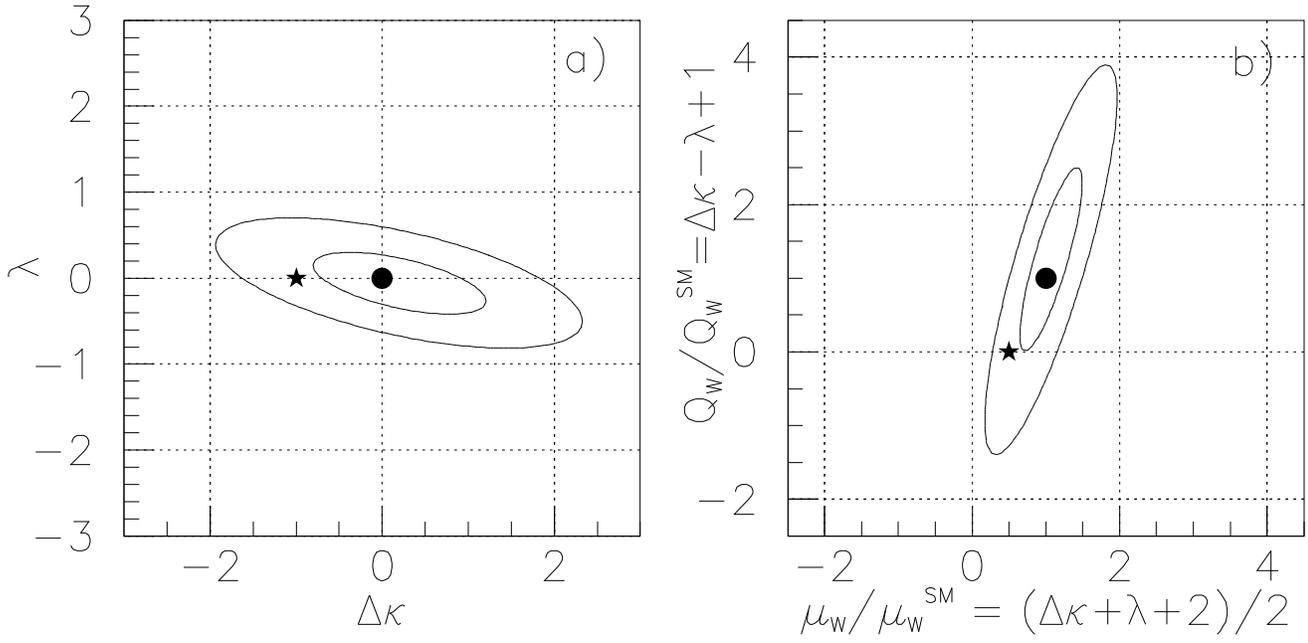}
\caption{Limits on (a) CP--conserving anomalous coupling
parameters $\Delta \kappa$ and $\lambda,$ and on (b) the magnetic dipole,
$\mu_W$, and electric quadrupole, $Q^e_W$, moments.
The ellipses represent the $68\%$ and $95\%$ CL exclusion contours.
The symbol,~$\bullet$, represents the Standard Model values, while
the symbol,~$\star$, indicates
the $U(1)_{EM}$--only coupling of the $W$ boson to a photon,
$\Delta\kappa=-1$ and $\lambda=0$ ($\mu_W=e/2m_W$ and $Q^e_W=0$).}
\label{fig-cont}
\end{figure}

To set limits on the anomalous coupling parameters,
a binned maximum likelihood fit was performed
on the  $E_T^\gamma$ spectrum for each of the $W(e\nu)\gamma$ and
$W(\mu\nu)\gamma$ samples, by calculating the
probability for the sum of the Monte Carlo prediction and the background
to fluctuate to the observed number of events (see Appendix~2 for more detail). 
The uncertainties in background estimate, efficiencies, acceptance
and integrated luminosity
were convoluted  in the likelihood function as Gaussian distributions.
A dipole form factor with a scale $\Lambda=1.5$ TeV
was assumed for the anomalous couplings in the Monte Carlo event generation.
The Monte Carlo events were generated at $11 \times 11$ grid points of the
CP--conserving anomalous coupling parameters, $\Delta\kappa_{\gamma}$ 
and $\lambda_{\gamma}$,
assuming that the CP--violating
anomalous coupling parameters $\tilde{\kappa}_{\gamma}$ and
$\tilde{\lambda}_{\gamma}$ are zero.
The limit contours for $\Delta \kappa_{\gamma}$ and
$\lambda_{\gamma}$ are shown in Fig.~\ref{fig-cont}.
The numerical values of the limits at the $95\%$ confidence level (CL) were
$$-1.6<\Delta\kappa_{\gamma}<1.8\ 
(\lambda_{\gamma}=0),~~~-0.6<\lambda_{\gamma}<0.6\ 
(\Delta\kappa_{\gamma}=0)$$ for $\hat{s}=0$ (i.e.~the static limit).
The $U(1)_{EM}$--only coupling of the $W$ boson to
a photon, which leads to  $\kappa_{\gamma}=0\ (\Delta\kappa_{\gamma}=-1)$ 
and $\lambda_{\gamma}=0$, and
thereby, $\mu_W=e/2m_W$ and $Q_W^e=0$\cite{TDLEE},
was excluded at the $80\%$ CL, while
the zero magnetic moment ($\mu_W=0$) was excluded at more than the $95\%$ CL.
Similarly, limits on CP--violating coupling parameters  were obtained
as $-1.7<\tilde{\kappa}_{\gamma}<1.7\ (\tilde{\lambda}_{\gamma}=0)$ and
$-0.6<\tilde{\lambda}_{\gamma}<0.6\ (\tilde{\kappa}_{\gamma}=0)$ 
at the $95\%$ CL. The form factor scale dependence of the results was studied.
It was found that the limits were insensitive to the values of the 
form factor scale for $\Lambda>200$~GeV and were well within the constraints 
imposed by S-matrix unitarity~\cite{gl23} for $\Lambda=1.5$~TeV.
A simultaneous fit to $E_T^{\gamma}$ and the $\Delta {\cal R}_\ell\gamma$
spectra was performed.
It was found that the results were within $3\%$ of those obtained from a fit to
the  $E_T^\gamma$ spectrum only.

\section{$WW\rightarrow$ Dileptons}
\label{sec-wwdilep}
In this section the results of a search for 
$p\bar{p} \rightarrow WW+X 
\rightarrow \ell \bar{\ell}'\bar{\nu}\nu '+X$,
where the leptons included muons and electrons, are presented. The signal
and background were estimated and an upper limit was set for the cross section 
of the SM process.  
Anomalous $WWZ$ and $WW\gamma$ couplings would have enhanced the 
expected $WW$ cross section by upsetting the 
cancellation\cite{delicate-cancellation} between the 
production diagrams and the trilinear diagram as seen in Fig.~\ref{fig-wwxs},
which shows the cross section $vs$ anomalous couplings for $\Lambda = 1000$ 
GeV. The  detection   efficiency also  increases  with  anomalous  couplings
because of the higher average $E_T$ of the $W$ bosons 
(see Fig.~\ref{fig-wzpt}), 
resulting in a higher average $E_T$ for leptons and in more central events.
This expected 
increase in the cross section and efficiency was exploited 
to set limits on the anomalous coupling parameters, $\lambda$ and 
$\Delta \kappa$.

The expected signature for $W$ boson pair production with subsequent decay to
dileptons was two high-$p_T$ isolated leptons in association with 
large \hbox{$\rlap{\kern0.25em/}E_T$}.
The major sources of background were the following: events with a 
$W + {\rm jet(s)}$ where a jet was misidentified as a lepton;
$W + \gamma$ events where a photon was misidentified as an electron;  
QCD multi-jet events where two jets were misidentified as leptons;
$Z \rightarrow \ell \ell$, $Z \rightarrow \tau\tau \rightarrow 
\ell \ell ' \nu \nu \bar{\nu}\bar{\nu}$ events;
and $t\bar{t} \rightarrow \ell \ell '+{\rm X}$ events. 
The event selection requirements were designed to reduce these backgrounds
while retaining high detection efficiency for signal events. 
The selection requirements were slightly different for the $ee$, $e\mu$, and
$\mu\mu$ channels because the electrons had a better $p_T$ resolution 
but a larger background contamination than muons.  
In what follows, the analyses of individual channels and our limits on the
cross section for $W$ boson pair production as well as on the anomalous
gauge boson trilinear couplings are presented.

\subsection{The $ee$ channel}
\begin{figure}
\epsffile{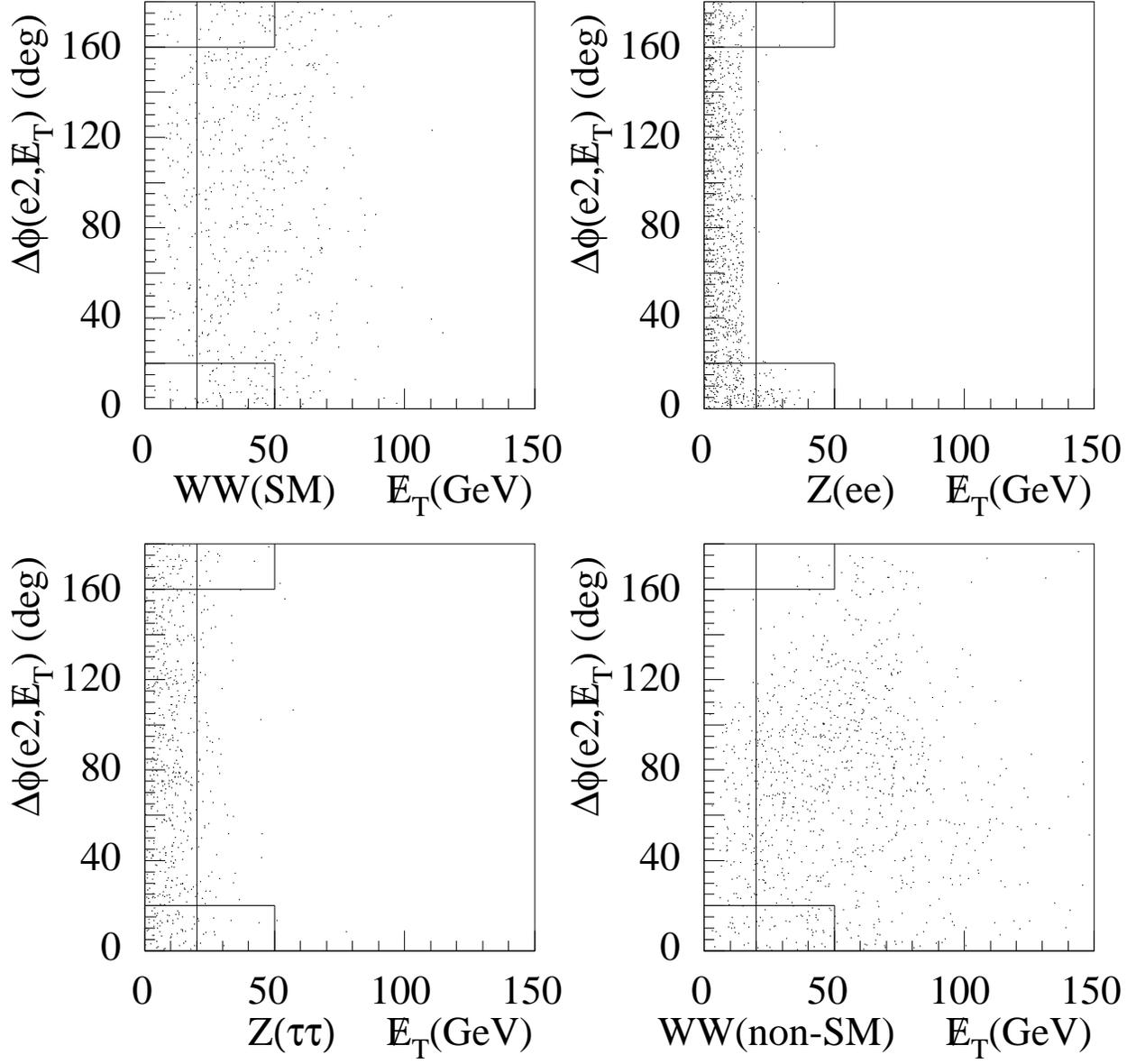}
\caption{$\Delta\phi(p_{T}^{e}, \hbox{$\rlap{\kern0.25em/}E_T$} )$ vs 
\hbox{$\rlap{\kern0.25em/}E_T$} distributions for
$WW \rightarrow (e\nu)(e\nu)$ with the SM couplings, $Z \rightarrow ee$,
$Z \rightarrow \tau\tau \rightarrow ee\nu \bar{\nu} \nu \bar{\nu}$ and
$WW \rightarrow ee\nu \bar{\nu})$ with the non-SM couplings.}
\label{fig-eecuts}
\end{figure}

\begin{figure}
\epsffile{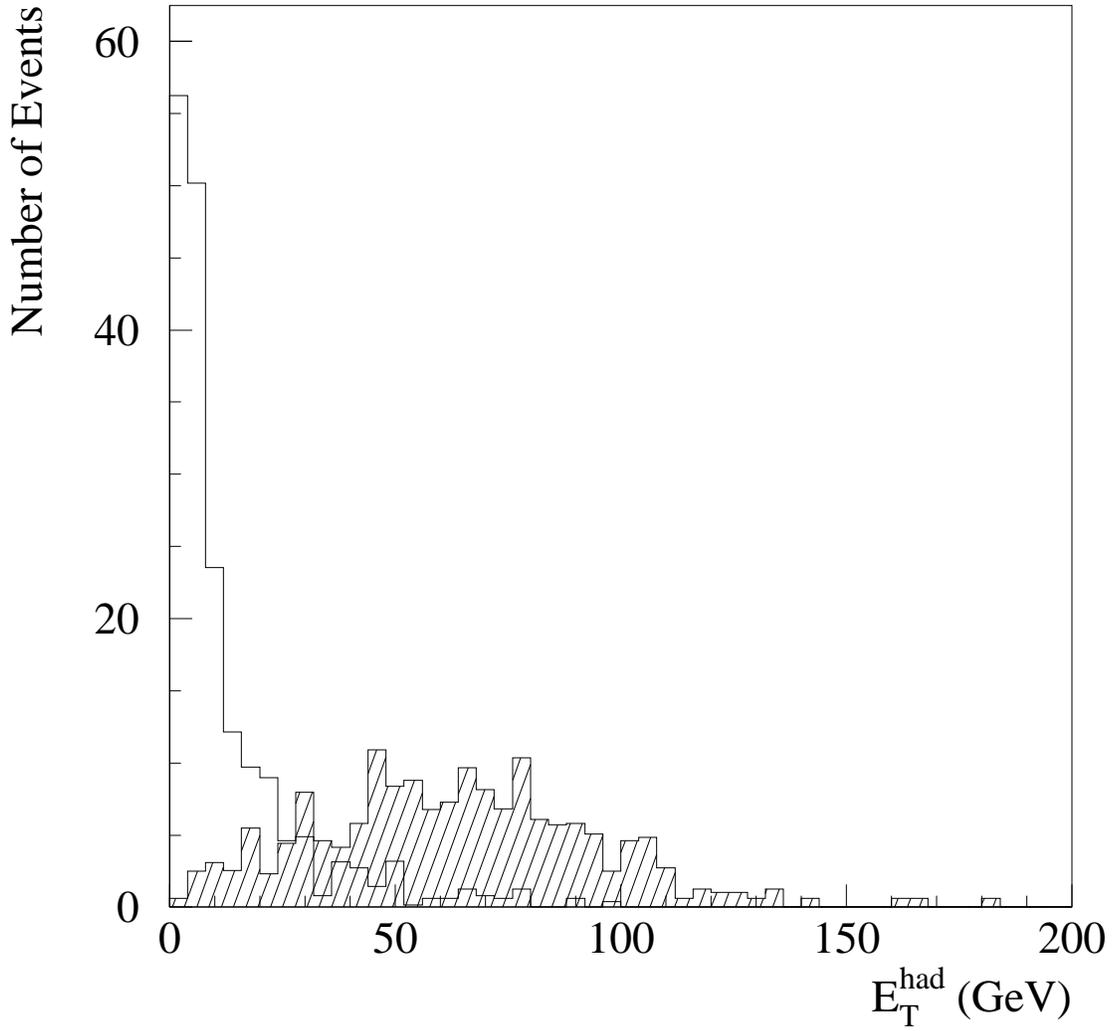}
\caption{$E_T^{{\rm{had}}}$ for Monte Carlo 
$WW$ (open histogram) and  $t\bar{t}$ events (shaded histogram) with 
${\rm M}_{top} = 160$ GeV/c$^2$  $(\int Ldt \sim 20$ fb$^{-1})$. 
Events with $E_T^{{\rm{had}}} \geq 40 $ GeV were rejected. }
\label{fig-ethad}
\end{figure}

The $WW \rightarrow ee\nu\bar{\nu}$ candidate events were selected
from the data sample recorded using the {\small ELE-2-HIGH}
trigger which required two EM clusters with ${E_{T} > 7}$ GeV at level 1
and two isolated EM clusters with ${E_{T} > 10}$ GeV at level 2 
(see Table~\ref{tab-trigs}). Candidate events containing 
two electrons that passed the ``Tight I" requirements were selected.
The ``Tight I" requirements discussed in Section~\ref{sec-pid},
and detailed in Table~\ref{tab-emid}, 
provided the largest rejection of fake electrons.
The following event selection requirements were then imposed. 
Both electrons were required to have a large transverse energy
($E_T \geq 20$ GeV);  at this stage  the remaining sample of 605 
events was comprised 
primarily of $Z$ bosons. The \hbox{$\rlap{\kern0.25em/}E_T$} of the event
was then required to be $\geq 20$ GeV.  These first two selection criteria 
strongly reduced the background due to QCD fakes.  
The dielectron invariant mass was required to be outside of
the $Z$ boson mass window (between $77$ and $105$ GeV/c$^2$).  
The \hbox{$\rlap{\kern0.25em/}E_T$} and dielectron invariant mass selections 
had very strong rejection $(> 100)$ of $Z\rightarrow ee$ decays. The 
background from $Z\rightarrow\tau\tau \rightarrow ee\nu\bar{\nu}$ which was 
not eliminated by the electron
$p_T$ thresholds was further reduced by requiring that  
the \hbox{$\rlap{\kern0.25em/}E_T$}
not be collinear with the direction of the lower energy electron;
it was required that $20^{\circ}\leq \Delta\phi(p_T^{e},
\hbox{$\rlap{\kern0.25em/}E_T$} )\leq 160^{\circ}$ 
for the  lower energy electron if  \hbox{$\rlap{\kern0.25em/}E_T$}
$\leq 50$ GeV.  Releasing this requirement for events with large  
\hbox{$\rlap{\kern0.25em/}E_T$} increased the acceptance for $W$ boson
pairs in a region where the $Z$ boson background was very small. 
This can be seen in Fig.~\ref{fig-eecuts} 
which shows $\Delta\phi(p_{T}^{e}, \hbox{$\rlap{\kern0.25em/}E_T$}) 
{\rm vs }\; \hbox{$\rlap{\kern0.25em/}E_T$}$ distributions for $W$ boson pairs,
$Z\rightarrow ee$ and $Z\rightarrow\tau\tau \rightarrow e e\nu\bar{\nu}$.
Finally, the sum of the $E_T$ of the recoiling hadrons 
($\vec{E}_T^{{\rm{had}}}$), defined as 
$-(\vec{E}_T^{l1}+\vec{E}_T^{l2}+\vec{\hbox{$\rlap{\kern0.25em/}E_T$}})$
was required to be less than 40 GeV in magnitude. 
The background from $t\bar{t}$ production was effectively eliminated
by this requirement. 
Figure~\ref{fig-ethad} shows a Monte Carlo ({\small PYTHIA} plus 
{\small D\O GEANT}) simulation of $E_T^{{\rm{had}}}$ for $\sim 20$ fb$^{-1}$ 
of SM $WW$ and $t\bar{t}$ events.
For $WW$ events, non-zero values of $E_T^{{\rm{had}}}$ were due to 
gluon radiation and detector resolution.  
For $t\bar{t}$ events, the most 
significant contribution was from $b$ quark jets from $t$ quark decays. 
This selection reduced the background from $t\bar{t}$ production
by a factor of more than four for a $t$ quark mass of 170 GeV/c$^2$. 
The efficiency of this selection criterion for SM $W$ boson pair 
production events was $0.95^{+0.01}_{-0.04}$ and decreased slightly with 
increasing $W$ boson pair invariant mass.  
The systematic uncertainty in the efficiency of this last selection criteria,
included in the uncertainties presented, 
was estimated from the difference between $\vec{E}_T^{{\rm{had}}}$ for
$Z$ boson data and Monte Carlo ( {\small PYTHIA} plus {\small D\O GEANT})
distributions.

Table~\ref{tab-ww_cuts_evts}  shows the numbers of events remaining after each 
selection cut. One event survived all the selection criteria.

\vbox{
\begin{table}
\begin{tabular}{|lc|} 
Event selection criteria          & Number of events surviving \\ \hline \hline
$E_T \geq 20$ GeV                 & 605 \\
\hbox{$\rlap{\kern0.25em/}E_T$} $> 20$ GeV  
                                  & 5   \\
$M_{ee} < 77 \, {\rm GeV/c}^{2}$ or $M_{ee} \geq 105 \, {\rm GeV/c}^{2}$ 
                                  & 3   \\
$\Delta\phi(p_{T}^{e2},\hbox{$\rlap{\kern0.25em/}E_T$} )$ cut 
                                  & 1 \\
$|\vec{E}_{T}^{had}| \leq 40$ GeV & 1 \\ 
\end{tabular}
\caption{The numbers of events after each selection cut 
for the $WW \rightarrow ee$ analysis.}
\label{tab-ww_cuts_evts}  
\end{table}}
\vbox{
\begin{table}
\begin{tabular}{|lc|} 
Fiducial Region & Efficiency \\ \hline \hline
CC -- CC & $0.526 \pm 0.041$ \\
CC -- EC & $0.368 \pm 0.044$ \\
EC -- EC & $0.257 \pm 0.058$ \\ 
\end{tabular}
\caption{The combined trigger and electron selection efficiency
for individual fiducial regions in the $WW\rightarrow ee$ analysis for 
SM $W$ pair production.}
\label{tab-eeeff}
\end{table}}

The integrated luminosity of the data sample was
$14.3\pm 0.8$ pb$^{-1}$.
The trigger efficiency was calculated with the {\small TRIGSIM}
simulation package to be ${\epsilon_{trig} = 0.989 \pm 0.002}$.
The measured electron selection efficiency was used to estimate the detection
efficiency for SM $WW \rightarrow ee\nu\bar{\nu}$ events.
The geometrical acceptance was
obtained from a {\small PYTHIA} and {\small D\O GEANT} Monte Carlo simulation.
These efficiencies for individual fiducial regions are listed in 
Table~\ref{tab-eeeff}.  The overall detection efficiency for SM $W$ pair 
events was estimated to be $\epsilon_{ee} = 0.094 \pm 0.008$.
The expected number of events $N_{ee}^{SM}$ was 
$N_{ee}^{SM} = 0.149 \pm 0.013 ({\rm stat}) \pm 0.019 ({\rm sys})$,
using the  next-to-leading order cross section \cite{ohnemus}, and 
branching fraction ${\rm Br}(W\rightarrow  e\nu) ( = 0.108 \pm 0.004)$
\cite{pdgbr}.

The backgrounds from $W\gamma$, Drell-Yan dilepton,
$Z\rightarrow \tau\tau \rightarrow ee\nu\nu\bar{\nu}\bar{\nu}$
and $t{\bar t}$ processes were
estimated using the {\small PYTHIA} and {\small ISAJET}
Monte Carlo event generators followed
by the {\small D\O GEANT} detector simulation.
The $t\bar{t}$ cross section estimates were from the
calculations of Laenen {\it et al.} \cite{topxsec}. 
The $t\bar{t}$ background was averaged 
for $M_{\rm top} = 160$, $170$, and $180$ GeV/c$^2$.
The production and decay of
$Z$ bosons was modeled using the double differential (in rapidity and $p_T$) 
cross section calculated at next-to-leading order\cite{AK}, and a fast 
detector simulation of the type discussed in Section~\ref{sec-detsimG}.
The line shape of the 
$Z$ boson was taken to be a relativistic Breit-Wigner function.  The
kinematic distributions were compared with the $Z$ boson data sample
and found to be consistent. 

The backgrounds from $W + {\rm jet(s)}$ with a jet misidentified as an 
electron and multijet events with two jets misidentified as electrons were 
called ``fake" background.
The size of this background was estimated with the following method.
Two sub-samples of data were derived from the full data set.
One was similar to the signal sample and contained two ``Tight I"
electrons each with $E_T \geq 20$ GeV.
The other was a sample of events with at least one {\it bad} electron
which had an H-matrix $\chi^{2} \ge 200$ and Isolation $f_{iso} > 0.15$ (the 
fake sample). A normalization factor ($F_{fake}$) of this fake
sample relative to the signal
sample was calculated using the number of events with 
\hbox{$\rlap{\kern0.25em/}E_T$} $ < 15$ GeV, which contained solely fake
electrons, in both the samples. 
All the event selection cuts were applied to the fake sample
and the number of remaining events ($N_{fake}$) was counted.
The fake background ($N_{fake}^{BG}$) was then computed from the products of
$F_{fake}$ and $N_{fake}$. The result was 
$N_{fake}^{BG} = 0.152 \pm 0.012 ({\rm stat}) \pm 0.076 ({\rm sys})$.
The total number of background events was estimated to be
$N_{ee}^{BG} = 0.222 \pm 0.020 ({\rm stat}) \pm 0.080 ({\rm sys})$.
Table~\ref{tab-wwdilepback} contains a summary of the expected background
in the $ee$ channel.
\vbox{
\begin{table}
\begin{tabular}{|lccc|} 
Background                      & $ee$         & $e\mu$ &$\mu\mu$ 
                                                                  \\ \hline
$Z\rightarrow$ $ee$ or $\mu\mu$ &$0.02\pm0.01$ & -----  &$0.068\pm0.026$ \\ 
$Z\rightarrow \tau \tau$        &$<10^{-3}$    &$0.11\pm0.05$ 
                                                        &$<10^{-3}$    \\
Drell-Yan dileptons             & $<10^{-3}$   & -----  &$<10^{-3}$    \\
$W\gamma$                       &$0.02\pm0.01$ &$0.04\pm0.03$
                                                        & -----        \\
QCD ($N_{fake}^{BG}$)           &$0.15\pm0.08$ &$0.07\pm0.07$
                                                        &$<10^{-3}$    \\ 
$t\bar{t}$                      &$0.03\pm0.01$ &$0.04\pm0.02$
                                                        &$0.009\pm0.003$
                                                                     \\ \hline 
Total                           &$0.22\pm0.08$ &$0.26\pm0.10$ 
                                                        &$0.077\pm0.026$ \\ 
\end{tabular}
\caption{Summary of the expected number of background events 
 to $WW\rightarrow e e\nu\bar{\nu}$, 
$WW\rightarrow e\mu \nu\bar{\nu}$ and $WW\rightarrow \mu\mu \nu\bar{\nu}$. 
The uncertainties include both statistical and systematic contributions.}
\label{tab-wwdilepback}
\end{table}}

\subsection{The $e\mu$ channel}
\begin{figure}
\centerline{\epsffile{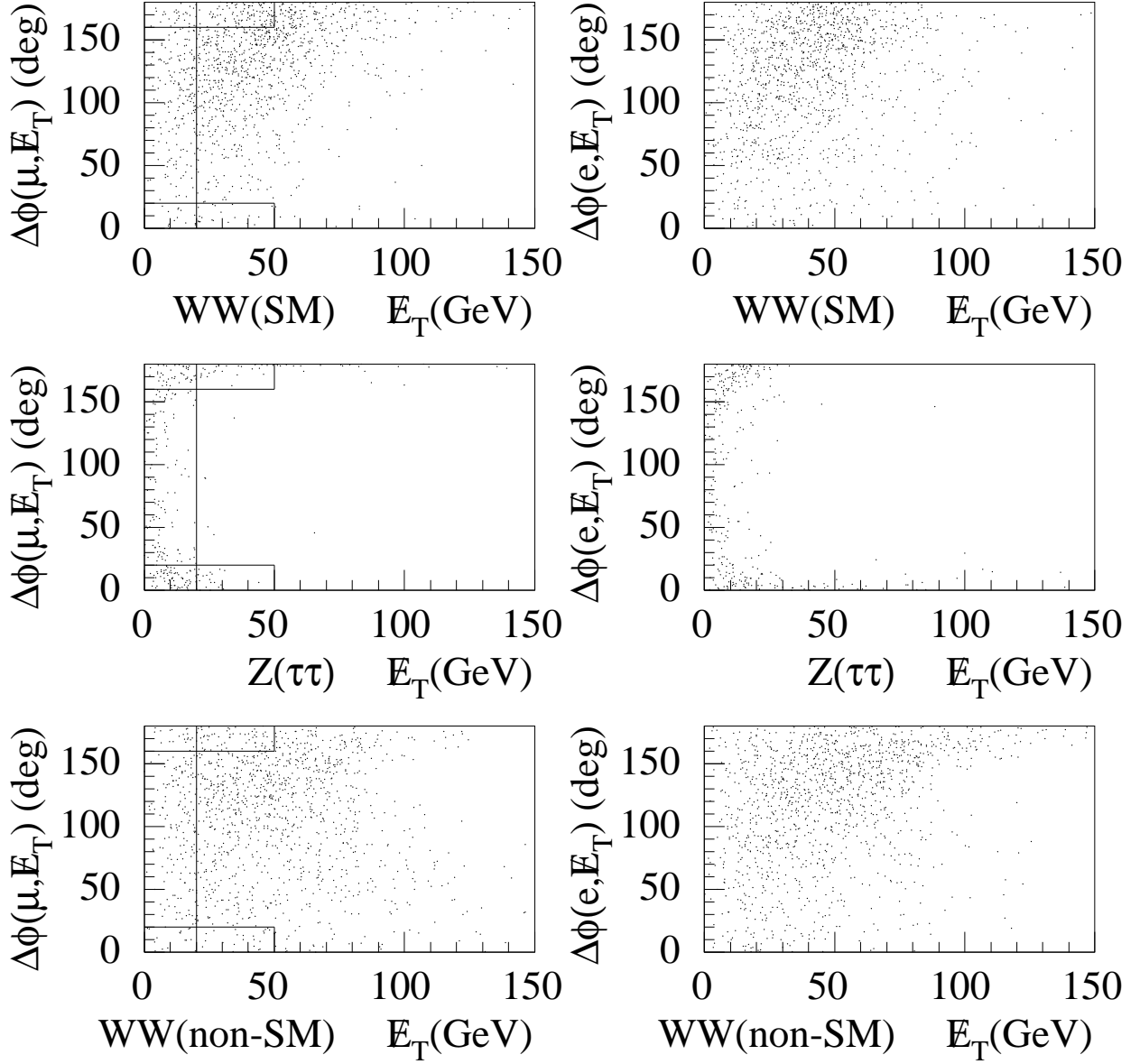}}
\vspace{0.5in}
\caption{$\Delta\phi(p_{T}^{\ell}, \hbox{$\rlap{\kern0.25em/}E_T$}
)$ vs \hbox{$\rlap{\kern0.25em/}E_T$}
distributions for
$WW \rightarrow e\mu $ with the SM couplings,
$Z \rightarrow \tau\tau \rightarrow e\mu \nu \bar{\nu} \nu \bar{\nu}$
and $WW \rightarrow e\mu$ with non-SM couplings.}
\label{fig-wweudphi}
\end{figure}

The $WW \rightarrow e\mu\nu\bar{\nu}$ candidate events were selected from the
data sample acquired with the {\small MU-ELE} trigger which required at least 
one EM tower with $E_T > 7$ GeV and one muon with $|\eta | < 1.7$ at level 1 
and one EM cluster with $E_T > 7$ GeV and one muon with $p_T > 5$ GeV/c at
level 2 (see Table~\ref{tab-trigs}).
The electron in a candidate event was required to pass ``Tight I"
criteria, providing the strictest rejection against fake electrons,
and the muon to pass the ``Loose I" criteria of Table~\ref{tab-muonid}.
The following event selection requirements were imposed. 
Both the electron and the muon were required to have a large transverse energy
(momentum) $E_{T}^{e} \geq 20$ GeV and $p_{T}^{\mu} \geq 15$ GeV/c. 
Both \hbox{$\rlap{\kern0.25em/}E_T$} and \hbox{$\rlap{\kern0.25em/}E_T
^{cal}$} of the event were required to be $\geq 20$ GeV. These first two
requirements provided large rejection of the background from multijet events. 
In order to reduce the backgrounds from ${Z \rightarrow \tau\tau \rightarrow 
e\mu\nu\nu\bar{\nu}\bar{\nu}}$, 
the \hbox{$\rlap{\kern0.25em/}E_T$} 
was required not to be collinear to the muon: 
$20^{\circ}\leq \Delta\phi(p_T^{\mu}, \hbox{$\rlap{\kern0.25em/}E_T$}
)\leq 160^{\circ}$ if \hbox{$\rlap{\kern0.25em/}E_T$} $\leq 50$ GeV.
Figure~\ref{fig-wweudphi} shows the $\Delta\phi(p_{T}^{\mu},
\hbox{$\rlap{\kern0.25em/}E_T )$} $ vs \hbox{$\rlap{\kern0.25em/}E_T$}
distributions. Finally,
the recoil hadronic $E_T$ ($\vec{E}_T^{{\rm{had}}}$), 
defined as $-(\vec{E}_T^{e}+\vec{E}_T^{\mu}+\vec{
\hbox{$\rlap{\kern0.25em/}E_T$}})$ 
was required to be less than 40 GeV in magnitude to reduce the 
background from $t{\bar t}$ production. Table~\ref{tab-wweunevt} shows 
the number of events remaining after each selection cut.
One event survived all the requirements but the last; 
this event is a candidate for $t{\bar t}$ production and has been discussed
extensively elsewhere\cite{417}. 
\vbox{
\begin{table}
\begin{tabular}{|lc|} 
Event selection cut & Number of events \\ \hline \hline
$E_{T}^{e} \geq 20$ GeV & 9 \\
$p_{T}^{\mu} \geq 15$ GeV & 6 \\
\hbox{$\rlap{\kern0.25em/}E_T$}  $\geq 20$ GeV & 1 \\
$\Delta\phi(p_{T}^{\mu},\hbox{$\rlap{\kern0.25em/}E_T$} )$ cut & 1 \\
$|\vec{E}_{T}^{had}| \leq 40$ GeV & 0 \\ 
\end{tabular}
\caption{The numbers of events remaining after each selection criteria for the 
$WW\rightarrow e\nu\mu\nu$ analysis.}
\label{tab-wweunevt}
\end{table}}

The integrated luminosity of the data sample was
$13.9\pm 0.8$ pb$^{-1}$.
The trigger efficiency was largely determined by the trigger efficiency
for the muons and was estimated using data as was discussed in 
Section~\ref{sec-pid}. 
The detection efficiency for the SM ${WW \rightarrow e\mu\nu\bar{\nu}}$ events,
including the muon selection efficiency, the geometrical
acceptance, and the event selection efficiency,
was estimated using the {\small PYTHIA} and {\small D\O GEANT}
Monte Carlo simulation.
The muon selection efficiency was implemented in the {\small D\O GEANT}
Monte Carlo
program by introducing the measured hit efficiencies and resolutions of
the muon chamber modules.
The measured electron selection efficiency was implemented as a multiplicative
factor after the detector simulation. 
The detection efficiencies, including lepton identification efficiencies,
of individual fiducial regions are listed in Table~\ref{tab-wweueff}.  
The uncertainty on the efficiency for the regions involving EF muons was
dominated by the statistics of the {\small GEANT} Monte Carlo simulation. 
The overall detection efficiency of the SM $W$ pair 
events was estimated to be $\epsilon_{e\mu} = 0.092 \pm 0.010$.
The expected number of events was estimated to be
$N_{e\mu}^{SM} = 0.283 \pm 0.031 ({\rm stat}) \pm 0.037 ({\rm sys})$.
\vbox{
\begin{table}
\begin{tabular}{|lc|} 
Fiducial Region & Efficiency \\ \hline \hline
CC -- CF & $0.43 \pm 0.08$ \\
CC -- EF & $0.21 \pm 0.14$ \\
EC -- CF & $0.30 \pm 0.09$ \\
EC -- EF & $0.15 \pm 0.15$ \\ 
\end{tabular}
\caption{The combined efficiencies of trigger, electron and muon selection,
and kinematic event selection for individual fiducial regions in the
$WW\rightarrow e\mu \nu\bar{\nu}$ analysis. The fiducial regions are those 
of the charged leptons.}
\label{tab-wweueff}
\end{table}}

The backgrounds from $W\gamma$,
$Z\rightarrow \tau\tau \rightarrow e\mu\nu\nu\bar{\nu}\bar{\nu}$ 
and $t{\bar t}$ were estimated using the {\small PYTHIA} and {\small ISAJET}
Monte Carlo event generators followed
by the {\small D\O GEANT} detector simulation.
The background due to a jet misidentified as an electron was estimated by a
different method from the $ee$ channel,
since the accuracy of the estimate was limited by statistics
when that method was applied to the $e\mu$ channel.
Instead, the inclusive $W\rightarrow \mu\nu$ data were used to estimate this
background. 
Each jet in an event was treated as an electron and the event selection
requirements were applied.
The events that survived the criteria were weighted by the probability of
a jet being misidentified as an electron.
The misidentification probabilities were measured from data to be
$P^{CC}({\rm jet}\rightarrow e) = (0.9 \pm 0.4) \times 10^{-4}$ for CC and
$P^{EC}({\rm jet}\rightarrow e) = (4.0 \pm 1.0) \times 10^{-4}$ for EC.
The total ``fake" background from this source was calculated to be
$N_{{\rm fake}}^{BG} = 0.074 \pm 0.016 ({\rm stat}) \pm 0.074 ({\rm sys})$.
The background due to a jet faking a muon was estimated to be
negligibly small.
The total number of background events was estimated to be
$N_{e\mu}^{BG} = 0.264 \pm 0.052 ({\rm stat}) \pm 0.084 ({\rm sys}).$
The backgrounds to $WW\rightarrow e\mu \nu \bar{\nu}$
are summarized in Table~\ref{tab-wwdilepback}.

\subsection{The $\mu\mu$ channel}
\begin{figure}
\centerline{\epsffile{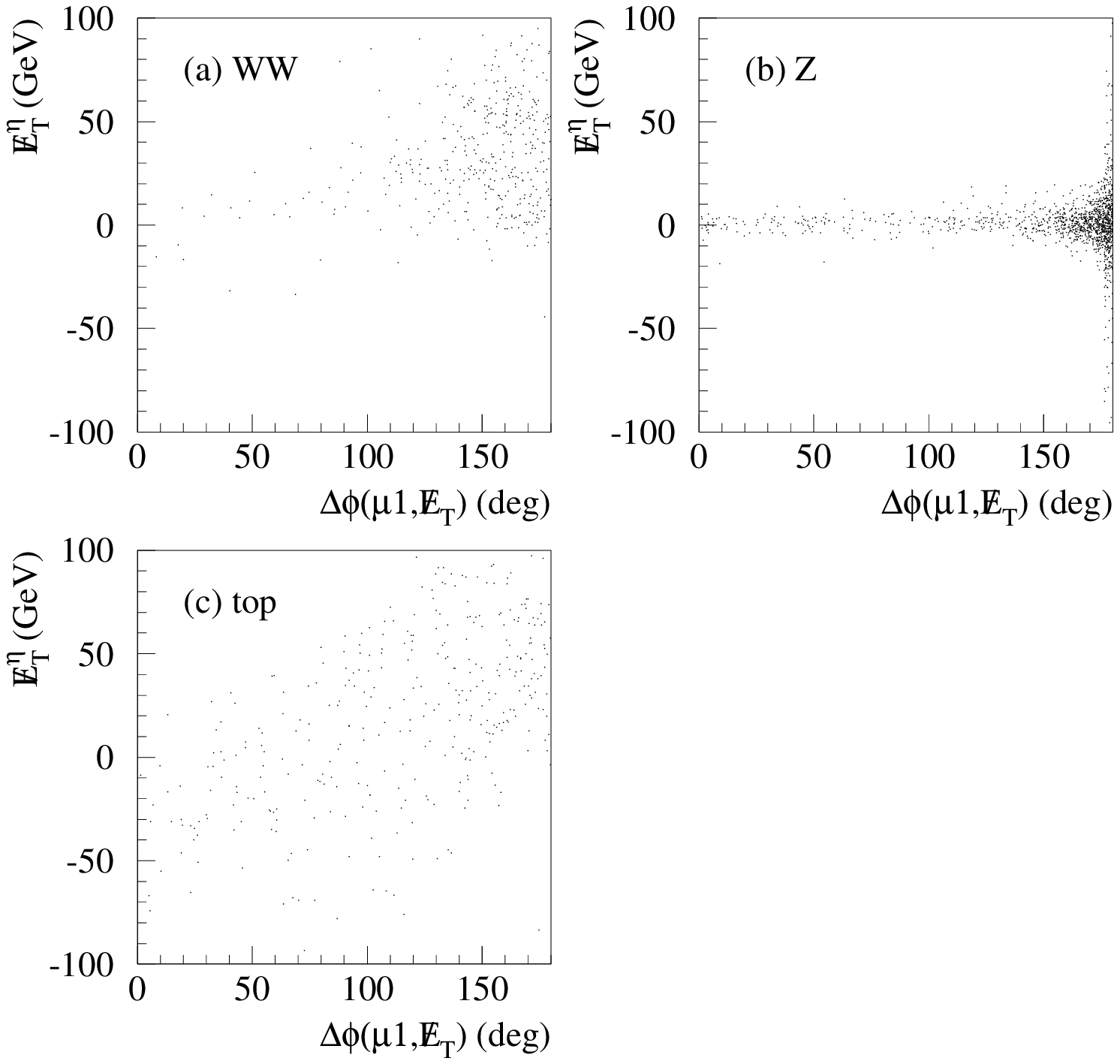}}
\caption{\hbox{$\rlap{\kern0.25em/}E_T
^{\eta}$} vs $\Delta\phi(p_{T}^{\mu1}, \hbox{$\rlap{\kern0.25em/}E_T$}
)$ distributions
for $WW \rightarrow \mu\mu \nu \bar{\nu}$ with SM couplings,
$Z \rightarrow \mu\mu$, and 
$t\bar{t}$ events. }
\label{fig-uucuts}
\end{figure}

The $WW \rightarrow \mu\mu\nu\bar{\nu}$ candidate events 
were selected from the 
data sample recorded with the {\small MU-MAX} trigger of 
Table~\ref{tab-trigs}. 
This trigger required at least one muon in $|\eta| < 1.7$ at levels 1 and
1.5 with $p_T > 7$ GeV/c (threshold determined by the hardware) 
and one muon with $p_T \geq 15$ GeV/c at level 2.
Candidate events with two muons that passed ``Tight I" requirements
were selected. 
The following event selection requirements were imposed. 
Both muons were required to have large transverse momentum:
$p_{T}^{\mu 1} \geq 20$ GeV/c and $p_{T}^{\mu 2} \geq 15$ GeV/c.
To reduce the background from $Z\rightarrow \mu\mu$ decays,
the \hbox{$\rlap{\kern0.25em/}E_T^{\eta}$} was required to be 
$\geq 30$ GeV, where
\hbox{$\rlap{\kern0.25em/}E_T^{\eta}$} was defined as the projection 
of the \hbox{$\rlap{\kern0.25em/}E_T$}
vector onto the bisector of the opening angle of the two muons in 
the transverse plane.  By selecting this component of the 
\hbox{$\rlap{\kern0.25em/}E_T$}, it was ensured that the 
\hbox{$\rlap{\kern0.25em/}E_T$}
was least sensitive to mismeasurements of the muon momentum. 
This selection requirement was also less sensitive to the momentum
resolution of the muons than was a dimuon invariant mass cut.
The \hbox{$\rlap{\kern0.25em/}E_T$}
was required not to be collinear to the higher momentum muon:
$\Delta\phi(p_T^{\mu 1}, \hbox{$\rlap{\kern0.25em/}E_T$}
)\leq 170^{\circ}$. This reduced the background from 
${Z \rightarrow \tau\tau \rightarrow \mu\mu\nu\nu\bar{\nu}\bar{\nu}}$. 
Figure~\ref{fig-uucuts} shows 
$\Delta\phi(p_{T}^{\mu},\hbox{$\rlap{\kern0.25em/}E_T$}
)$ vs \hbox{$\rlap{\kern0.25em/}E_T$} distributions.
The recoil hadronic $E_T$
($\vec{E}_T^{{\rm{had}}}$), 
defined as $-(\vec{E}_T^{\mu 1}+\vec{E}_T^{\mu 2}+\vec{
\hbox{$\rlap{\kern0.25em/}E_T$}})$ 
was required to be less than 40 GeV in magnitude, rejecting $t\bar{t}$
as in the other two channels.  Even though the $p_T$ resolution of the muons
was worse than that of the electrons, the resolution of this variable was the 
same in all three channels since the mismeasurement of the leptons
cancels when taken in a vector sum with the 
\hbox{$\rlap{\kern0.25em/}E_T$}. 
Table~\ref{tab-wwuucuts} shows the numbers of events after each selection cut.
No event survived all the selection cuts.
\vbox{
\begin{table}
\begin{tabular}{|lc|} 
Event selection cut & Number of events \\ \hline \hline
$p_{T}^{\mu1} \geq 20$ GeV/c & 102 \\
$p_{T}^{\mu2} \geq 15$ GeV & 88 \\
\hbox{$\rlap{\kern0.25em/}E_T$}$^{\eta} \geq 30$ GeV & 0 \\
$\Delta\phi(p_{T}^{\mu1},\hbox{$\rlap{\kern0.25em/}E_T$}
) \le 170^{\circ}$ & 0 \\
$|\vec{E}_{T}^{had}| \leq 40$ GeV & 0 \\ 
\end{tabular}
\caption{The number of events remaining after each selection criteria 
for the $WW \rightarrow \mu\mu \nu\bar{\nu}$ analysis.}
\label{tab-wwuucuts}
\end{table}}
\vbox{
\begin{table}
\begin{tabular}{|lc|} 
Fiducial Region & Efficiency \\ \hline \hline
CF -- CF & $0.023 \pm 0.006$ \\
CF -- EF & $0.009 \pm 0.002$ \\
EF -- EF & $0.0010 \pm 0.0006$ \\ 
\end{tabular}
\caption{The combined efficiencies of trigger and muon selection
for individual fiducial regions in the $WW \rightarrow \mu\mu \nu\bar{\nu}$
analysis. CF--CF is, for instance, the case that both
muons were in the central region. }
\label{tab-wwuueff}
\end{table}}

The integrated luminosity of the data sample was
$12.2\pm 0.7$ pb$^{-1}$.
The trigger efficiency was measured using data.
The detection efficiency for the SM $WW \rightarrow \mu\mu\nu\nu$ events
that included the muon selection efficiency, the geometrical
acceptance and the event selection efficiency
was estimated using the {\small PYTHIA} and {\small D\O GEANT}
Monte Carlo simulation.
The muon selection efficiency was implemented in the {\small D\O GEANT}
Monte Carlo
program by introducing the measured hit efficiencies and resolutions of
the muon chamber modules as in the $e\mu$ channel.
The efficiencies of individual fiducial regions are listed in
Table~\ref{tab-wwuueff}.
The overall detection efficiency for the SM $W$ pair events was estimated to
be $\epsilon_{\mu\mu} = 0.033 \pm 0.006$.
The expected number of events was estimated to be
$N_{\mu\mu}^{SM} = 0.045 \pm 0.004 ({\rm stat}) \pm 0.006 ({\rm sys}).$

The backgrounds from Drell-Yan dilepton,
$Z\rightarrow \tau\tau \rightarrow \mu\mu\nu\nu\bar{\nu}\bar{\nu}$
and $t{\bar t}$ processes were
estimated using the {\small PYTHIA} and {\small ISAJET}
Monte Carlo event generators followed
by the {\small D{\O}GEANT} detector simulation.
The background from $Z\rightarrow \mu\mu$ was estimated using the 
same fast simulation program as in the $ee$ channel.
The fake background due to a jet faking a muon was 
negligibly small.
The total number of background events was estimated to be
$ N_{\mu\mu}^{BG} = 0.077 \pm 0.023 ({\rm stat}) \pm 0.012 ({\rm sys}).$

\subsection{Limit on the cross section for $W$ boson pair production}
The results from the analyses of the $ee$, $e\mu$ and $\mu\mu$ channels are
summarized in Table~\ref{tab-wwsum}.
For the three channels combined, the expected number of events for 
SM $W$ boson pair production, based on a cross section of 
$9.5 \pm 1.0$ pb \cite{ohnemus}, was $0.47\pm0.07$. 
In approximately 14 pb$^{-1}$ 
of data, one event was found with an expected 
background of $0.56\pm 0.13$ events.
\vbox{
\begin{table}
\begin{tabular}{|lccc|c|} 
Channel        & $ee$   & $e\mu$   & $\mu\mu$  & Total   \\ \hline 
Efficiency     &$0.094 \pm 0.008$ 
                        &$0.092 \pm 0.010$ 
                                   &$0.033 \pm 0.006$
                                               &        \\
$N^{SM}$       &$0.15 \pm 0.01 \pm 0.02$ 
                        &$0.28 \pm 0.03 \pm 0.04$ 
                                   &$0.045 \pm 0.004 \pm 0.006$ 
                                               & $0.47 \pm 0.03 \pm 0.06$\\
$N^{BG}$       &$0.22 \pm 0.02 \pm 0.06$ 
                        &$0.26 \pm 0.05 \pm 0.08$
                                   &$0.077 \pm 0.023 \pm 0.012$ 
                                               & $0.56 \pm 0.06 \pm 0.10$ \\
                                                                 \hline
$N^{observed}$ & 1      & 0        & 0         & 1        \\ 
\end{tabular}
\caption{The summary of $WW \rightarrow$ dileptons analyses including
 the efficiency, number of SM events expected, expected backgrounds, and 
number of candidates observed.}
\label{tab-wwsum}
\end{table}}

The 95\% confidence level upper limit on the $W$ boson pair 
production cross section was estimated  based on one observed event, 
taking into account the expected background of $0.56\pm0.13$ events.  
Poisson-distributed numbers of 
events were convoluted with Gaussian uncertainties on the detection 
efficiencies, background and luminosity.  
For SM $W$ boson pair production, the upper limit for the cross section was 
$87$ pb at the 95\% confidence level.  

\subsection{Limits on the trilinear gauge boson couplings}
The limit on the $W$ boson pair production cross section
can be translated into limits on the anomalous gauge boson couplings.
The Monte Carlo program of Ref. \cite{HWZ} followed by a fast detector 
simulation  was used to estimate\cite{HJthesis} the detection efficiency for
$W$ boson pair production as a 
function of the coupling parameters $\lambda$ and $\Delta \kappa$.
It was assumed that the $W$ boson
couplings  to the photon and to the $Z$ boson were equal: $\lambda
\equiv \lambda_{\gamma}=\lambda_Z$ and $\Delta \kappa \equiv \Delta 
\kappa_{\gamma} = \Delta \kappa_Z$.
The form factor scale $\Lambda = 900$ GeV was chosen.  This was the 
highest value of $\Lambda$ that produced anomalous coupling limits within 
the corresponding unitarity bound in this analysis.  For smaller values of 
$\Lambda$, the anomalous coupling limits are looser (see Appendix 1).
The MRSD$-^{\prime}$ parton distribution functions were used in the event 
generation.

\begin{figure}
\epsffile{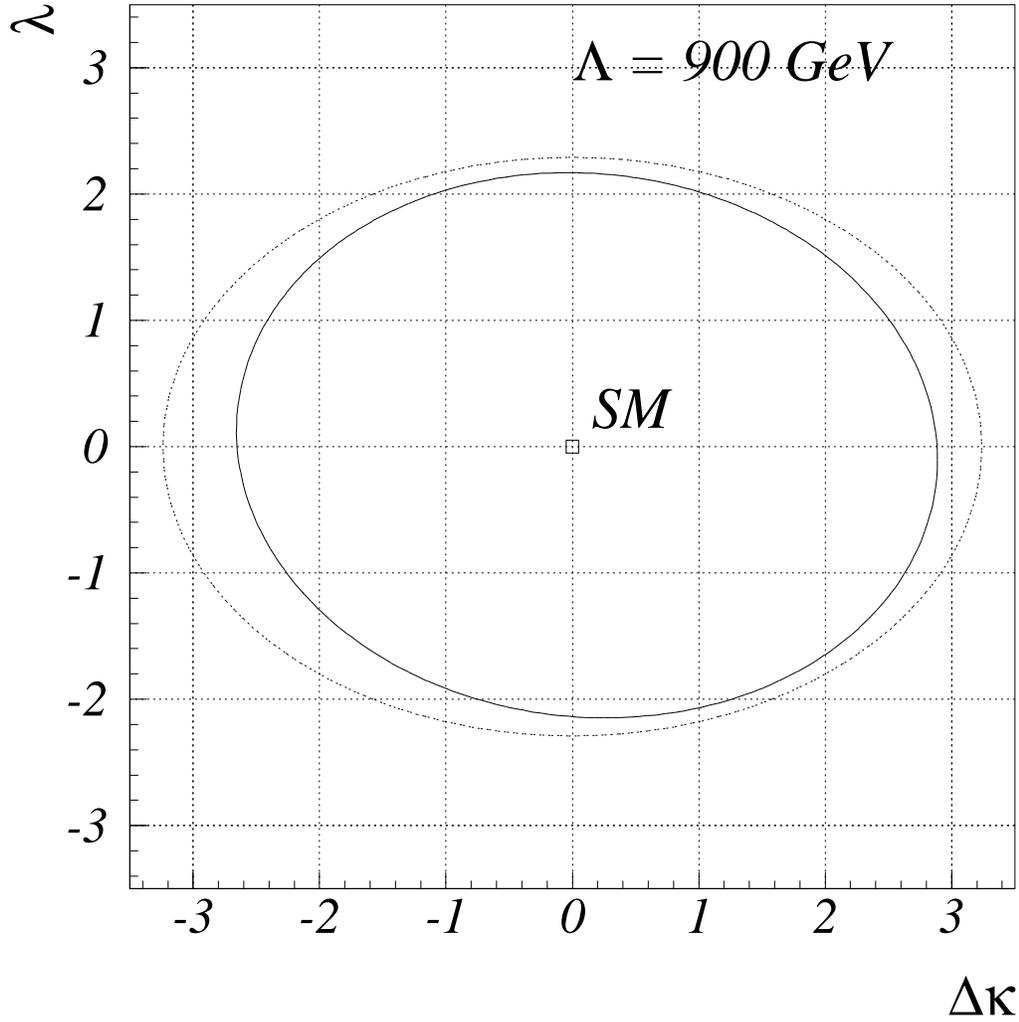}
\caption{95\% CL limits on the CP-conserving anomalous couplings $\lambda$ and 
$\Delta \kappa$, assuming that $\lambda_{\gamma} = \lambda_{Z}$ and 
$\Delta \kappa_{\gamma} = \Delta \kappa_Z$.
The dotted contour is the unitarity limit for the form factor scale 
$\Lambda = 900$ GeV which was used to set the coupling limits.}
\label{fig-DILEP_CONT}
\end{figure}

The number of $W$ boson pair events expected at each point in a grid of
$\lambda$ and $\kappa$, including SM production,
was fitted to the following equation which reflected
the general Lagrangian form of gauge boson self-interactions:
\begin{eqnarray*}
N(\Delta\kappa, \lambda) = & a_{1} + a_{2}\Delta\kappa
+ a_{3}(\Delta\kappa)^{2} \\ 
& + a_{4}\lambda + a_{5}(\lambda)^{2} + a_{6}\lambda \Delta\kappa,
\end{eqnarray*}
where the $a_i$ were parameters determined from the fit.
A 95\% CL limit contour on the coupling parameters $\Delta\kappa$ and 
$\lambda$ was formed by intersecting the parabolic 
surface of expected number of events 
with the plane of the 95\% CL upper limit on the observed 
number of events (with the background subtracted), fluctuated by 
Gaussian uncertainties on the detection efficiencies, backgrounds, and 
luminosity and by the Poisson uncertainty on the statistics of the 
observation. The 95\% CL upper limits on
the coupling parameters are shown in Fig.~\ref{fig-DILEP_CONT} (solid line).  
Also shown in Fig.~\ref{fig-DILEP_CONT} (dotted line) is the contour of the 
unitarity constraint on the  coupling limits for the form factor scale
$\Lambda = 900$ GeV. 
This value of $\Lambda$ was chosen so that
the observed coupling limits lie within this ellipse.
The limits on the CP-conserving anomalous coupling parameters were 
$-2.6<\Delta\kappa<2.8$ ($\lambda$ = 0) and 
$-2.1<\lambda<2.1$ ($\Delta\kappa$ = 0). 
The limits for the CP-violating 
parameters, $\tilde{\kappa}$ and $\tilde{\lambda}$, were similar. 
The limits on $\lambda$ and $\Delta \kappa$ exhibited almost no 
correlation, in contrast to limits from the $W\gamma$ analyses presented in 
Section~\ref{sec-wg} and in Refs.~\cite{wgua2,wgcdf}. 

\section{Search for Anomalous $WW$ and $WZ$ Production in the 
${\lowercase {e}}\nu$ jet jet channel}
\label{sec-wwev}
In this section, a search for anomalous $WW$ and $WZ$ production is 
presented.  The method was to identify $WW$ and $WZ$ candidates where 
one $W$ boson decayed to an electron and a neutrino and the $Z$ boson
or other $W$ boson decayed to two jets.  The expected cross section times 
branching fraction for the SM $WW$ and $WZ$ processes $(\sim 1.6 \; {\rm pb })$
was much smaller than that expected from the $W$ boson plus dijet background 
$ (\sim 76 \; {\rm pb})$ \cite{wjetsxscdf}, 
which had similar characteristics.  Rather than 
isolating the SM signal,  limits were set on the anomalous couplings by 
comparing the characteristics of the events with those expected from 
non-SM couplings.  Figure~\ref{fig-wzpt} shows the $p_T$ of the $W$ bosons 
for SM $WW$ production and for $WW$ production with anomalous 
trilinear couplings.  Anomalous couplings lead to a dramatic increase
in the cross section at high $p_T(W)$. 
To exploit this, the $p_T$ of $W$ bosons in the candidate events was 
measured, the contribution of the backgrounds to that spectrum was estimated,
and the data were compared with the sum of the background plus the 
expectations for the signal for various anomalous 
couplings. 

\subsection{Event Selection and Efficiency} 
\begin{figure}
\centerline{\epsffile{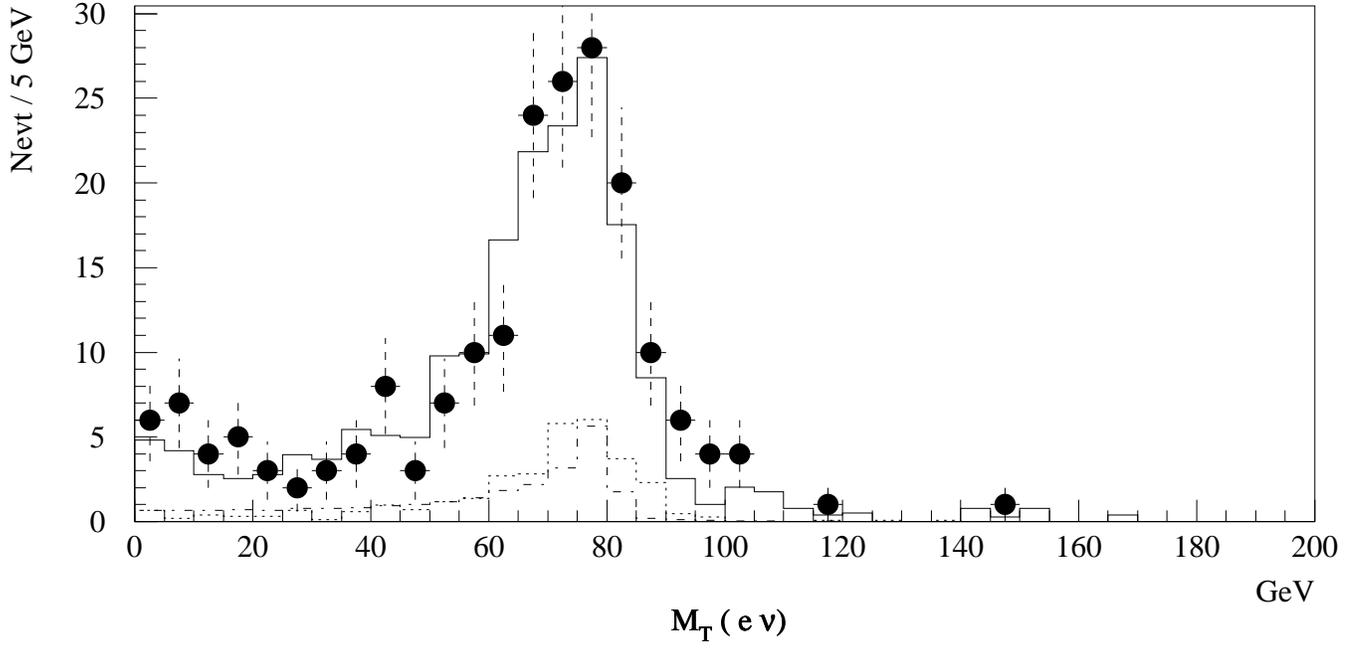}}
\caption{ The distribution of the transverse mass of the electron and 
\hbox{$\rlap{\kern0.25em/}E_T$} for the data (points), major 
backgrounds (solid line), 10 times SM $WW$ signal (dotted), and $WW$ with 
$\Delta \kappa =2$ $\lambda=0$ 
(dot-dash).  The backgrounds are normalized as described 
in the text.}
\label{fig-wwwz-mt}
\end{figure}

\begin{figure}
\centerline{\epsffile{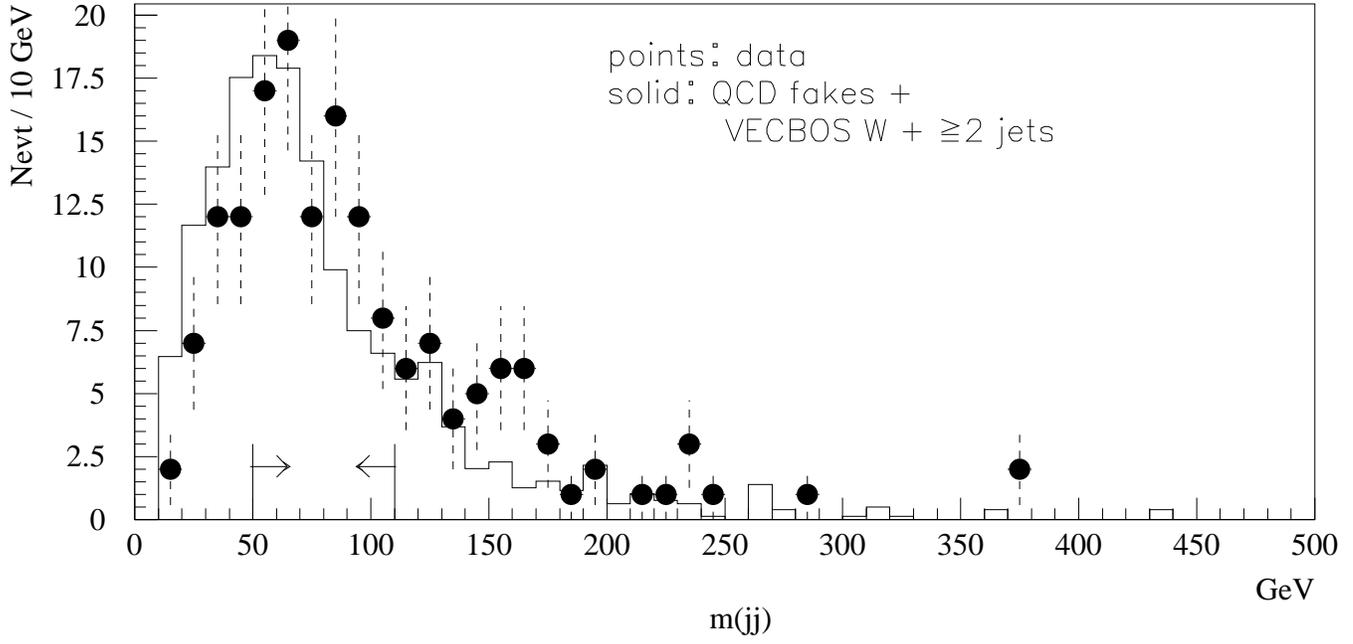}}
\caption{ The distribution of the dijet invariant mass 
for the data and major backgrounds. The backgrounds are normalized as described 
in the text. The arrows indicate the region accepted by the dijet mass 
selection criterion. }
\label{fig-wwwz-mjj}
\end{figure}

The $WW(WZ)\rightarrow e\nu jj$ candidates were selected by searching 
the data which passed the {\small ELE-HIGH} trigger (see 
Table~\ref{tab-trigs}) for events with a high $E_T$ electron 
accompanied by significant \hbox{$\rlap{\kern0.25em/}E_T$}
and at least two jets consistent with $W\rightarrow jj$ or $Z\rightarrow jj$.   
Events with electrons which satisfied the ``Tight IV" criteria within 
$|\eta|<2.5$ and with $E_T>25$ GeV were chosen. 
The \hbox{$\rlap{\kern0.25em/}E_T$} was required to be greater than 25 GeV,
and at least two jets were demanded, each with $E_T>20$ GeV and $|\eta|<2.5$.
A small jet cone size, ${\cal{R}}<0.3$, was used to ensure that 
the two jets from the $W$ or $Z$ decay,
close together for high-$p_T$ $W$ or $Z$ bosons, were resolved into 
distinct jets.  
After the jets were identified, a cleanup algorithm was applied to
remove events with ``fake" jets due to noisy cells or badly mismeasured 
jets, which occured  primarily in the intercryostat region.  
Figure~\ref{fig-wwwz-mt} shows the transverse mass of the electron and 
\hbox{$\rlap{\kern0.25em/}E_T$}, $M_T^{e\nu}$, for the 
candidates which survived the preceding selection criteria.  
$M_T^{e\nu}$ was required to be greater than 40 GeV/c$^2$.
The dijet invariant mass distribution of these events is shown in 
Fig.~\ref{fig-wwwz-mjj}.  In case there were more 
then two jets with $E_T > 20$ GeV in the fiducial region, the combination
yielding the largest invariant mass was taken to be the dijet mass of the 
candidate $W$ or $Z$ boson. Requiring the dijet invariant mass to be 
$50 < m_{jj} < 110 $ GeV/c$^2$ yielded 84 candidate events.

\begin{figure}
\centerline{
  \epsffile{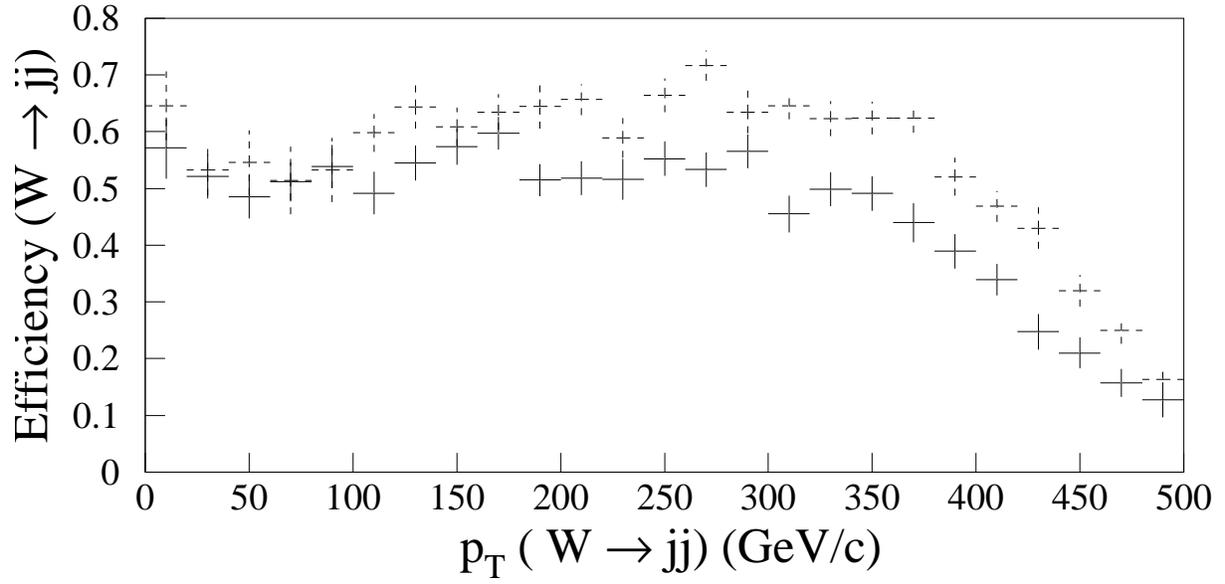}}
\vspace{1.0in}
\caption{Efficiency for reconstructing the dijets and for the dijet
mass selection for $W\rightarrow jj$ vs. $p_T(W)$. The solid crosses are the 
results from {\small ISAJET}. The dashed crosses are results from 
{\small PYTHIA}. }
\label{fig-wweffy_jets}
\end{figure}

The efficiency for identifying two separated jets depended on the $p_T$ 
of the $W$ boson.
For $p_T(W) < 125 $ GeV/c, the efficiency was dominated by the 
jet $E_T$ threshold. For $p_T(W) >350$ GeV, the efficiency 
was dominated by the probability for the two jets to merge into one in the 
reconstruction process (hence the use of the small cone size).  Using 
the {\small ISAJET} and {\small PYTHIA} event generators, followed by the 
detailed detector simulation, {\small D\O GEANT}, and the 
Shower Library described 
in Section~\ref{sec-detsimG}, the efficiency for reconstructing 
$W\rightarrow jj$ was estimated as a function of $p_T(W)$, including 
the jet-finding efficiency and the efficiency for the dijet mass requirement.
The $Z\rightarrow jj$ reconstruction efficiency 
was obtained in a similar manner. 
From the Monte Carlo it was determined that the use of the two highest $E_T$
jets to form the dijet mass was the correct assignment 90\% of the time.
Figure~\ref{fig-wweffy_jets} shows the efficiency for the dijet reconstruction
of $W\rightarrow j j$ as a function of $p_T(W)$ for events 
generated with {\small ISAJET} and {\small PYTHIA}.   
For the efficiency, the results from the {\small ISAJET} simulation were used 
because they were smaller than the efficiencies determined from {\small
PYTHIA}; the difference (9\%) was included in the systematic uncertainty 
for the efficiency.

The overall efficiency was calculated for SM and anomalous couplings using
the fast detector simulation described in Section~\ref{sec-mc} along 
with the $WW$ $(WZ)$ generator\cite{HWZ}.  
The reconstruction efficiencies for $W$ and $Z$ boson decays to dijets 
were incorporated as look-up tables. 
The $p_T$ distribution
of the $WW$ and $WZ$ systems was included in the simulation by using the 
observed $p_T(Z)$ spectrum from the inclusive $Z\rightarrow ee$ sample.  
The uncertainty in the 
absolute jet energy scale and in the jet corrections, $10$\%, 
was included in the systematic uncertainty by recalculating the results 
shifting the jet energies within their uncertainty.
Other sources of uncertainty included:
6\% for the uncertainty in the resolution of the 
\hbox{$\rlap{\kern0.25em/}E_T$} and 4\% for the uncertainty in the electron
identification efficiency.  All of the uncertainties in the efficiency are 
listed in Table~\ref{tab-wwwz-eff-unc}.  
\vbox{
\begin{table}
\begin{tabular}{|cc|}
 Source                    & Uncertainty  (\%)               \\ \hline \hline
Statistical                & $1$                             \\
Electron Efficiency        & $4$                             \\
\hbox{$\rlap{\kern0.25em/}E_T$} Smearing
                           & $6$                             \\
Jet Energy Scale           & $6$                             \\
Jet Reconstruction Efficiency 
                           & $9$                             \\ \hline \hline
Total                      & $13$                            \\
\end{tabular}
\caption{Summary of systematic uncertainties for the $WW/WZ\rightarrow 
e\nu jj$ analysis.} 
\label{tab-wwwz-eff-unc}
\end{table}}
The total efficiency for the detection of SM $WW$ and $WZ$ events was 
estimated to be $0.15\pm0.02$
and $0.16\pm0.02$, respectively.  Therefore the total number of expected 
(SM) signal was $3.2\pm0.6$ events where $2.8\pm0.64$ events were $WW$ and 
$0.4\pm0.1$ events 
were $WZ$, including the uncertainties in the efficiency and luminosity. 

\subsection{Background Estimate}
The background  included contributions from the following: 
$W + \geq 2$ jets; $t\bar{t}$ 
production with subsequent decay to $W^+W^-b\bar{b}$, where the top mass was 
assumed to be $180$ GeV/c$^2$; $WW(WZ)$ production 
with $W\rightarrow \tau\nu$ followed 
by $\tau\rightarrow e\nu\bar{\nu}$; $ZX\rightarrow ee X$, where one electron
was mismeasured or not identified; and multijet events, where one or
more jets was misidentified as an electron and there was significant  
\hbox{$\rlap{\kern0.25em/}E_T$}
due to mismeasurement or the presence of neutrinos.  

\begin{figure}
\centerline{
 \epsffile{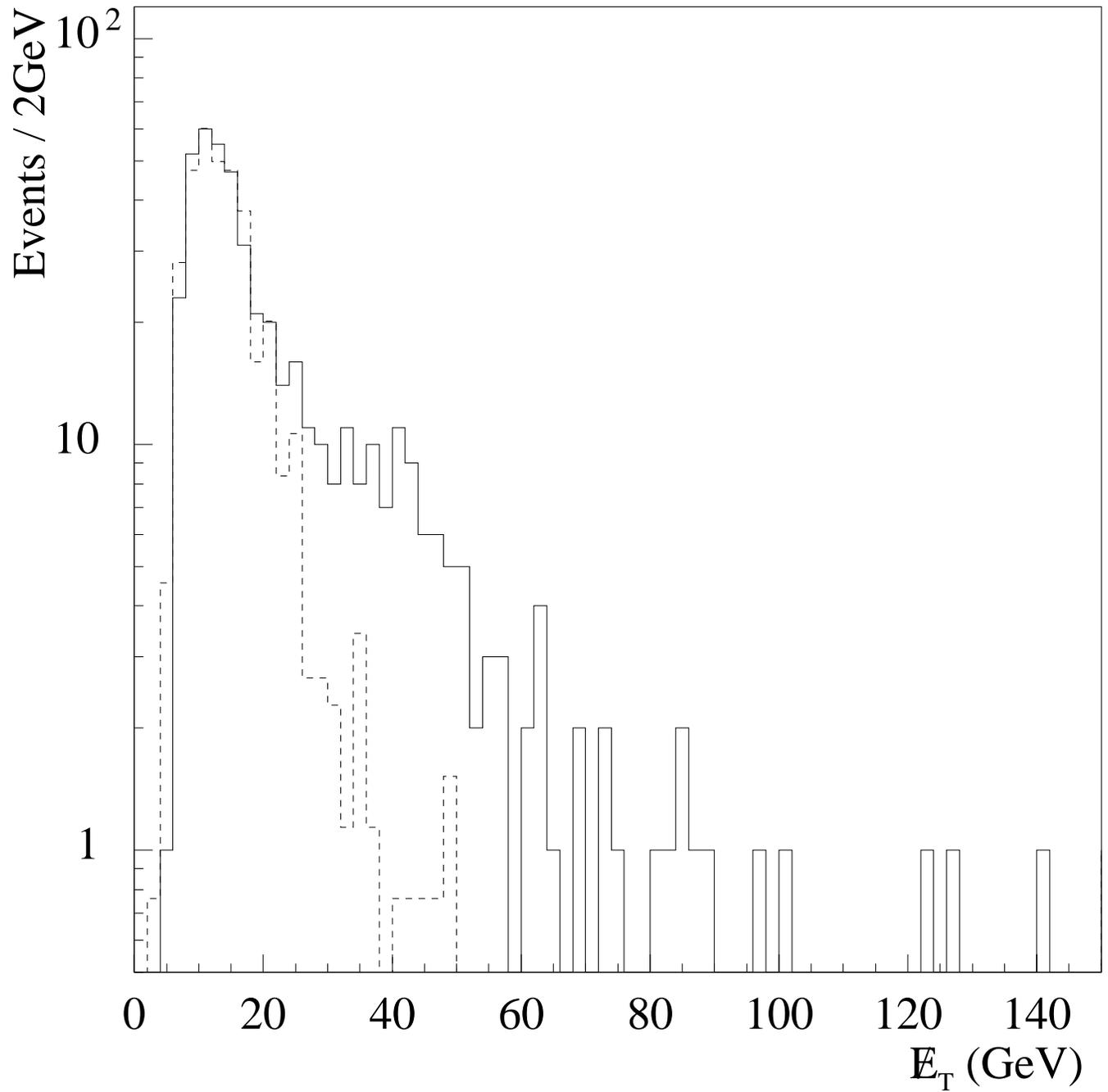}}
\vspace{0.5in}
\caption{Distribution of the \hbox{$\rlap{\kern0.25em/}E_T$} for 
the $WW/WZ$ candidates (solid) and the QCD fake sample (dashed) 
before the dijet mass selection. }
\label{fig-wwwz-qcd-fake-met}
\end{figure}

The multijet background was estimated following the same procedure
used in the $WW\rightarrow ee\nu\bar{\nu}$ channel.  The background sample
was comprised of events which contained a jet with an EM fraction greater 
than 0.9 within the electron fiducial region and a matching track. 
However, these electron candidates satisfied at least one of the following 
three electron ``anti-identification" criteria:  ${\rm f_{iso}} > 0.15$, 
H-matrix $\chi^2 > 250$, or track match significance, $TMS > 10$.
The number of events in the region 
$0 < \hbox{$\rlap{\kern0.25em/}E_T$} < 15$ GeV/c was used to normalize the 
fake sample to the signal sample.  This was done after all selections 
except for the dijet mass cut. Then the dijet mass selection 
criterion was applied to the 
fake sample to determine the number of background events. 
The possible signal contamination of the fake sample was included as a 7\% 
systematic uncertainty in the normalization of the fake background. 
A systematic uncertainty of  3.4\% comes from the variation in the fake event 
normalization when the upper end of the normalization region was varied in 
the range 12 to 18 GeV.  An uncertainty of 4\% arose from variation of the
\hbox{$\rlap{\kern0.25em/}E_T$} threshold of the signal over  the range 
22 to 28 GeV.  Figure~\ref{fig-wwwz-qcd-fake-met} shows the 
\hbox{$\rlap{\kern0.25em/}E_T$}
for the QCD background and for the signal candidates before the 
dijet mass selection. 

The backgrounds from $t\bar{t}$, $WW, WZ\rightarrow \tau\nu jj$ and 
$ZX\rightarrow eeX$ were estimated from {\small{D\O GEANT}} simulation of
{\small{PYTHIA}} and {\small {ISAJET}} events.  The background from 
$W+\geq $ 2 jets came from {\small VECBOS}\cite{vecbos} generated events
carried through a hybrid detector simulation which combined the 
{\small {D\O GEANT}} detector simulation with the 
parton-based jet Shower Library.
The normalization of the $W+\geq $ 2 jets background, made before the
dijet mass selection, was determined from the comparison of the number of 
candidate events outside the dijet mass window with that expected from 
the {\small {VECBOS}} Monte Carlo, after subtracting the multijet backgrounds
and expected SM signal.  
The systematic uncertainty in the $W+\geq $ 2 jets background included 
contributions from uncertainty in the fake normalization amounting to
7\%; variation when the dijet mass window was increased in width to
$40 < m_{jj} < 120$ GeV/c$^2$, amounting to 10\%; and variation in the 
background when the Monte Carlo jet energy scale was increased by 10\%,
amounting to 11\%.  The cross section for the resulting $W+\geq $ 2 jets 
background agreed within 1.5\% of the {\small {VECBOS}} expectation. 
The background estimate is summarized in Table~\ref{tab-wwwzback}. 
\vbox{
\begin{table}
\begin{tabular}{|c|c|c|}
Background Source & \multicolumn{2}{c|}{Number of Expected Events} \\ 
                  & Before Dijet Mass Cut  & After Dijet Mass Cut \\ \hline 
                                                                     \hline
$W+\geq $ 2 jets  & $125.4\pm25.9$         & $62.2\pm13.0$             \\ 
$t\bar{t}$        & $3.42\pm0.47$          & $0.87\pm0.12$             \\ 
$WW, WZ\rightarrow \tau\nu jj$
                  & $0.24\pm0.02$          & $0.22\pm0.02$             \\ 
$ZX\rightarrow eeX$
                  & $0.00^{+0.34}_{-0.00}$ & -                         \\ 
multijets         & $30.0\pm 4.5$          & $12.2\pm2.6$   \\ \hline \hline
Total Background  & $159 \pm 26 $        & $75.8\pm13.3$  \\ \hline \hline
SM $WW + WZ$ Prediction 
                  & $3.4\pm 0.6$           & $3.2\pm0.6$    \\ \hline \hline
Data              & 166                    & 84                      \\ 
\end{tabular}
\caption{Summary of $WW (WZ)\rightarrow e\nu jj$ backgrounds and data.}
\label{tab-wwwzback}
\end{table}}

\begin{figure}
\centerline{\epsffile{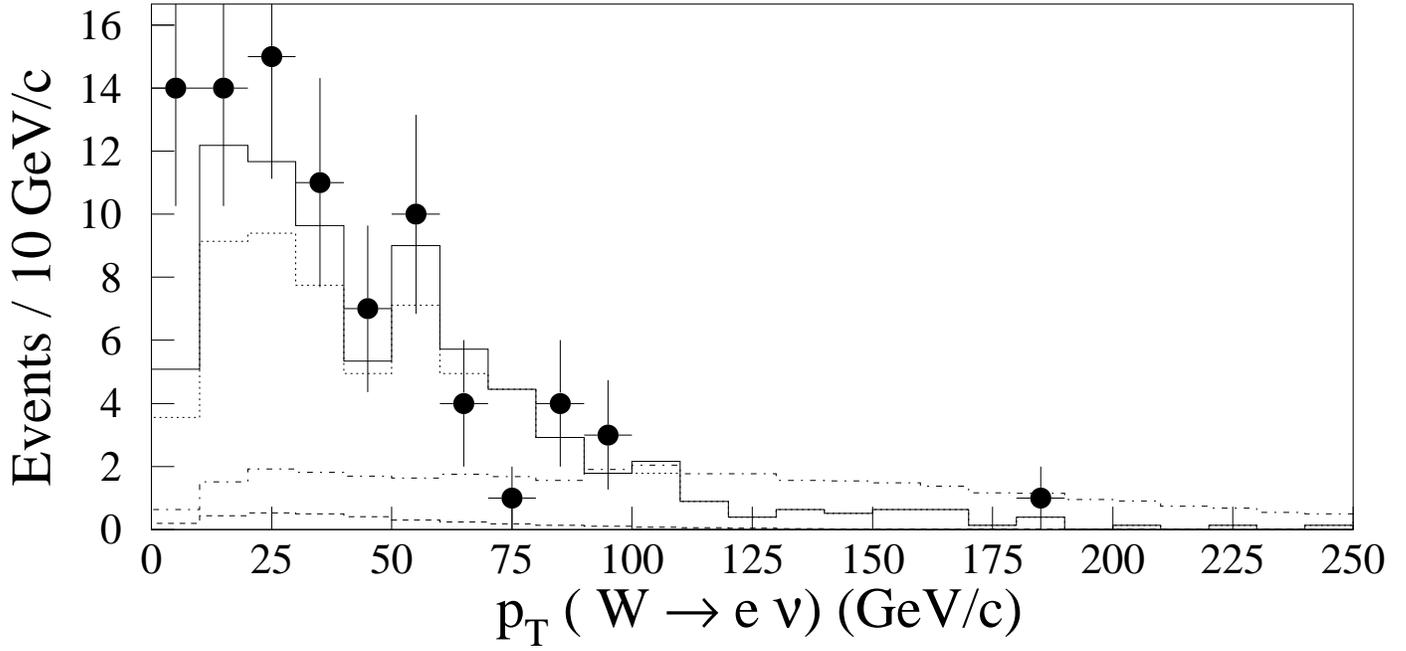}}
\vspace{2.25in}
\caption{$p_T$ distributions of the $e\nu$ system: data (points), 
$W+\geq $ 2 jets background (dotted), total background (solid), and 
Monte Carlo predictions for the SM (dashed) and non-SM couplings 
$\Delta \kappa =2$ $\lambda=0$ (dot-dashed) $WW$ production.}
\label{fig-wwwz_pt}
\end{figure}

The distributions in $p_T(e\nu)$  of the final event sample, and for the 
$W+\geq $ 2 jets background, the total background, the 
SM $WW$ and $WZ$ Monte Carlo, and the 
$WW$ and $WZ$ Monte Carlo for a non-SM value 
of the couplings $(\Delta\kappa_Z = \Delta\kappa_\gamma = 2, \lambda_Z = 
\lambda_\gamma = 1.5)$ are shown in Fig.~\ref{fig-wwwz_pt}.  The $p_T$
spectrum was consistent with that expected from the background. 
The highest $p_T$ event, important in setting anomalous coupling limits,
had $p_T(e\nu)=186\pm 20$ GeV/c. 
There were no other candidates with $p_T(e\nu)\geq 100$ GeV/c.  

Using the detection efficiencies for SM $WW$ and $WZ$ production and the 
background-subtracted signal, and assuming the SM ratio of cross sections
for $WW$ and $WZ$ production, an upper limit at the 95\% confidence
level (CL) on the cross section $\sigma (p\bar{p}\rightarrow W^+W^-X)$ of
$183$ pb was determined.

\subsection{Determination of Limits on Anomalous Couplings}
The absence of an excess of events with  high $p_T(W)$ excluded large
deviations of the trilinear couplings from the SM values.  
The $p_T$ spectrum expected at each element 
in a 225 point grid in $\lambda$ and  
$\Delta \kappa$ space, centered around and including the SM values, was 
obtained 
using Monte Carlo simulation.  The assumptions on the $\Lambda$ scale and
on the relation between the $WW\gamma$ and $WWZ$ couplings 
affected the 
$p_T$ spectra.  
Unequal width bins were used in order to
evenly distribute the events, particularly at the ends of the 
spectrum.  
An analytic form of the prediction of the number of events in each 
$p_T$ bin was obtained with a quadratic function of the coupling parameters, 
similar to that used in the dilepton analysis, fit to the number of 
events for each pair of anomalous couplings.
The difference between the estimated number of events and the fit was 
calculated for each pair of anomalous couplings for a particular $p_T$ 
bin and found to be less than 10\%. This value was included as a 
systematic uncertainty in the fitting procedure.
To set limits on the anomalous couplings, a binned likelihood fit was 
performed on the $p_T(W)$ spectrum of the expected signal plus background 
for $p_T(W) > 25$ GeV/c.  In each $p_T$ bin, the probability was calculated
for the predicted number of events to fluctuate to the observed number 
of events.  The uncertainties in the efficiency, background estimates, and 
total luminosity were convoluted in the likelihood function using Gaussian
distributions.  This likelihood
fit procedure is described in detail in Appendix~2.

\begin{figure}
\centerline{
 \epsffile{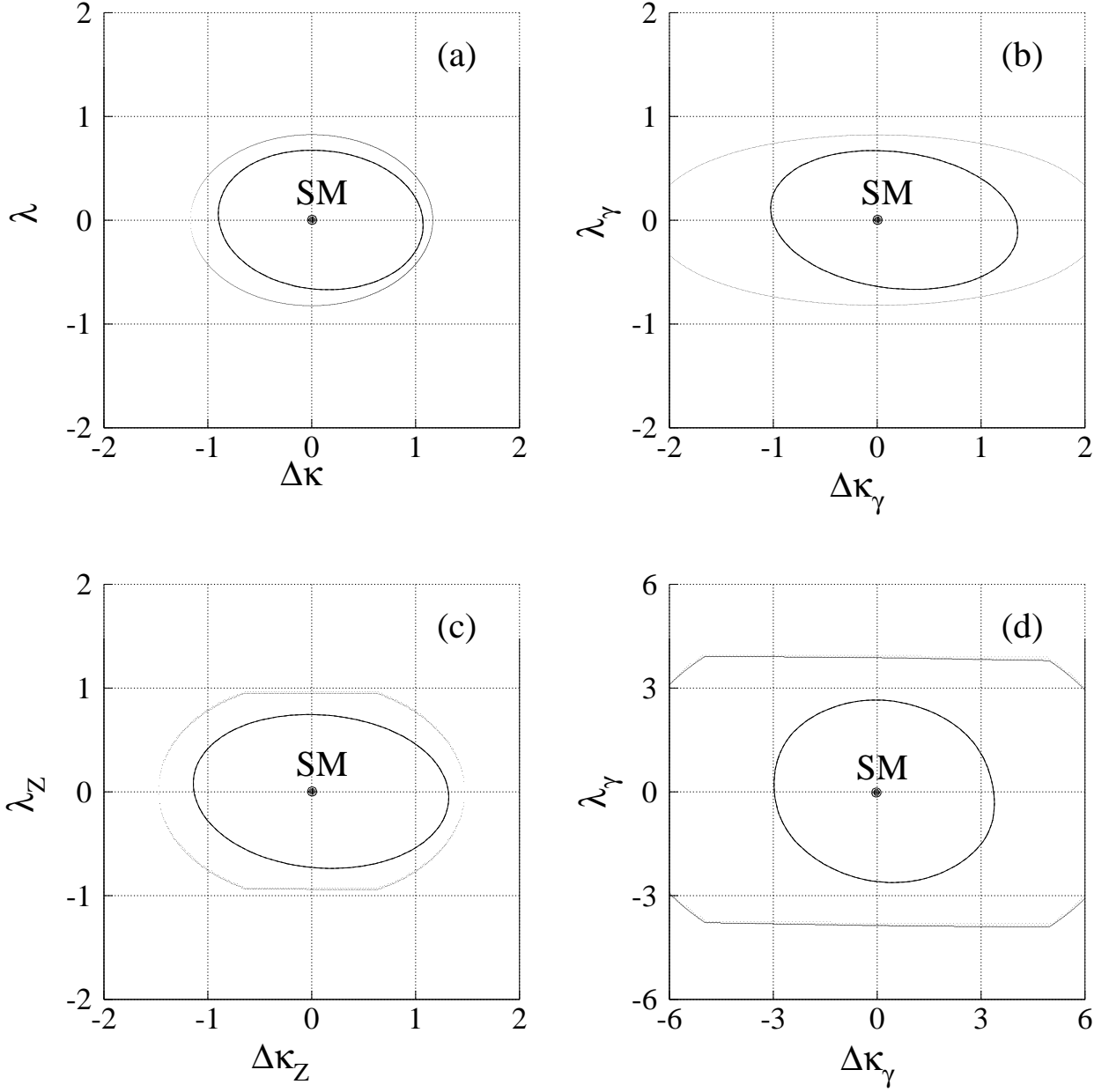}}
\caption{Contour limits on anomalous coupling parameters at the 95\% CL 
(inner curves) and limits from $S$-matrix unitarity (outer curves) for the 
assumptions a) $\lambda_{\gamma} = \lambda_Z$ and $\Delta \kappa_{\gamma} =
\Delta \kappa_{Z}$,  b) HISZ relations, c) SM $WW\gamma$ couplings and d)
SM WWZ couplings.  $\Lambda = 1500$ GeV was used for (a), (b), and (c). 
$\Lambda = 1000$ GeV was used for (d). }
\label{fig-wwwz-lam-dk}
\end{figure}

Limits were obtained on the coupling parameters under four sets of 
assumptions on the relations among the coupling parameters. 
For all four assumptions, the most likely point in the $\lambda - 
\Delta \kappa$ grid was the SM point. 
For the assumption $\Delta \kappa \equiv \Delta \kappa_{\gamma} = 
\Delta \kappa_{Z}$, and $\lambda \equiv \lambda_{\gamma} = \lambda_Z$ with 
$\Lambda = 1500$  GeV, 
the contours for the 95\% CL limit on
$\lambda$ and $\Delta \kappa$, with $\Lambda = 1500$ GeV, are shown 
in Fig.~\ref{fig-wwwz-lam-dk}(a).  The 95\% CL limits were
$-0.9\leq \Delta\kappa \leq 1.1 \;\; (\lambda = 0 )$ and 
$-0.6 \leq \lambda \leq 0.7 \;\; (\Delta \kappa = 0)$.
As in the $WW\rightarrow$ dileptons analysis, the limits on $\lambda$ and 
 $\Delta \kappa$ exhibited almost no correlation.
Under the HISZ relations\cite{HISZ}, which 
parameterize the $WWZ$ couplings in terms of the $WW\gamma$ couplings:
$\Delta \kappa_Z = 0.5\, \Delta\kappa_{\gamma}\, (1-\tan^2{\theta_{w}}), \;
 \Delta g_Z = 0.5\, \Delta\kappa_{\gamma}/\cos^2{\theta_w}, \; \lambda_Z = 
\lambda_{\gamma}$, the 95\% CL coupling
limit contours with $\Lambda = 1500 \;{\rm{GeV}}$ are shown in 
Fig.~\ref{fig-wwwz-lam-dk}(b).  The limits were
$-1.0\leq \Delta\kappa_{\gamma} \leq 1.3 \;\; (\lambda_{\gamma} = 0 )$ 
and $ -0.6 \leq \lambda_{\gamma} \leq 0.7 \;\; (\Delta \kappa_{\gamma} = 0)$.
Under the assumption that the $WW\gamma$ couplings have the SM value, the 
95\% CL upper limit contour, in $\lambda_Z$ and $\Delta \kappa_Z$,
is shown in Fig.~\ref{fig-wwwz-lam-dk}(c).
The 95\% CL limits were
$-1.1\leq \Delta\kappa_{Z} \leq 1.3 \;\; (\lambda_{Z} = 0 )$ and 
$-0.7 \leq \lambda_{Z} \leq 0.7 \;\; (\Delta \kappa_{Z} = 0)$.
Under the assumption that the $WWZ$ couplings have the SM value, the
95\% CL upper limit contour, in $\lambda_{\gamma}$ and $\Delta 
\kappa_{\gamma}$,  is shown in Fig.~\ref{fig-wwwz-lam-dk}(d).
Here the $\Lambda$ scale was 1000 GeV.
The 95\% CL limits were
$-2.8\leq \Delta\kappa_{\gamma} \leq 3.3 \;\; (\lambda_{\gamma} = 0 )$ and 
$-2.5 \leq \lambda_{\gamma} \leq 2.6 \;\; (\Delta \kappa_{\gamma} = 0)$.
The limits from $S$-matrix unitarity are also shown in 
Fig.~\ref{fig-wwwz-lam-dk}(a)--(d) for each assumption.  The unitarity 
limits were ellipses for (a) and (b) due to the form of 
Equation~\ref{eq-int-ul}, shown in Appendix 1. 
However, for (c) and (d), the intersections of the $W\gamma$ and $WW/WZ$ 
unitarity contours are shown in the Figure.

Because this analysis accounted for the background in fitting the  
spectrum for $p_T(W) > 25$ GeV/c, it was sensitive to anomalous 
couplings at both large and small $\hat{s}$.   All of the results of the 
fits were insensitive to the $p_T(W)$ threshold when varied between 
25 and 130 GeV/c.  In contrast, the analysis in Ref.~\cite{CDFWZ}, which 
required $p_T(W) > 130$ GeV/c, loses sensitivity at small $\hat{s}$; 
deviations from the SM restricted to $\hat{s}<500$ GeV could have been 
missed\cite{UCLA}. 

\section{Combined $W\gamma$ and $WW/WZ$ Anomalous Coupling Results}
\label{sec-cl}
The $WW\rightarrow {\rm dileptons}$ counting experiment and $WW/WZ$ $p_T$ 
spectrum analysis are sensitive to the same $WW\gamma$ couplings as the
$W\gamma$ photon spectrum analysis. 
The three analyses can be combined to form tighter limits on anomalous 
couplings. In this section, the procedure and results of the combined 
fit are discussed.  

The likelihood method used in the $WW/WZ$ $p_T$ analysis and $W\gamma$ 
photon spectrum analysis was used in the combined analysis. The joint 
log-likelihood was the sum of the log of the probabilities, as discussed in 
Appendix~2.
The likelihood was formed from the Monte Carlo $WW/WZ$ $p_T$ spectrum 
and Monte Carlo $W\gamma$ photon spectrum, expected background, and 
observed number of events in each channel with identical binning as was used 
in the separate analyses.  The expected number of 
$WW\rightarrow {\rm dilepton}$ events was recalculated for $\Lambda =
1500$ GeV (equivalent to use of a single bin for all $p_T(W)$); while the 
$\lambda$ and $\Delta \kappa$ limits would have violated 
unitarity for this value of $\Lambda$, the combined limit does not. 
Common systematic uncertainties, including lepton identification efficiency
(4\% for all channels with an electron in the final state and 12\% 
for all channels with a muon in the final state),
integrated luminosity (5.4\%), and choice of parton distribution 
function (9.1\%), were treated as discussed in Appendix~2. 
The limits are insensitive to a change in the size of the 
common systematic uncertainty by as much as a factor of two.
The statistical uncertainties of the data dominate the uncertainty in  
the analysis. 

\begin{figure}
\epsffile{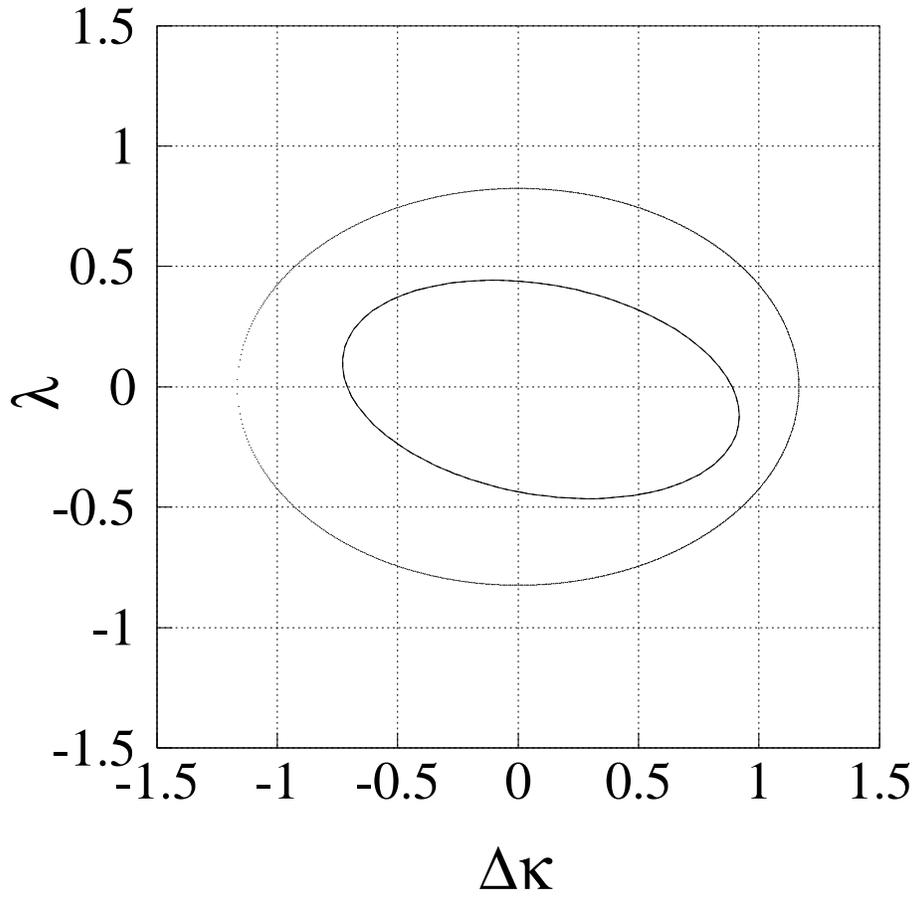}
\vspace{1.5in}
\caption{95\% CL limits (inner contour) on $\lambda$ and $\Delta \kappa$, 
assuming the $WW\gamma$ and $WWZ$ couplings are equal and $\Lambda = 1500$ GeV,
from the combined $W\gamma$, $WW$, and $WZ$ results.  
The outer contour is the limit from $s$-matrix unitarity. }
\label{fig-CombinedCL}
\end{figure}

The following results were obtained. For the assumption that the $WW\gamma$
couplings are equal to the $WWZ$ couplings and with $\Lambda = 1500$ GeV,
the 95\% CL limits were
$-0.71\leq \Delta\kappa \leq 0.89 \;\; (\lambda = 0 )$ and 
$-0.44 \leq \lambda \leq 0.44 \;\; (\Delta \kappa = 0)$.
Figure~\ref{fig-CombinedCL} shows the 95\% CL limit contour for
$\lambda$ and $\Delta\kappa$ along with the unitarity contour.

\section{$Z \gamma$ Production}
\label{sec-zg}
A measurement of the $ZZ\gamma$ and $Z\gamma\gamma$ couplings  
using $p\bar p \to \ell\ell\gamma +  X~ (\ell=e,\mu, \nu )$  events is 
discussed in this section. 

The signature for $Z\gamma$ events was two high-$p_T$ leptons 
($e^+e^-$, $\mu^+\mu^-$ or $\nu\bar{\nu}$),  and a photon.   
The leptons would not necessarily have combined to give the $Z$ boson mass.
In initial state radiation and anomalous coupling events, of the type shown 
in Figs.~\ref{fig-vpair}(a)--(c), the dilepton invariant mass for the 
electron 
and muon decay channels would be at the $Z$ boson mass.  However, for events
with bremsstrahlung radiation from a charged lepton, as shown in 
Fig.~\ref{fig-zgrad}, the two leptons would have a pair mass below that of
the $Z$ boson.
Furthermore, photons radiated from the leptons would have tended to be close
to the leptons.   
The neutrino decay channels had several important differences. 
Besides having a higher branching fraction than 
the electron and muon decay channels (20.0\% for three generations of 
neutrinos vs. 3.37\% for $ee$ or $\mu\mu$), the $Z\rightarrow \nu\nu$ 
decays are inferred with high efficiency in the detector through the 
\hbox{$\rlap{\kern0.25em/}E_T$} measurement.
The radiative diagrams do not contribute to the neutrino decay channel. 
Thus, the cross-section changes more quickly with anomalous couplings 
than the cross section for the electron and muon channels.  
The signature for these events was a 
photon recoiling against the \hbox{$\rlap{\kern0.25em/}E_T$} of the undetected
neutrinos.  The main disadvantage of the neutrino channel was that the 
backgrounds were larger than in the other channels.

\subsection{The $ee$ channel}
The $ee\gamma$ sample was selected from events which satisfied the {\small
ELE-2-MAX} trigger described in Table~\ref{tab-trigs}. 
The data set corresponded  to an integrated luminosity of 
$14.3 \pm 0.8\; {\rm pb}^{-1}$.
From this sample, candidate events were required to have two electrons 
in the fiducial region with $E_T >17$ GeV.  At least 
one electron had to satisfy the ``Tight II" requirements (see 
Table~\ref{tab-emid}) while the other satisfied the ``Loose I"
requirements.  This combination of tight and loose electron selection was 
possible because the backgrounds from fake electrons were small, relative to 
the expected signal, when the photon was required to pass the ``Loose"
requirements (see Table~\ref{tab-phoid}) within the fiducial region.
After trigger, fiducial region, and particle selection 
criteria were applied, 10 events with two electrons with $E_T>17$ GeV and 
a photon with $E_T > 5$ GeV survived. Final selected events 
were required to have $E_T^e > 25$ GeV and a photon separated from each 
electron by $\Delta {\cal R}_{\ell \gamma} > 0.7$ with $E_T^{\gamma} > 
10$ GeV.   These last two requirements reduced the contribution
of radiative events.  Four events survived in the final sample.
Table~\ref{tab-zges} indicates the number of 
events surviving the last few selection criteria.  For details on the 
characteristics of individual events and for event displays see 
Ref.~\cite{glthesis}. 
\vbox{
\begin{table}
\begin{tabular}{|c|c|} 
Selection Criteria             & No. of Surviving Events  \\ \hline 
Starting Sample                & 77                       \\ 
Fiducial and Particle ID       & 15                       \\ 
Trigger Criteria               & 10                       \\ 
$E_T^{ele} > 25$ GeV           & 10                       \\ 
$\Delta {\cal R}_{e\gamma}>0.7$& 8                        \\ 
$E_T^{\gamma}> 10$ GeV         & 4                        \\ 
\end{tabular}
\caption{Number of $ee\gamma$ candidates which passed the selection criteria.}
\label{tab-zges}
\end{table}}

The trigger efficiency for SM $Z\gamma$ production was estimated using the 
$Z\rightarrow ee$ event sample. 
It was found to be $0.98\pm 0.01$.  
The acceptance for SM $Z\gamma$ production and 
for production via anomalous $ZZ\gamma$ and $Z\gamma\gamma$ couplings 
was estimated using the event generator of Ref.~\cite{gl20} combined with the
fast detector simulation discussed in Section~\ref{sec-mc}. 
${\rm MRSD-^{\prime}}$ structure functions~\cite{MRSD} were used in the 
 event generation and the cross section was scaled by a k-factor of 1.34.
The geometric acceptance for SM production was 53\%. Averaged over $E_T$ for
SM production, the photon identification efficiency was also $0.53\pm0.05$. 
With the particle identification criteria, 
the kinematic, and the fiducial requirements on the electrons and photons 
described above, the selection efficiency for SM $Z\gamma$ production was 
$0.17\pm0.02$ and the expected cross section times efficiency was 
$0.20\pm0.02$ pb.

The background included contributions from 
$Z + {\rm{jet(s)}}$ production where one
of the jets mimicked an electron or photon, multijet production where more than
one jet was misidentified as a photon or electron, and $\tau\tau\gamma$
production followed by decay of each $\tau$ to 
$e\bar{\nu}_{e}\nu_{\tau}$. 

Processes where jets mimicked photons, jets mimicked electrons, and 
double and triple fakes contributed to the QCD background.
The background and its  $E_T$-dependence were  estimated by counting the 
 number of $ee + {\rm jet(s)}$ and  $e\gamma + {\rm jet(s)}$ events,  
 with the electrons  and photons passing the signal  cuts and with jet 
 transverse  energy above 10  GeV and 25 GeV, respectively.   The probabilities
 for jets to mimic EM objects were determined with a procedure similar to that
 described in Section~\ref{sec-wgbacksec} and observed to be approximately 
 $E_T$  independent (see Fig.~\ref{fig-fake}). Table~\ref{tab-zgfp} contains 
 the probabilities for a jet to mimic  photons and electrons in the CC and EC, 
 and the final probabilities with the direct photon 
 contribution removed.   Multiplying  these probabilities by the number of  
 jets in  these  samples led to a background of  $0.43 \pm 0.06$ QCD events.
\vbox{
\begin{table}
\begin{tabular}{|c|c|c|c|}
Type of Fake  & CC & EC & Avg. after direct $\gamma$ correction
                                                 \\ \hline \hline
Jet$\rightarrow \gamma$  
              & $(0.84\pm0.08)\times 10^{-3}$ 
                   & $(0.90\pm0.11)\times 10^{-3}$ 
                        & $(0.65\pm0.18)\times 10^{-3}$          \\ \hline
Jet$\rightarrow e_{{\rm{Tight II}}}$ 
              & $(0.62\pm0.07)\times 10^{-3}$ 
                   & $(1.5\pm0.2)\times 10^{-3}$ 
                        & $(0.84\pm0.10))\times 10^{-3}$          \\ \hline
Jet$\rightarrow e_{{\rm{Loose I}}}$  
              & $(1.7\pm0.1)\times 10^{-3}$ 
                   & $(1.6\pm0.2)\times 10^{-3}$ 
                        & $(1.5\pm 0.2)\times 10^{-3}$           \\
\end{tabular}
\caption{Probability for a jet to mimic a photon or electron,  averaged over 
$E_T$,  for the $Z\gamma\rightarrow ee \gamma$ and $Z\gamma\rightarrow \mu\mu
\gamma$ electron and photon selection criteria. }
\label{tab-zgfp} 
\end{table}}

\begin{figure}
\centerline{
  \epsffile{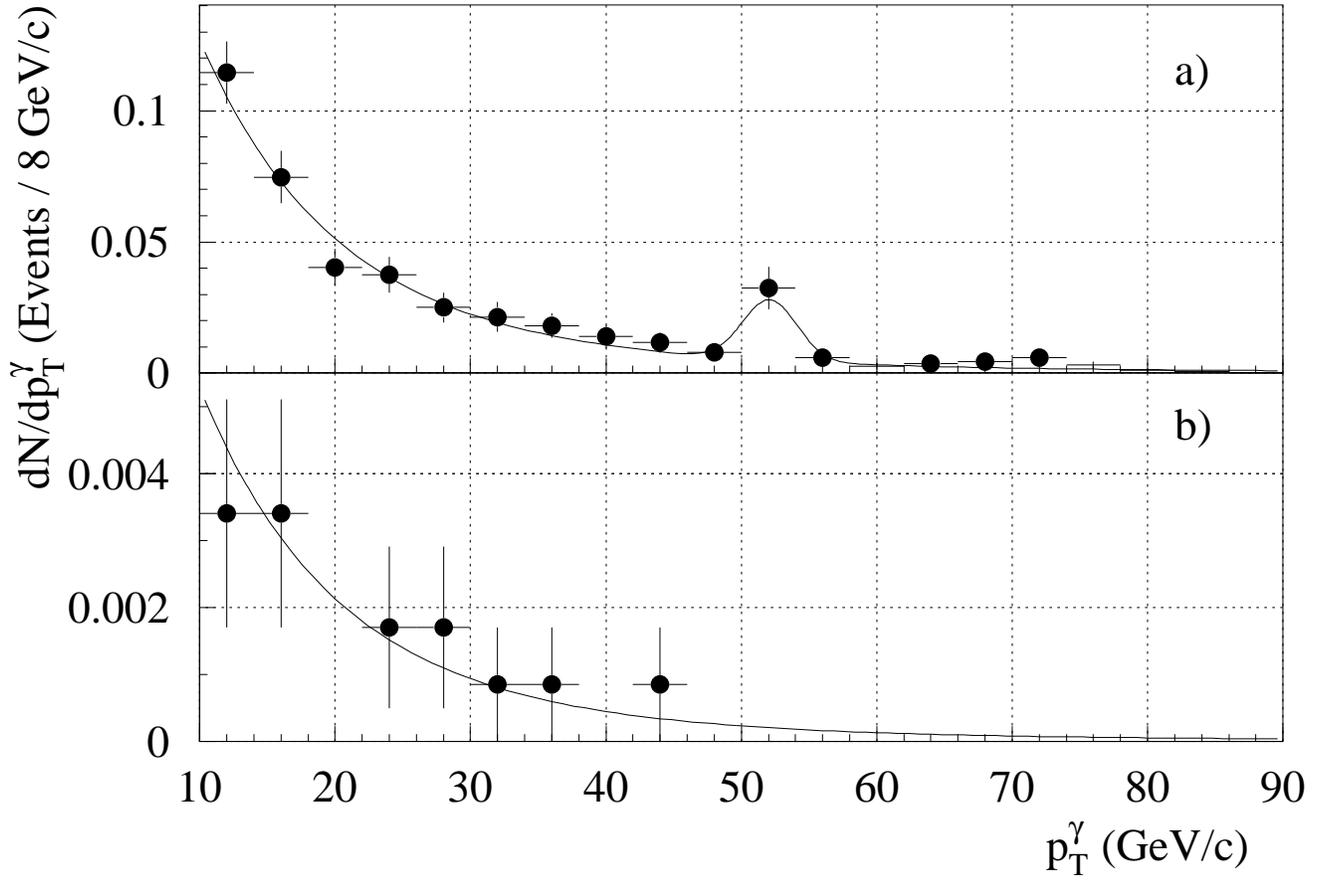}}
\caption{a) QCD background in the $ee\gamma$ channel. The bump around 50 GeV/c
is due to the electrons from $Z$ boson decays where a jet mimics an electron
and an electron mimics a photon. b) QCD background in the $\mu\mu\gamma$
channel.  The fit is described in the text.}
\label{fig-Zgfakespt}
\end{figure}

The $E_T$ spectra of the jets allowed the background to be calculated 
as a function of $E_T$. This is shown in Fig.~\ref{fig-Zgfakespt}(a). 
$Z$ boson events where an electron was misreconstructed as a 
photon and a jet was misreconstructed as the lost electron contributed to 
a bump in the fake photon at $E_T\sim 50$ GeV.
Thus, the shape of the fake photon sample was parameterized with an exponential 
function plus a Gaussian. The fit is also shown 
in Fig.~\ref{fig-Zgfakespt} (a). 

The $\tau\tau\gamma$ background
was estimated with a sample of {\small{ISAJET}}
events passed through the {\small{D\O GEANT}} detector simulation and the 
offline reconstruction algorithm.  
The total $\tau\tau\gamma$ background increased with increasing  
anomalous couplings because more $Z\gamma \rightarrow \tau\tau \gamma$ 
events would have been produced along with the $Z\rightarrow ee\gamma$ events.
After normalization with the production
cross section and $\tau$ branching fractions, the expected fraction
of the cascading tau decays in the final $Z\gamma$ sample was 
$f_e=(0.10\pm0.05\%)$, where the uncertainty came from the expected 
difference in acceptance from using $Z\rightarrow \tau\bar{\tau} 
\rightarrow ee\nu\nu \bar{\nu}\bar{\nu}$ Monte Carlo to simulate a background
which included photons which radiated from a charged lepton, 
as well as the uncertainty in cross section and branching ratios. 
For the SM  couplings the  $\tau\bar{\tau}\gamma$ background was negligible.

To summarize, four $Z\gamma \rightarrow ee\gamma$ candidates were 
observed.  The total background expected was $0.43\pm0.06$ events.  
This corresponds to an  observed signal of $3.57^{+3.15}_{-1.91}  \pm 0.06$
events, where the  first uncertainty is statistical and the second is 
the uncertainty in the background. 
The observed signal agrees with the SM prediction of 
$2.8 \pm 0.3 \pm 0.2$ events, where the  first 
uncertainty reflects systematics of the Monte Carlo model and the second
is the uncertainty in the luminosity.

\subsection{ The $\mu\mu$ channel}
The   $\mu\mu\gamma$  sample was  selected  from events which satisfied the 
{\small MU\_ELE} trigger described in Table~\ref{tab-trigs}. This data 
set corresponded  to an  integrated luminosity of $13.7\pm 0.7 {\rm pb}^{-1}$.
At least two muons and one photon were required in the event. 
One muon was required to satisfy the ``Tight III" requirements and the 
other to satisfy the ``Loose II" requirements of Table~\ref{tab-muonid}.  
It was required that 
$p_T^{\mu_{1}} > 15$~GeV$/c$ and $p_T^{\mu_{2}} > 8$~GeV$/c$, where $\mu_1$
and $\mu_2$ are the higher and lower $p_T$ muons respectively.  
The photon, satisfying the ``Loose" requirements, was required to 
have $E_T > 10$ GeV and to be separated from both muons
by $\Delta {\cal R}_
{\mu\gamma} > 0.7$, as in the electron channel.  Two candidates
for $Z\gamma\rightarrow \mu\mu \gamma$ passed these selection criteria.
Details on the characteristics of the candidates and event displays
are in Ref.~\cite{glthesis}. Table~\ref{tab-zguugev} indicates the number 
of events which survived the last few selection criteria, after the particle
identification and kinematic selection were applied.
\vbox{
\begin{table}
\begin{tabular}{|c|c|} 
Selection Criteria             & No. of Surviving Events  \\ \hline 
Particle ID and                &                          \\
Kinematic  Selection           & 4                        \\ 
Trigger Criteria               & 3                       \\ 
$\Delta {\cal R}_{\mu \gamma}>0.7$& 2                        \\ 
\end{tabular}
\caption{Number of $\mu\mu \gamma$ candidates which passed the 
selection criteria.}
\label{tab-zguugev}
\end{table}}

\begin{figure}
\epsffile{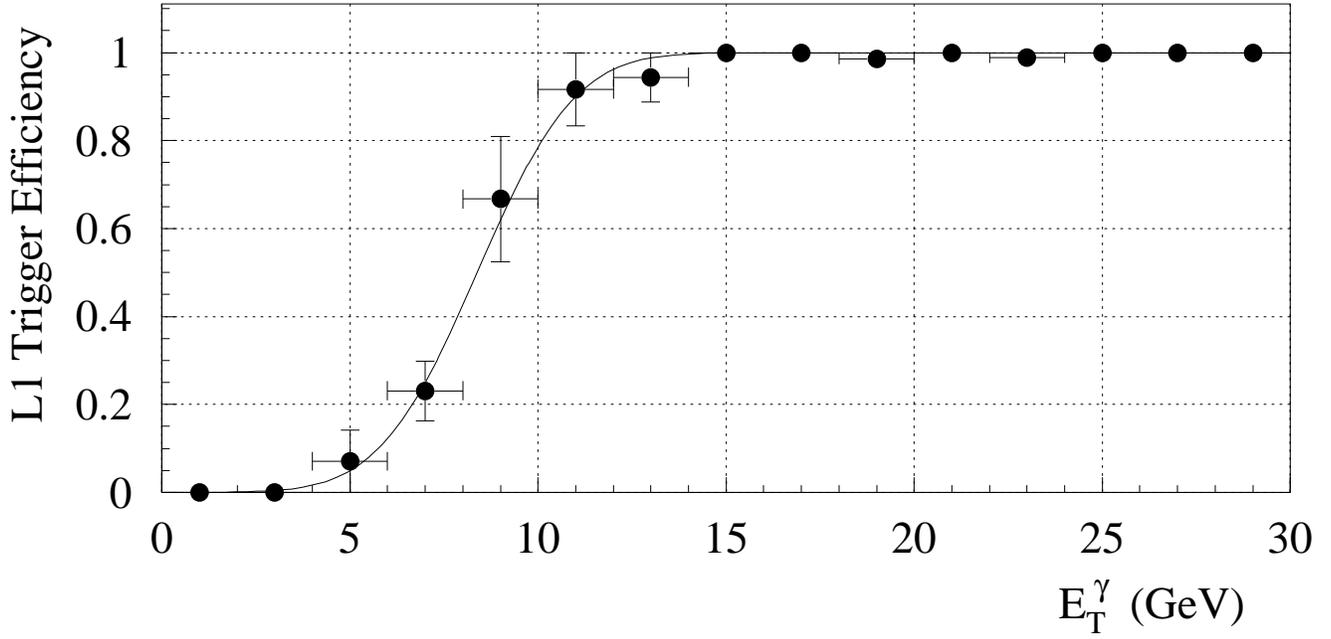}
\caption{Efficiency as a function of photon $E_T$ for the level 1 EM trigger 
with threshold at 7 GeV. The minimum allowed photon $E_T$ is 10 GeV.}
\label{fig-etrig7gev}
\end{figure}

The efficiencies were calculated as a function of $ZZ\gamma$ and 
$Z\gamma\gamma$ couplings with the event generator of Ref.~\cite{gl20}
combined with the parameterized detector simulation.  
${\rm MRSD-^{\prime}}$ structure functions~\cite{MRSD} were used in the 
event generation and the cross section was scaled by a k-factor of 1.34.
The level 1 muon
trigger efficiency (with two chances to trigger on each event), the 
photon trigger efficiency curve shown in Fig.~\ref{fig-etrig7gev},
and the efficiencies for particle identification criteria as discussed in
Section~\ref{sec-pid}, were included.
The detector acceptance was 20\% for SM $Z\gamma\rightarrow \mu\mu\gamma$
production. The overall efficiency for SM production, for events satisfying
the kinematic criteria, was $0.06\pm0.01$.  
The cross section times efficiency for SM production was $0.17\pm0.03$ pb. 
The efficiency increased with anomalous couplings because the muons became 
more central, increasing their acceptance.

The background consisted of $\mu\mu + {\rm{jet(s)}}$ events, where
the jet was misidentified as a photon, and $Z\rightarrow \tau\tau$ cascade 
decays including a final state photon.  
The backgrounds  where a jet mimicked a photon
included Drell-Yan production with associated jets,
$Z+{\rm{jets}}$ production and cosmic ray muons (already small because of 
the tight muon identification criteria) in coincidence with jet 
events.  The contribution to the QCD background from $b\bar{b}$ production was 
expected to be negligible due to the muon isolation requirements and the 
muon $p_T$  threshold.  
The QCD background was estimated from a data sample containing 
a pair of muons satisfying the same muon identification criteria as the 
signal sample.  This sample contained all of  the $\mu\mu + \rm{jet}$  
backgrounds in the same proportions as the signal sample. The procedure was 
to count the number of jets of $E_T>10$ GeV, in events which pass the 
selection criteria, and multiply that number by the probability for a jet
to mimic a photon from Table~\ref{tab-zgfp}. Because different triggers 
were used in collecting the signal and background samples, a scaling 
factor $1.3\pm 0.3$ was necessary to account for differences in the trigger
efficiency and integrated luminosities.  Fifteen jets passed the selection 
criteria;  eight jets were in the CC and seven in the EC.  The result  
was $N^{\mu}_{QCD} = 0.02\pm0.01$ expected background events. 
The $p_T$ spectrum of the jets in the background sample 
is shown in Fig.~\ref{fig-Zgfakespt}(b).  The fit made to the $E_T$ spectrum
of the electron channel fakes was used to parameterize the $E_T$ spectrum 
of the fakes in the muon channel, with the appropriate normalization,
because of the much higher statistics of the former.

The $\tau\tau\gamma$ background was estimated using the same procedure as 
was used for the 
electron channel.  The expected fraction of the cascade $\tau$ decays in the 
final $Z\gamma\rightarrow\mu\mu\gamma$ sample was  $1.4\pm0.5\%$.
For SM couplings, the $\tau\tau\gamma$ background was $0.03\pm0.01$ events.

\begin{figure}
\epsffile{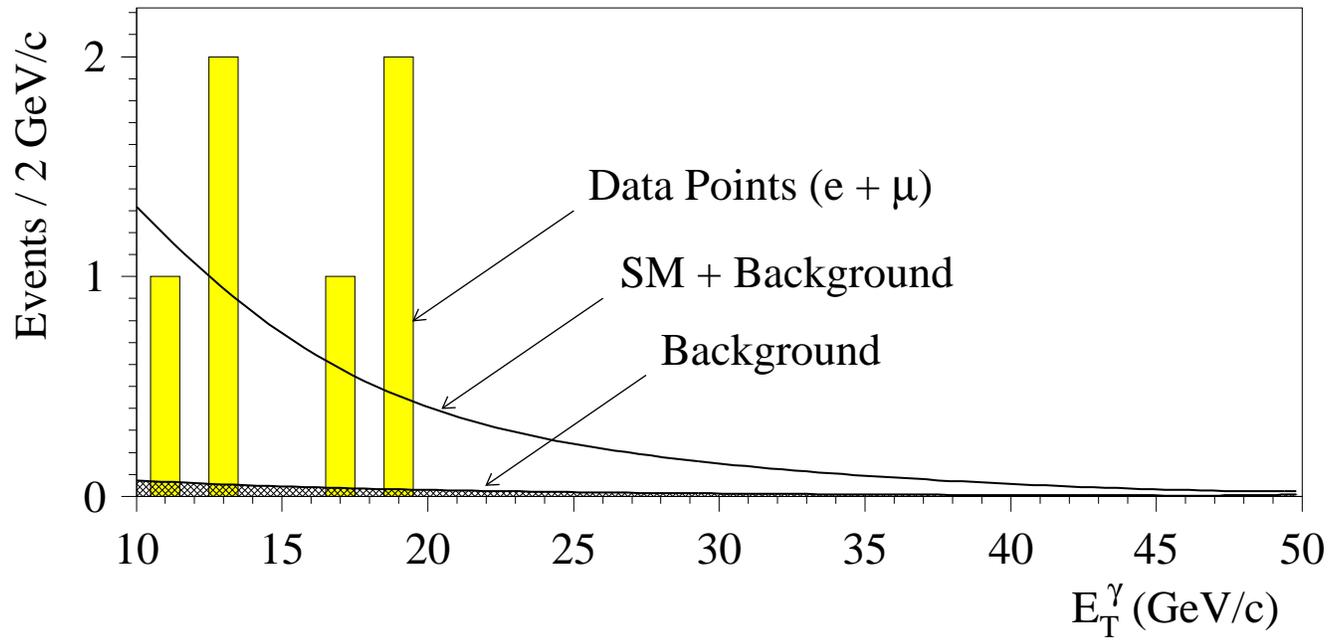}
\caption{Transverse energy spectrum of photons in $ee\gamma$ and $\mu\mu\gamma$
events. The shadowed bars correspond to the data, the hatched curve represents
the total for background, and the solid line shows the sum of SM predictions
and background. }
\label{fig-zgchlep}
\end{figure}

To summarize, two $Z\gamma \rightarrow \mu\mu\gamma$ candidates were 
observed.  The total background expected was $0.05\pm0.01$ events.  
This corresponds to an  observed signal of $1.95^{+2.62}_{-1.29}  \pm 0.01$
events, where the  first uncertainty is statistical and the second is 
the uncertainty on the background.  The observed signal agrees with the 
SM prediction of $2.3 \pm 0.4 \pm 0.1$ events, where the  first 
uncertainty reflects systematics of the Monte Carlo model and the second
is the uncertainty in the luminosity. 
The photon $E_T$ spectrum for the combined $ee$ and $\mu\mu$ data,
expected signal and expected background are shown in Fig.~\ref{fig-zgchlep}.

\subsection{The $\nu\bar{\nu}$ channel }
\label{sec-zgvvg}
The $Z + \gamma \rightarrow \nu\bar{\nu}\gamma$ signature was a single photon 
which recoiled against the \hbox{$\rlap{\kern0.25em/}E_T$} of the unmeasured
neutrino pair.  The nature of the backgrounds 
for this channel was very different from the
electron and muon channels in that they were larger and included contributions
from sources to which the previous channels were immune. 
One background resulted from unreconstructed cosmic ray and Tevatron 
beam related muons which deposited energy in the electromagnetic calorimeter 
through bremsstrahlung. A second important 
background, occurring at moderately high $E_T(\gamma)$, 
came from $W\rightarrow e\nu$ where the electron was misidentified 
as a photon due to a missing track.   These backgrounds forced the analysis 
to use much stronger
particle identification criteria and tighter kinematic selection than the
$Z\gamma$ analyses presented above. 

The candidate sample was selected from events passing the {\small{ELE-HIGH}} 
trigger of Table~\ref{tab-trigs}.  
A selection on the ``event quality" removed events 
with noisy cells in the calorimeter, second EM objects with $E_T>5$ GeV, or
when the calorimeter was recovering due to a large pulse from Main Ring 
associated energy deposition.  Both 
the photon $E_T$ and the \hbox{$\rlap{\kern0.25em/}E_T$} were required to 
be greater than 40 GeV to reduce the background from $W$ boson decays. 
Events with muons in the central region were rejected
to reduce cosmic ray backgrounds. 
Events with jets of $E_T$ greater than 15 GeV were also rejected; 
by limiting the $p_T$ boost of the events, the kinematic range of 
\hbox{$\rlap{\kern0.25em/}E_T$} and $E_T^e$ from the $W$ boson background 
was reduced. 
The strictest photon requirement, ``Tight",
was used to reduce the backgrounds from cosmic rays, from beam related muons, 
and from $W$ boson decays.
Lastly, the calorimeter was searched, in a road about the line defined by 
the vertex located by the {\small EMVTX} package (see Section~\ref{sec-emvtx})
and the energy-weighted center of 
the photon shower in the CC, for energy deposition consistent with the passage
of an unreconstructed cosmic ray which might have radiated the photon. 
This algorithm (MTC) tracked the muon energy deposition through the 
longitudinally and azimuthally segmented towers of the calorimeter.

Four events remained after all selection criteria were applied. The 
photons in the four events had $E_T$ of 41, 41, 46, and 68 GeV.  
Table~\ref{tab-zgvvgevts} shows the number of events remaining after 
each of the selection criteria.  
\vbox{
\begin{table}
\begin{tabular}{|c|c|}
Selection Criteria     & Number of Events Remaining  \\ \hline \hline
Trigger, Event Quality and Kinematics 
                       & 1887                        \\ 
$|\eta_{\gamma}|<1.0$ or $1.5<|\eta_{\gamma}|<2.5$ 
                       & 1637                        \\ 
``Loose" Photon Criteria
                       & 1448                        \\ 
$N_{\mu}^{CF} = 0$     & 1098                        \\ 
No Jet with $E_T>15$ GeV
                       & 480                         \\ 
``Tight" Photon Criteria
                       & 5                           \\ 
MTC selection          & 4                           \\
\end{tabular}
\caption{Event selection criteria for the $Z\gamma\rightarrow \nu\nu 
\gamma$ analysis. }
\label{tab-zgvvgevts}
\end{table}}

The {\small ELE-HIGH} trigger was completely efficient for $E_T(\gamma)>40$
GeV. The efficiency for the photon identification criteria (excluding the 
fiducial requirements), the event quality,  the muon and jet vetoes, 
and the MTC selection criteria were estimated using the $Z\rightarrow ee$ 
candidates collected using the same trigger as the signal. 
The efficiency for the jet veto was cross checked with a 
Next-to-Leading logarithm $Z\gamma$ Monte Carlo 
generator\cite{ohnemuszg}. The calculated efficiency loss agreed with the 
measurement. Table~\ref{tab-zgvvgeff} contains a summary of  these 
efficiencies. The efficiency for the fiducial selection came from the 
event generator of Ref.~\cite{gl20} combined with the parameterized detector 
simulation.
\vbox{
\begin{table}
\begin{tabular}{|c|ccc|}
Selection Criteria         &  CC    & Combined        & EC        \\ \hline
                                                                     \hline
Event Quality              &        & $0.981\pm0.002$ &           \\ 
Photon ID Criteria         & $0.61\pm0.02$ 
                                    &                 & $0.66\pm0.05$       \\
$N_{\mu}^{CF} = 0$         &        & $0.988\pm0.002$ &           \\ 
No Jet with $E_T>15$ GeV   &        & $0.84\pm0.02$   &           \\ 
MTC Selection              & $0.97\pm0.02$ 
                                    &                 & --        \\ \hline
Total                      & $0.48\pm0.02$ 
                                    &                 & $0.54\pm0.04$ \\ 
\end{tabular}
\caption{Efficiency, excluding photon fiducial requirements, for the 
$Z\gamma\rightarrow \nu\nu \gamma$ analysis.  The uncertainties 
indicated are statistical only. }
\label{tab-zgvvgeff}
\end{table}}

The cosmic ray and beam halo backgrounds, due to unreconstructed muons which 
radiated a photon as they passed through the calorimeter, was 
estimated using cosmic ray and beam halo muons identified in the data.

Two samples of this kind of background event which radiated a photon into 
the CC were identified.  The first sample was identified by applying all the
selection criteria except for the {\small HITSINFO} criteria of the 
``Tight" photon ID requirements, the CF muon veto, and the MTC requirement.
The event was required to have a reconstructed muon.
The rejections for the {\small HITSINFO} and the MTC selection criteria were 
determined from this sample.  The reconstruction efficiency for the muon 
background $(\epsilon_{{\rm cosmic}\; \mu})$ 
were estimated from a second sample of events, dominated by cosmic ray 
muons, which passed the same selection criteria as the former sample 
(excluding the requirement that the muon was reconstructed) and failed 
the {\small EMVTX} criteria.  The inefficiency was 
$1- \epsilon_{{\rm cosmic}\; \mu} = 0.34\pm 0.03$. 
The background to $Z\gamma$ was then determined from the number of events 
in the former sample, modified by the rejection provided by the 
{\small HITSINFO} and MTC criteria and by the factor 
$(1-\epsilon_{cosmic\;\mu})/\epsilon_{cosmic\;\mu}$. 
The resulting expected background was $1.4\pm0.6$ events for CC photons.  

For the muon background events with a photon in the EC, a two-sample 
procedure analogous to that described above for the CC was used to 
estimate the background due to beam-halo muons.  It was measured to be 
$0.38\pm0.23$ events, where the large fractional uncertainty is due to the 
low statistics of one of the tagged samples.

\begin{figure}
\epsffile{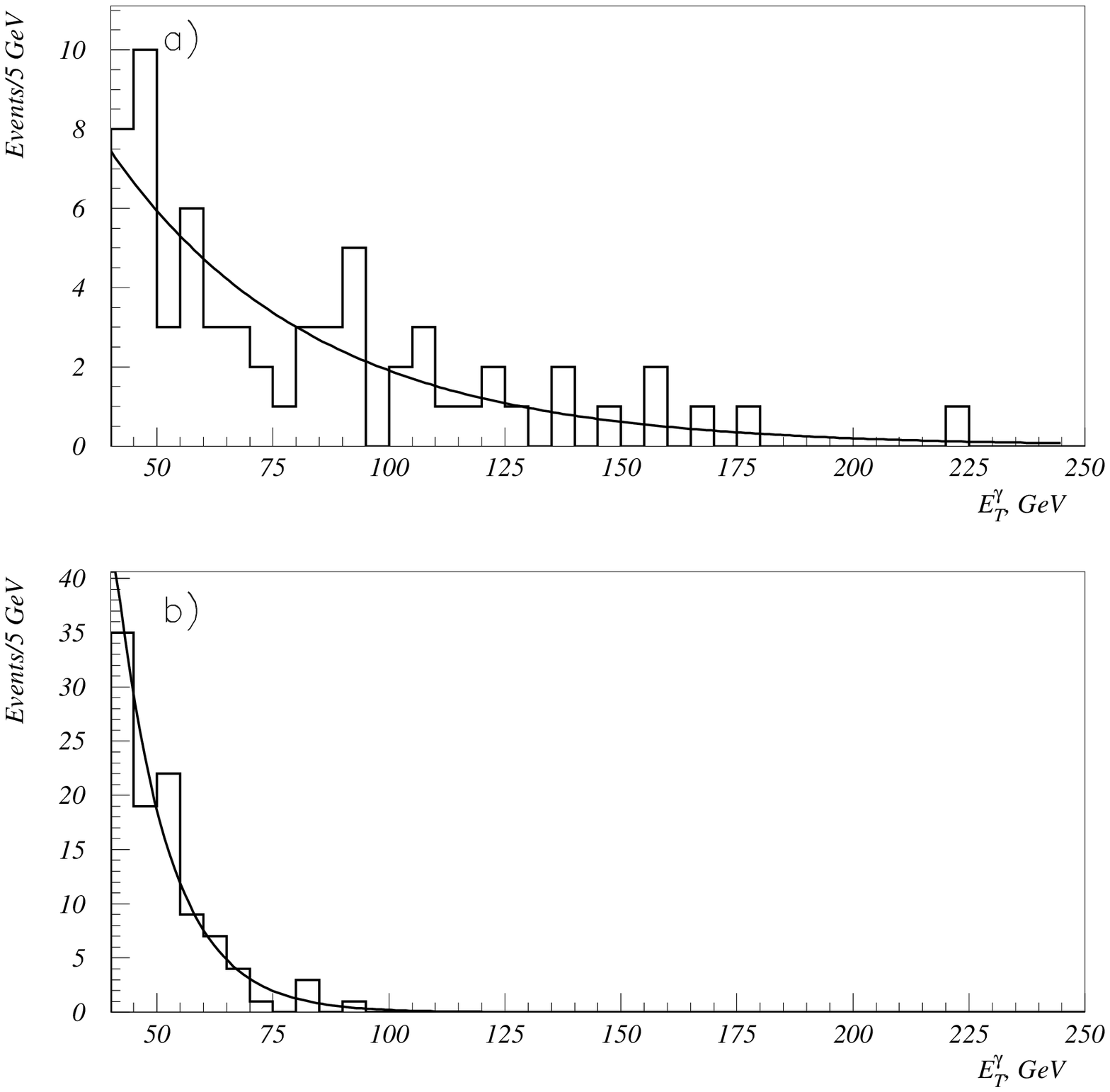}
\caption{Shape of the bremsstrahlung photon spectrum for the background due 
to cosmic ray and beam halo muons in the a) CC and b) EC calorimeters. 
The solid lines are the resulting fitted parameterizations.}
\label{fig-MUBCK}
\end{figure}

The muon background rejection 
was found to be independent of the photon $E_T$ for both the CC and EC 
regions, so the background events were used to make a parameterization of 
the photon spectrum.  The doubly-identified background muons were used for the
CC. For the EC, the larger of the two available background samples  was used.
The result of the fit 
is shown in Fig.~\ref{fig-MUBCK} together with the background data.

\begin{figure}
\epsffile{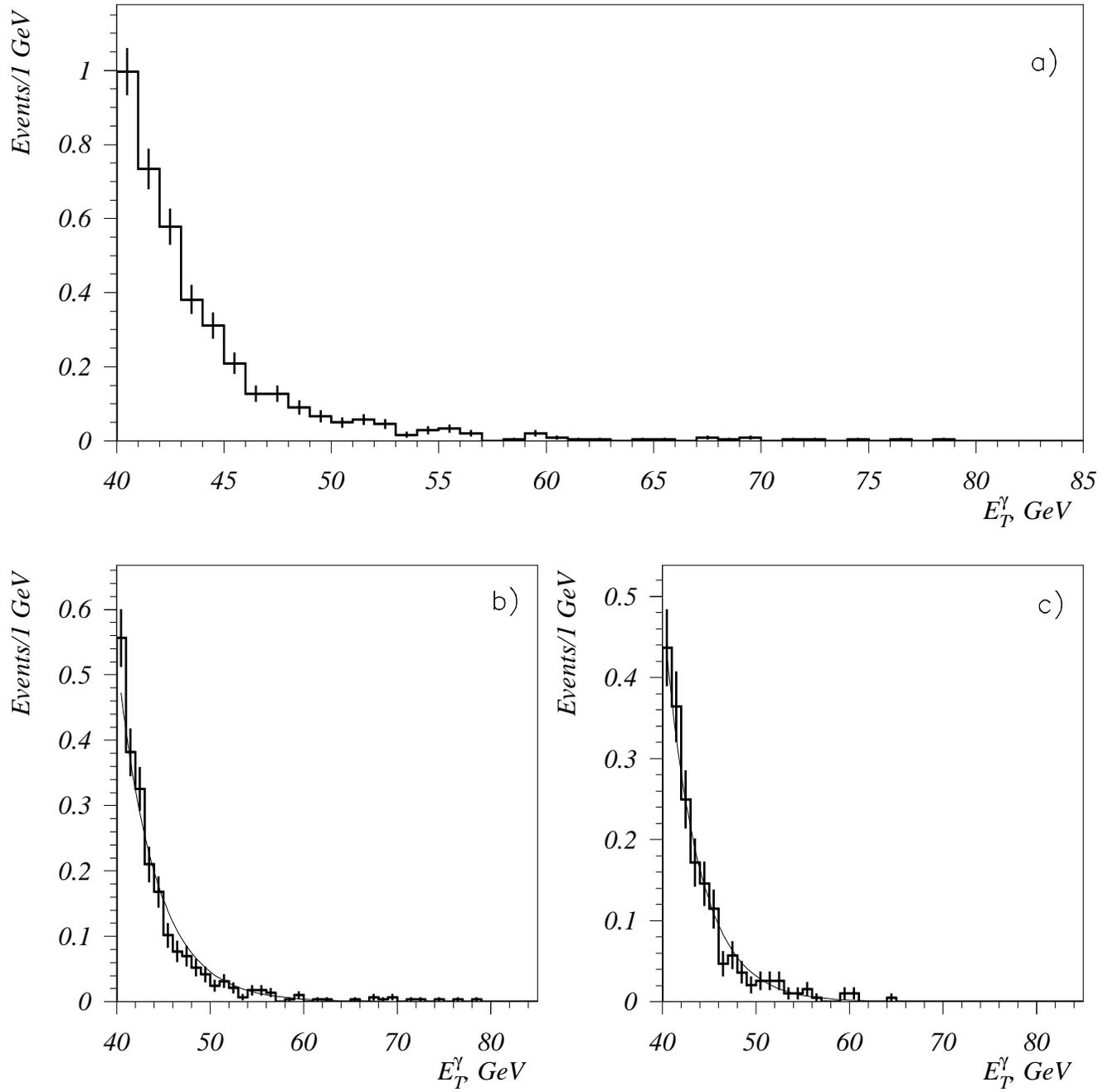}
\caption{Shape of the photon spectrum for the background due to $W\rightarrow
e\nu$  in the $Z\gamma\rightarrow \nu\nu\gamma$ analysis. (a) is the spectrum 
in the CC and EC. (b) and (c) are the individual spectra in the CC and EC 
where the fits to the background are shown (lines). }  
\label{fig-wevback}
\end{figure}

Another background comes from  $W\rightarrow e\nu$ events where the 
electron is misidentified as a photon due to tracking inefficiency.
The kinematic requirements combined with the jet veto rejected some of the 
misidentified $W$ boson decays, but additional rejection was  required. 
It came from the {\small HITSINFO} criteria of the ``Tight" photon
selection.
A procedure similar to that used to estimate the background due to 
muon bremsstrahlung was used in this case. A nearly pure
sample of tagged $W\rightarrow e\nu$ events 
with $E_T^{e} $ and \hbox{$\rlap{\kern0.25em/}E_T$} both greater than $25$ GeV
was obtained by applying all of the 
event selection criteria except for the {\small HITSINFO} and MTC (CC only)  
criteria and by requiring an electron with a good track match significance,
$TMS\leq 10$ (see Section \ref{sec-tms}), instead of a photon.   The rejection,
$R_{H}$, due to {\small HITSINFO} came from a sample of $W\rightarrow e\nu$ 
events which failed the electron tracking requirement. 
Including small corrections to account for the fraction of these 
mistracked $W$ events, lost from this sample because of 
an overlapping random track,  and to account for cosmic ray bremsstrahlung
in this background sample, 
$R_{H}$ was found to be $48\pm 12 \; (43\pm 14)$ for the CC (EC).
With the measured efficiency of the track finding, 
the track match significance and the 
MTC criteria ($\epsilon_T$, $\epsilon_{\sigma}$, and $\epsilon_{MTC}$,
respectively), the $W\rightarrow e\nu$ 
background was simply the number of tagged $W \rightarrow e\nu$ events times 
$(1-\epsilon_T)/(\epsilon_T \, \epsilon_{\sigma}\, \epsilon_{MTC}\, R_{H})$. 
This was found to be $2.2\pm0.6 \; (1.8\pm0.6)$ events for the CC (EC).
The $E_T$ spectrum of the expected background 
is shown in Fig.~\ref{fig-wevback}.  
Parameterizations for the expected $E_T^{\gamma}$ spectrum of background 
were derived from the $W\rightarrow e\nu$ events. They are shown in
Fig.~\ref{fig-wevback}.

The possible QCD backgrounds included: multijet production, where a jet was 
misidentified as a photon and the \hbox{$\rlap{\kern0.25em/}E_T$} results from 
mismeasurement of a jet or from neutrinos in a jet; direct photon production 
(jet + photon) where a jet contributes to \hbox{$\rlap{\kern0.25em/}E_T$};
and $Z+{\rm{jets}}\rightarrow \nu\bar{\nu}+ {\rm{jets}}$ 
where a jet is misidentified as a photon.   
However, the size of the QCD backgrounds fell rapidly as the photon
$E_T$ and \hbox{$\rlap{\kern0.25em/}E_T$} thresholds were raised. 
The backgrounds were found to be negligible for $E_T^{\gamma}$ and 
\hbox{$\rlap{\kern0.25em/}E_T \geq 35$} GeV. 

 The total background was $3.6\pm0.8$ $(2.2\pm0.6)$ for the CC (EC). 
It is summarized in Table~\ref{tab-zgvvgback}. 
\vbox{
 \begin{table}
 \begin{tabular}{|c|c|c|} 
 Background         & $ N_{bck}^{CC}$ & $ N_{bck}^{EC}$\\  \hline \hline
 Muon Bremsstrahlung& $1.4\pm0.6$     & $0.38\pm0.23$  \\ \hline
 $W\rightarrow e\nu$& $2.2\pm0.6$     & $1.8\pm0.6$    \\ \hline
 QCD Sources        & negligible      & negligible     \\ \hline \hline
 Total              & $3.6\pm 0.8$    & $2.2\pm 0.6$   \\ 
 \end{tabular}
 \caption{Number of expected background events in the CC and EC for the 
 $Z\gamma\rightarrow \nu\bar{\nu} \gamma$ analysis. }
 \label{tab-zgvvgback}
\end{table}}

\begin{figure}
\epsffile{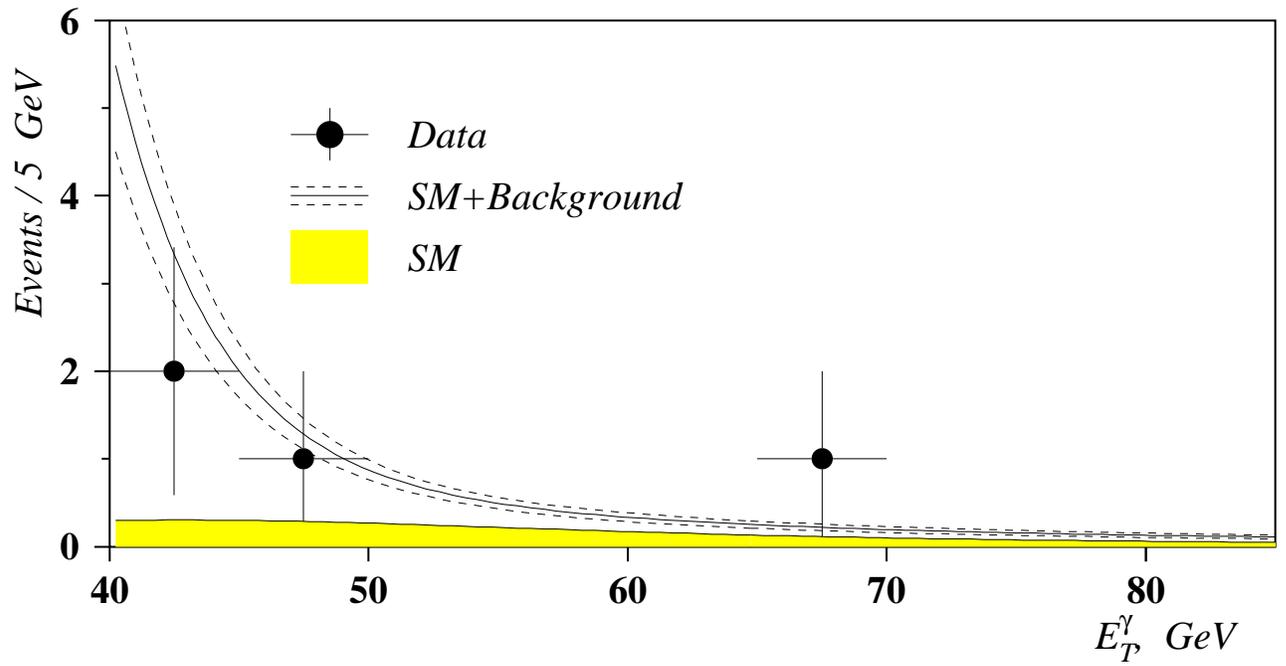}
\caption{The photon $E_T$ spectra for the $Z\gamma\rightarrow \nu\nu\gamma$ 
data (points), the background (solid line), the expected signal 
(shaded), and the sum of the expected signal and background with uncertainties
(dotted). }
\label{fig-gl15}
\end{figure}

The expected numbers of signal events for SM and for anomalous couplings
were estimated using the leading order event generator of Ref.~\cite{gl20} 
combined with the parameterized detector simulation, including a 
$p_T(Z)$ spectrum from the $Z$ boson data to mimic the effects of the 
jet veto on the acceptance. The energy scale for the underlying event
was determined by comparing the $p_T$ as determined from the electrons and
hadronic recoil in low $p_T$ $Z\rightarrow ee$ events.  
As in the charged lepton analyses, the cross section was scaled by a 
   $k$-factor of 1.34 and the ${\rm MRSD-^{\prime}}$ structure
   functions~\cite{MRSD} were used in the event generation.
A $12\%$ uncertainty in the cross
section resulted primarily from the choice of parton distribution functions,
modeling of the jet veto, modeling of the detector, and the detector 
efficiency.  Table~\ref{tab-zgnusig} presents a summary of the expected
signal and background as well as the number of events seen with photons
in the CC and EC.  The SM signal was expected to be $1.8\pm0.2$ events
with a $5.8\pm 1.0$ event background.   For comparison, the expected 
number of signal events for anomalous couplings was approximately a factor of 
9 higher for $h_{30}^Z = 3, \, h_{40}^Z = 1$. 
Four candidates were observed, consistent with the SM expectations. 
The photon spectra expected for the signal and background, as well as that 
seen in the data are shown in Fig.~\ref{fig-gl15}.
\vbox{
\begin{table}
\begin{tabular}{|c|c|c|c|} 
Region    & $N_{SM}$   & $ N_{bck}$    & Data    \\  \hline \hline
CC $(|\eta_{\gamma}|\leq 1.1)$ 
          & $1.4\pm0.2$& $3.6\pm0.8$   & 3       \\ \hline
EC $(1.5\leq |\eta_{\gamma}|\leq 2.5)$
          & $0.39\pm0.05$
                       & $2.2\pm0.6$   & 1       \\ \hline \hline
Total     & $1.8\pm 0.2$
                       & $5.8\pm 1.0$  & 4       \\ 
\end{tabular}
\caption{Number of expected $\nu\bar{\nu}\gamma$ events assuming SM 
couplings, number of expected background events, and observed signal. }
\label{tab-zgnusig}
\end{table}}

\subsection{Limits on Anomalous $ZZ\gamma$ and $Z\gamma\gamma$ Couplings}
To set limits on the anomalous coupling parameters, the observed 
$E_T$ spectrum of the photons in the three 
channels was fit with the MC predictions plus the estimated background
(summarized in Table~\ref{tab-zgsummary}).
The binned likelihood method described in Appendix~2 was used.  
To exploit 
the prediction that anomalous couplings lead to an excess of events 
with high $E_T$ photons, a high $E_T$ bin with no events was used in the fit. 
Common systematic uncertainties, including photon identification 
efficiency, integrated luminosity, choice of parton distribution 
functions, and choice of $p_T(Z)$ distribution were treated as discussed 
in Appendix~2.
\vbox{
\begin{table}
\begin{tabular}{|c|c|c|c|} 
Channel        & $ee$   & $\mu\mu$ & $\nu\bar{\nu}$           \\ \hline 
$N^{SM}$       &$2.8 \pm 0.3 \pm 0.2$ 
                        &$2.3 \pm 0.4 \pm 0.1$ 
                                   &$1.8 \pm 0.2$             \\
$N^{BG}$       &$0.43 \pm 0.06$ 
                        &$0.05 \pm 0.01 $
                                   &$5.8 \pm 1.0$             \\ \hline
$N^{observed}$ & 4      & 2        & 4                        \\ 
\end{tabular}
\caption{Summary the of $Z\gamma \rightarrow$ dileptons analyses including
 the number of SM events expected, expected backgrounds, and 
number of candidates observed.}
\label{tab-zgsummary}
\end{table}}

The form-factor scale dependence of the result was studied.  The chosen value 
of $\Lambda = 500$ GeV was close to the sensitivity limit of the experiment for
$h_{20}$ and $h_{40}$ for the $ee + \mu\mu$ channels; 
for larger values of $\Lambda$ partial wave unitarity was violated for certain
values of the coupling parameters allowed at 95\% CL. 
With the $\nu\bar{\nu}$ and combined analysis, 
$\Lambda$ could be extended to 750 GeV without violating unitarity.  In that 
case, tighter limits on anomalous couplings could be obtained.

\begin{figure}
\epsffile{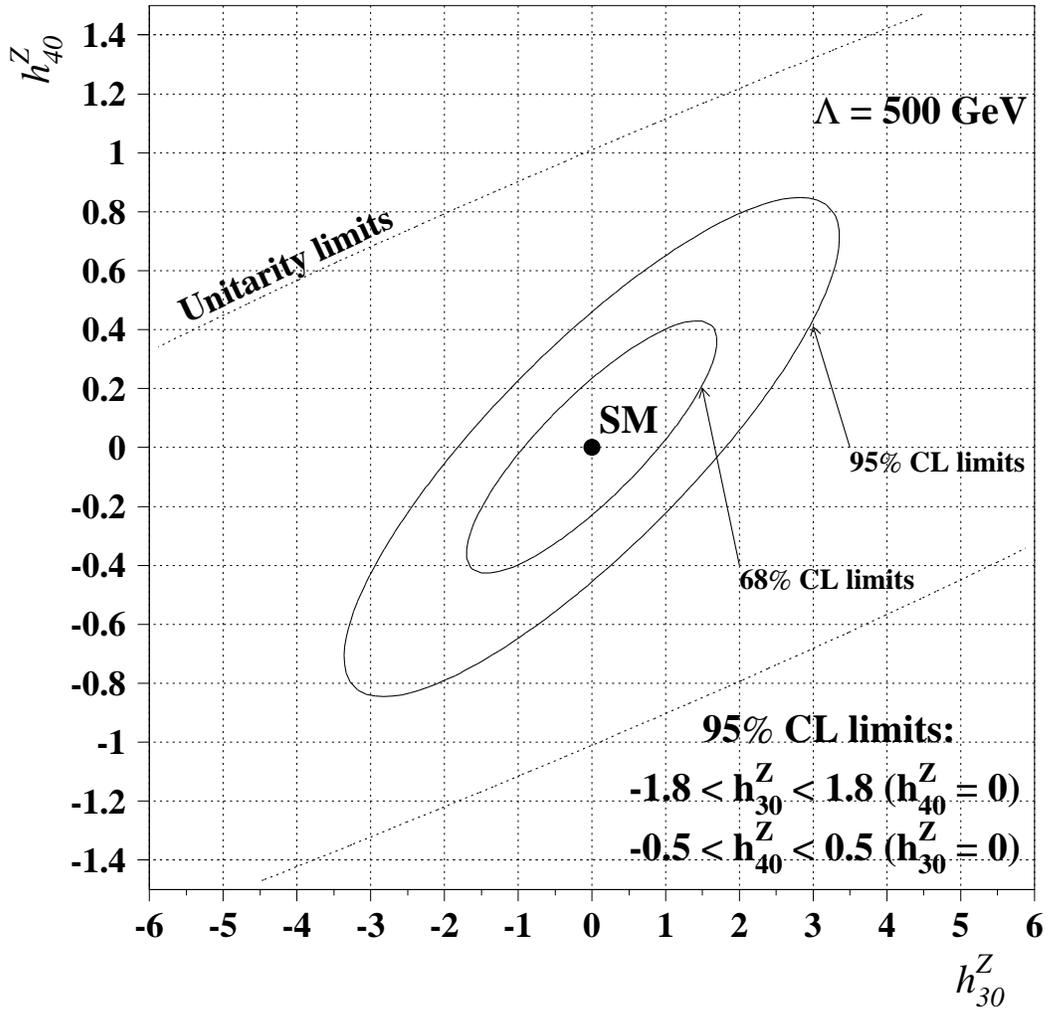}
\vspace{-3.0in}
\caption{Limits on the $CP$-conserving anomalous $ZZ\gamma$ coupling 
  parameters $h_{30}^Z$ and $h_{40}^Z$. The solid ellipses represent 68\% and 
  95\% CL exclusion contours for the $ee$ and $\mu\mu$ combined analysis. The 
  dashed curve shows limits from partial wave unitarity for 
  $\Lambda = 500$ GeV. }
\label{fig-Zglims1}
\end{figure}

\begin{figure}
\epsffile{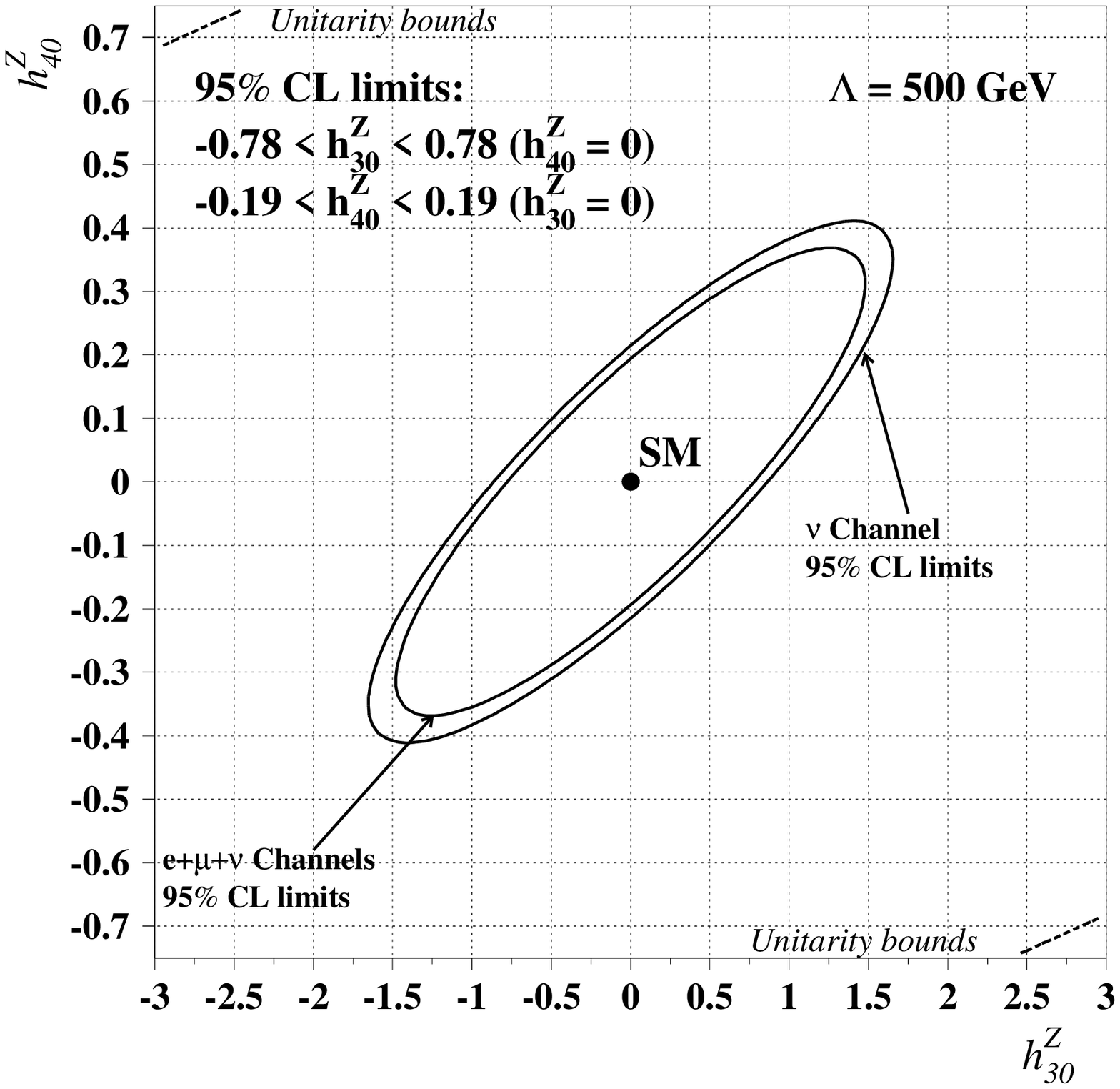}
\caption{Limits on the $CP$-conserving anomalous $ZZ\gamma$ 
coupling parameters $h_{30}^Z$ and $h_{40}^Z$. The solid ellipses represent 
the 95\% CL exclusion contours for the $\nu\bar{\nu}$ and for combined
$ee$, $\mu\mu$, and $\nu\bar{\nu}$ analyses. The dashed curve shows limits from
partial wave unitarity for $\Lambda = 500$ GeV.}
\label{fig-Zglims2}
\end{figure}

\begin{figure}
\epsffile{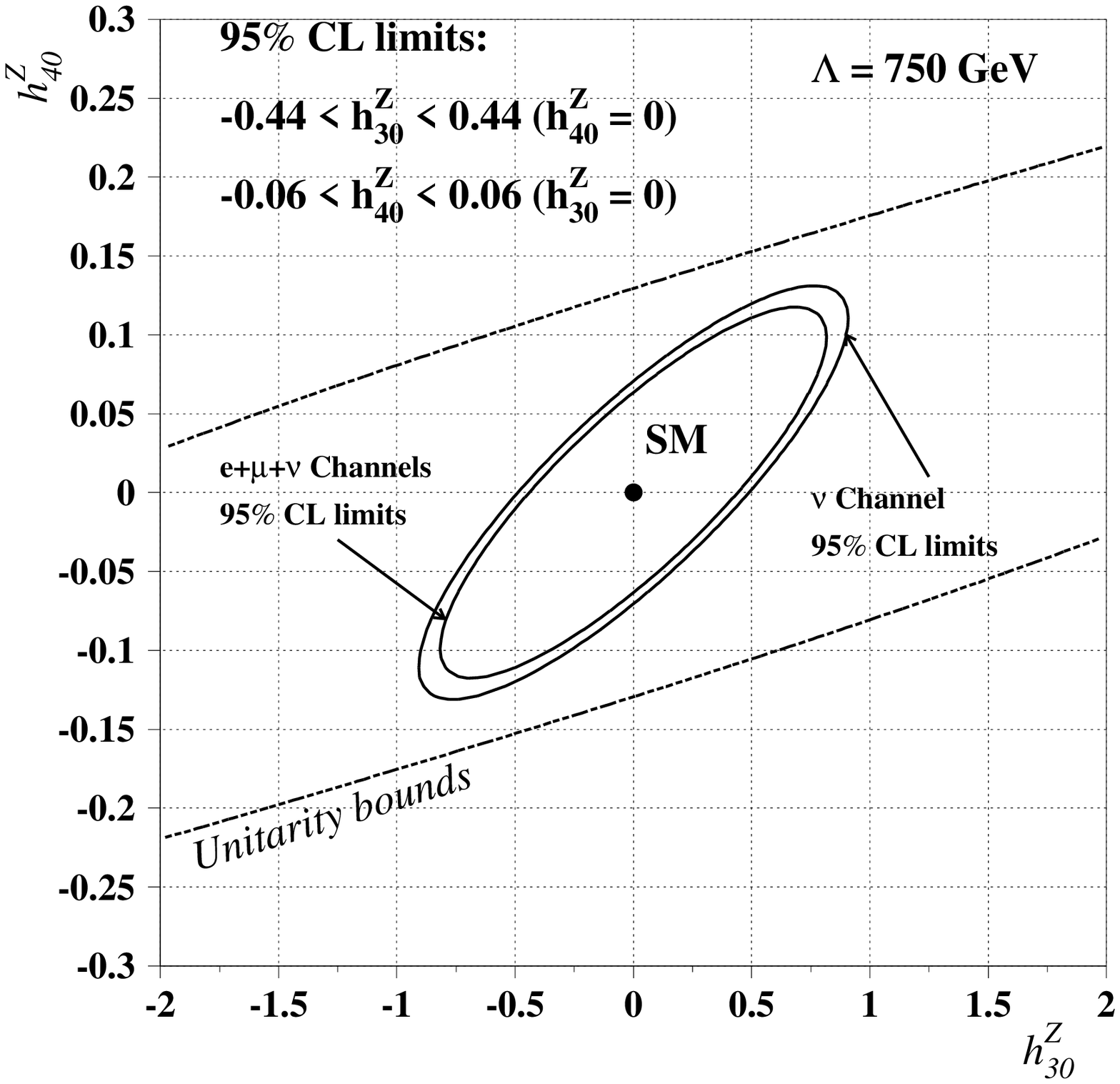}
\caption{Limits on the $CP$-conserving anomalous $ZZ\gamma$ 
coupling parameters $h_{30}^Z$ and $h_{40}^Z$. The solid ellipses represent 
the 95\% CL exclusion contours for the $\nu\bar{\nu}$ and for the combined
$ee$, $\mu\mu$, and $\nu\bar{\nu}$ analysis. The dashed curve shows limits from
partial wave unitarity for $\Lambda = 750$ GeV.}
\label{fig-Zglims3}
\end{figure}

 Figures~\ref{fig-Zglims1} to \ref{fig-Zglims3} show the coupling 
 limits for the $CP$-conserving $ZZ\gamma$ parameters. The shapes of the 
 $Z\gamma\gamma$ limit contours were similar.   Figure~\ref{fig-Zglims1} 
 shows the results of the fit for the $ee$ and $\mu\mu$ channels at 68\% and 
 95\% CL.  Figure~\ref{fig-Zglims2} shows the results of the fit for the 
 $\nu\bar{\nu}$  channels and for the three channels combined at 95\% CL.  
 The form factor scale  $\Lambda = 500$ GeV was used in these two figures.  
 Figure~\ref{fig-Zglims3}  shows the 95\% CL limits for $\Lambda = 750$ GeV.
 The 95\% confidence level limits on CP-conserving couplings are given in 
 Table~\ref{tab-zgacl}.  Shown are the 
 limits for the $ee+\mu\mu$ channels, the $\nu\bar{\nu}$ channel, and the 
 limits from the three analyses combined.  Limits on the $CP$-violating 
 couplings were numerically identical to the corresponding $CP$-conserving 
 couplings with the single exception that $-0.86<h_{10}^{Z}<0.87$ (to be 
 compared with $-0.87<h_{30}^{Z}<0.87$) for the $\nu\bar{\nu}\gamma$ analysis. 
\vbox{
\begin{table}
\begin{tabular}{|c|c|c|} 
$ee + \mu\mu$             & $\nu\bar{\nu}$& Combined Limits \\ \hline \hline
\multicolumn{3}{|c|} {Limits with $\Lambda = 500$ GeV}      \\ \hline
$-1.8<h_{30}^{Z}<1.8     $& $-0.87<h_{30}^{Z}<0.87     $
                                         &$-0.78<h_{30}^{Z}<0.78    $\\ \hline
$-1.9<h_{30}^{\gamma}<1.9$& $-0.90<h_{30}^{\gamma}<0.90$
                                         &$-0.81<h_{30}^{\gamma}<0.81$\\ \hline
$-0.5<h_{40}^Z<0.5       $& $-0.21<h_{40}^Z<0.21       $
                                         &$-0.19<h_{40}^{Z}<0.19    $\\ \hline 
$-0.5<h_{40}^{\gamma}<0.5$& $-0.22<h_{40}^{\gamma}<0.22$
                                         &$-0.20<h_{40}^{\gamma}<0.20
                                                                    $\\ \hline
\multicolumn{3}{|c|} {Limits with $\Lambda = 750$ GeV}      \\ \hline
 -                        & $-0.49<h_{30}^{Z}<0.49     $
                                          &$-0.44<h_{30}^{Z}<0.44    $\\ \hline
 -                        & $-0.50<h_{30}^{\gamma}<0.50$
                                         &$-0.45<h_{30}^{\gamma}<0.45$\\ \hline
 -                        & $-0.07<h_{40}^Z<0.07       $
                                          &$-0.06<h_{40}^{Z}<0.06    $\\ \hline 
 -                        & $-0.07<h_{40}^{\gamma}<0.07$
                                          &$-0.06<h_{40}^{\gamma}<0.06
                                                                     $\\ 
\end{tabular}
\caption{Limits on CP-conserving $ZZ\gamma$ and $Z\gamma\gamma$ anomalous 
coupling parameters for the $ee+\mu\mu$, $\nu\nu$ and combined $Z\gamma$ 
analyses. These axes limits are at 95\% confidence level with $\Lambda = 500$  
and $750$ GeV.}
\label{tab-zgacl}
\end{table}}

\section{Conclusions}
\label{sec-conclusions}
Four gauge boson pair production processes and corresponding trilinear 
gauge boson coupling parameters were studied using the data from 
1.8 TeV ${\bar p}p$ collisions collected with the D{\O} detector during 
1992-1993 Tevatron collider run at Fermilab.
The data sample corresponded to an integrated luminosity
of approximately $14~{\rm pb}^{-1}$.

Searches were made for deviations from the SM.  This would have been manifest 
as an enhancement in the production cross section and $E_T$ spectrum of the 
bosons. In the analyses of the $W\gamma$ final states, 
the photon $E_T$ spectrum was compared with the expectations of the 
SM and used to produce limits on anomalous $WW\gamma$ couplings.  A limit on 
the cross section for $WW\rightarrow$ dilepton led to a limit on anomalous 
$WW\gamma$ and $WWZ$ couplings.  In the $WW/WZ\rightarrow e\nu \,
{\rm jet \, jet}$ analysis, the $E_T$ spectrum of the $W$ and $Z$ bosons 
was used to produce limits on anomalous $WW\gamma$ and $WWZ$ couplings.
The $W\gamma$, $WW$ and $WW/WZ$ analyses were combined to produce limits
on $WW\gamma$ and $WWZ$ couplings. 
Finally, in the analysis of $Z\gamma$ final states, the photon $E_T$ spectrum 
was compared to the expectations of the SM and used to produce limits on 
anomalous  $ZZ\gamma$ and $Z\gamma\gamma$ couplings.  No deviations from 
the SM were observed. 

The $W\gamma$ analysis yielded 23 candidate events where the $W$ boson
was identified by its leptonic decay products, a high $p_T$ electron
(11 events) or muon (12 events), and a neutrino inferred by large
\hbox{$\rlap{\kern0.25em/}E_T$} in the event.
The expected backgrounds for the electron and muon channels were
$2.0 \pm 0.9$ and $4.4 \pm 1.1$ events, respectively.
Using the acceptance for the SM $W\gamma$ production events, the $W\gamma$
cross section (for photons with $E_T^{\gamma}>10$ GeV and $\Delta
 {\cal R}_{\ell\gamma}>0.7$) was calculated from the combined $e + \mu$
sample to be
$\sigma(W\gamma) = 138_{-38}^{+51} (stat) \pm 21 (syst)$ pb.
A binned maximum likelihood fit was performed on the $E_T^\gamma$ spectrum
for each of the $W(e\nu)\gamma$ and $W(\mu\nu)\gamma$ samples to set limits
on the anomalous coupling parameters.
The limits on the $CP$-conserving anomalous coupling parameters at the
95\% CL were
$-1.6 < \Delta\kappa_\gamma < 1.8 \; (\lambda_\gamma = 0)$,
$-0.6 < \lambda_\gamma < 0.6 \; (\Delta\kappa_\gamma = 0)$
 using a form factor scale of $\Lambda = 1.5$ TeV.
The $U(1)_{EM}$-only couplings of the $W$ boson to a photon, which lead to
$\kappa_\gamma = 0$ and $\lambda_\gamma = 0$, and thereby $\mu_W = e/2m_W$ and
$Q_W^e = 0$, was excluded at the 80\% CL, while zero magnetic moment
($\mu_W = 0$) was excluded at the 95\% CL.

The search for $WW$ events where both of the $W$ bosons decay leptonically
to $e  \nu$ or $\mu  \nu$ yielded one candidate event with an expected
background of $0.56 \pm 0.13$ events.
The upper limit on the cross section for SM $W$ boson pair production
was estimated to be 87 pb at the 95\% CL.
The limit on the cross section was translated into limits on the anomalous
coupling parameters.
The limits on the CP-conserving anomalous coupling parameters were
$-2.6 < \Delta\kappa < 2.8 \; (\lambda = 0)$ and
$-2.1 < \lambda < 2.1 \; (\Delta\kappa = 0)$ at the 95\% CL,
using a form factor scale of $\Lambda = 900$ GeV, where
$\Delta\kappa \equiv \Delta\kappa_\gamma = \Delta\kappa_Z$ and
$\lambda \equiv \lambda_\gamma = \lambda_Z$ were assumed.

The analysis of the $WW$ and $WZ$ production events
in the electron + jets channels, where one
$W$ boson decayed into $e  \nu$ and the second $W$ boson or $Z$ boson
decayed into two jets, yielded 84 candidate events with an expected
background of $75.8 \pm 13.3$ events, while the SM predicted
$3.2 \pm 0.6$ signal events.
A maximum likelihood fit was performed on the $p_T$ spectrum of the $W$ boson,
computed from the $E_T$ of electron and the 
\hbox{$\rlap{\kern0.25em/}E_T$}, 
to set limits on the anomalous couplings.
The limits on the CP-conserving anomalous coupling parameters were
$-0.9 < \Delta\kappa < 1.1 \; (\lambda = 0)$ and
$-0.6 < \lambda < 0.7 \; (\Delta\kappa = 0)$ at the 95\% CL,
using a form factor scale of $\Lambda = 1.5$ TeV, where
$\Delta\kappa \equiv \Delta\kappa_\gamma = \Delta\kappa_Z$ and
$\lambda \equiv \lambda_\gamma = \lambda_Z$ were assumed.

The $W\gamma$ production process is sensitive only to the $WW\gamma$ coupling
parameters.  $WZ$ production is sensitive only to the $WWZ$ couplings.  
On the other hand, $W$ pair production is sensitive both to
the $WW\gamma$ and the $WWZ$ coupling parameters.
Using assumptions on the relationship between the $WW\gamma$ and $WWZ$ coupling
parameters, these three analyses were combined to set the tightest limits
on the coupling parameters.
A maximum likelihood fit was performed on the three sets of data
simultaneously using a common form factor scale of $\Lambda = 1.5$ TeV
and the assumption that
$\Delta\kappa \equiv \Delta\kappa_\gamma = \Delta\kappa_Z$ and
$\lambda \equiv \lambda_\gamma = \lambda_Z$.
The limits obtained at the 95\% CL were
$-0.71 < \Delta\kappa < 0.89 \; (\lambda = 0)$ and
$-0.44 < \lambda < 0.44 \; (\Delta\kappa = 0)$.  These are the tightest 
limits on $WW\gamma$ and $WWZ$ couplings presently available.

$Z\gamma$ final states in ${\bar p}p$ collisions are produced from
the $Z$ boson--quark and photon--quark couplings in the SM. 
The $ZZ\gamma$ and $Z\gamma\gamma$ couplings, which can produce $Z\gamma$ final
states, are absent in the SM.
The $Z\gamma$ analysis yielded a total of 10 candidate events; 4 events
with $Z\rightarrow ee$, 2 events with $Z\rightarrow \mu\mu$ and 4 events
with $Z\rightarrow \nu\nu$.
The expected backgrounds for the $ee$ and $\mu\mu$ channels were
$0.43 \pm 0.06$ events and $0.05 \pm 0.01$ events, respectively.
The expected background for the $\nu\nu$ channel was $5.8\pm1.0$ events.
The sum of the SM prediction and the expected background was $2.8\pm0.3\pm 0.2$
$(2.3\pm0.4\pm0.1)$ events for the electron (muon) decay modes and 
$7.6\pm1.0$ events for the neutrino decay mode. 
A maximum likelihood fit was performed on the $p_T$ spectrum of the photons
to set limits on the anomalous coupling parameters.
The 95\% CL axes limits on the CP-conserving coupling parameters are
$-0.78 < h_{30}^Z <0.78 \; (h_{40}^Z = 0) ; 
 -0.19 < h_{40}^Z < 0.19\; (h_{30}^Z = 0),
-0.81 < h_{30}^\gamma <0.81 \; (h_{40}^\gamma = 0) ;
-0.20 < h_{40}^\gamma < 0.20 \; (h_{30}^\gamma = 0)$,
using a form factor scale of $\Lambda = 500$ GeV and
$-0.44 < h_{30}^Z <0.44 \; (h_{40}^Z = 0) ; 
 -0.06 < h_{40}^Z < 0.06\; (h_{30}^Z = 0),
-0.45 < h_{30}^\gamma <0.45 \; (h_{40}^\gamma = 0) ;
-0.06 < h_{40}^\gamma < 0.06 \; (h_{30}^\gamma = 0)$,
using a form factor scale of $\Lambda = 750$ GeV. 
The limits obtained in this measurement are the  most  stringent  limits  
on  anomalous $ZV\gamma$ couplings currently available.

All of these limits on the anomalous coupling parameters will be significantly
improved when the analyses of the data taken during the
1994-1995 Tevatron collider run, which corresponded to approximately
$80~{\rm pb}^{-1}$, are completed.

We thank U.~Baur, J.~Ohnemus, and D.~Zeppenfeld for providing us with 
Monte Carlo generators and much helpful advice. 
We thank the staffs at Fermilab and collaborating institutions for their
contributions to this work, and acknowledge support from the 
Department of Energy and National Science Foundation (U.S.A.),  
Commissariat  \` a L'Energie Atomique (France), 
State Committee for Science and Technology and Ministry for Atomic 
   Energy (Russia),
CNPq (Brazil),
Departments of Atomic Energy and Science and Education (India),
Colciencias (Colombia),
CONACyT (Mexico),
Ministry of Education and KOSEF (Korea),
CONICET and UBACyT (Argentina),
and the A.P. Sloan Foundation.

\section{Appendix 1: Trilinear gauge boson coupling parameters}
\subsection{$WW\gamma$ and $WWZ$ coupling parameters}
The tree-level Feynman diagrams for $q{\bar q}\rightarrow W\gamma$,
$q{\bar q}\rightarrow WW$ and $q{\bar q}\rightarrow WZ$ production processes
are shown in Figs.~\ref{fig-vpair}, \ref{fig-wgrad} 
and \ref{fig-zgrad}.
A formalism has been developed to describe the $WW\gamma$ and $WWZ$
vertices for the most general gauge boson 
self-interactions \cite{lagrangian,HWZ}.
The Lorentz invariant effective Lagrangian for the gauge boson
self-interactions
contains fourteen dimensionless coupling parameters, seven each for
$WW\gamma$ and $WWZ$:

\begin{eqnarray*}
 {\cal L}_{WWV}/g_{WWV} = 
   ig_1^V \left( W^{\dag}_{\mu\nu}W^{\mu}V^{\nu}
  -  W^{\dag}_{\mu}V_{\nu}W^{\mu\nu} \right) \\ + 
i\kappa_V W^{\dag}_{\mu}W_{\nu}
  V^{\mu\nu} + i\frac{\lambda_V}{M_W^2} W_{\lambda\mu}^{\dag}W_{\nu}^{\mu}
  V^{\nu\lambda} \nonumber \\
  -g^V_4W^{\dag}_{\mu}W_{\nu}(\partial^{\mu}V^{\nu} + \partial^{\nu}V^{\mu}) \\
  +g^V_5\epsilon^{\mu\nu\rho\alpha} \left( W^{\dag}_{\mu} 
  \stackrel{\leftrightarrow}{\partial}_{\rho}
  W_{\nu} \right)V_{\sigma} \nonumber \\
  + i\tilde{\kappa}_V
   W_{\mu}^{\dag}W_{\nu}\tilde{V}^{\mu\nu}
   +\frac{i\tilde{\lambda}_V}{M_W^2}W^{\dag}_{\lambda\mu}W^{\mu}_{\nu}
   \tilde{V}^ {\nu\lambda} ,
\end{eqnarray*}
where $W^{\mu}$ denotes the $W^-$ field, $W_{\mu\nu}=\partial_{\mu}W_{\nu}-
\partial_{\nu}W_{\mu}$, $V_{\mu\nu} = \partial_{\mu}V_{\nu}-
\partial_{\nu}V_{\mu}$, $\tilde{V}_{\mu\nu}=\frac{1}{2} \epsilon_
{\mu\nu\rho\alpha}V^{\rho\alpha}$, and $(A\stackrel{\leftrightarrow}\partial
_{\mu}B)=A(\partial_{\mu}B)-(\partial_{\mu}A)B$, $V=\gamma$, $Z$ and
$M_W$ is the mass of the $W$ boson.
The overall couplings $g_{WWV}$ are
$g_{WW\gamma} = -e$ and $g_{WWZ} = -e ({\rm cot} \theta _{w})$ as in the SM,
where $e$ and $\theta_w$ are the positron charge and the weak mixing
angle.
The couplings $\lambda_V$ and $\kappa_V$ conserve $C$ and $P$. 
The couplings $g_4^V$ are odd under $CP$ and $C$, $g_5^V$ are
odd under $C$ and $P$, and $\tilde{\kappa}_V$ and $\tilde{\lambda}_V$ are odd
under $CP$ and $P$.  

In the SM, all the couplings are zero with the 
exception of $g_1^V$ and $\kappa_V$
($g_1^{\gamma} = g_1^Z = \kappa_{\gamma} = \kappa_Z = 1$).
Electromagnetic gauge invariance restricts
$g_1^{\gamma}, g_4^{\gamma}$ and $g_5^{\gamma}$ to the SM
values of 1, 0, and 0.
The $SU(2)_{L} \times U(1)_{Y}$ gauge invariance requires
$\lambda = \lambda_{\gamma} = \lambda_{Z}$ and
$\tilde{\lambda} = \tilde{\lambda}_{\gamma} = \tilde{\lambda}_{Z}$.
If the photon and $Z$ boson couplings are assumed to be equal, then
$g_1^{Z} = g_1^{\gamma} = 1$ and
$g_4^{Z} = g_4^{\gamma} = g_5^{Z} = g_5^{\gamma} = 0$. 

With the non--SM coupling parameters, the amplitudes for the gauge boson
pair production grow with energy, eventually violating tree--level
unitarity.
Using dipole form factors for anomalous couplings,
\begin{eqnarray*}
\Delta\kappa(s) = \frac{\Delta\kappa}{(1+ \hat{s}/\Lambda^2)^2}
\end{eqnarray*}
with a form factor scale, $\Lambda$, the unitarity is restored.
The scale, $\Lambda$, is  constrained by
\begin{equation}
\label{eq-int-ul}
\Lambda \leq \left[ \frac{6.88}{(\kappa - 1)^2 + 2\lambda^2
+ 2\tilde{\lambda}^2 }\right]^{1/4}
\rm{TeV},
\end{equation}
if the photon and $Z$ boson couplings are assumed to be equal.

The $CP$--conserving $WW\gamma$ coupling parameters are related\cite{KIM} 
to the magnetic dipole moment ($\mu_W$)
and electric quadrupole moment ($Q_W^e$) of the $W$ boson:
\begin{eqnarray*}
\mu_{W^+}=\frac{e}{2M_W}(1+\kappa_{\gamma}+\lambda_{\gamma})
\end{eqnarray*}
\begin{eqnarray*}
Q_{W^+}^{e}=-\frac{e}{M_W^2}(\kappa_{\gamma}-\lambda_{\gamma}).
\end{eqnarray*}
The $CP$--violating $WW\gamma$ coupling parameters are related to
the electric dipole moment ($d_W$) and magnetic quadrupole moment 
($Q_W^m$) of the $W$ boson:
\begin{eqnarray*}
d_{W^+} = \frac{e}{2M_W}(\tilde{\kappa}_{\gamma}+\tilde{\lambda}_{\gamma})
\end{eqnarray*}
\begin{eqnarray*}
Q_{W^+}^{m} = -\frac{e}{m^2_W}(\tilde{\kappa}_{\gamma} - \tilde{\lambda}
_{\gamma}).
\end{eqnarray*}
The $CP$--violating $WW\gamma$ couplings $\tilde{\lambda}_\gamma$ and
$\tilde{\kappa}_\gamma$ are tightly constrained by measurements of the
neutron electric dipole moment to
$|\tilde{\kappa}_{\gamma}|,|\tilde{\lambda}_{\gamma}| < 10^{-3}$
\cite{nedm}.

\subsection{$ZZ\gamma$ and $Z\gamma\gamma$ Couplings}
Theoretical calculations of the tree-level cross section for $Z\gamma$ 
production for SM and anomalous couplings have been performed\cite{gl20}. 
Assuming only electromagnetic gauge invariance and Lorentz invariance,
the vertex function for the $ZZ\gamma$ and $Z\gamma\gamma$ interaction
can be described with the following form:
\begin{eqnarray*}
\Gamma^{Z\gamma V}_{\alpha\beta\mu} = & 
            C_V ( h_1^V(q_2^{\mu}g^{\alpha\beta}-q_2^{\mu\beta} ) \\
    & + \frac{h_2^V}{m^2_Z}P^{\alpha}((P\cdot q_2)g^{\mu\beta}  - 
                                                 q_2^{\mu}P^{\beta}) \\
    & + h_3^V\epsilon^{\mu\alpha\beta\rho}q_{2\rho}                  
      + \frac{h_4^V}{m_Z^2}P^{\alpha}\epsilon^{\mu\beta \rho\sigma}P_{\rho}
                                                   q_{2\sigma} ) ,
\end{eqnarray*}
where $V$ indicates a photon or $Z$ boson, $C_Z$ is $(P^2-q_1^2)/M_Z^2$ and 
$C_{\gamma}$ is $P^2/M_Z^2$, $q_1$ and $q_2$ are the momenta of the outgoing
particles and $P$ is the momentum of the virtual boson.
An overall 
normalization factor of $g_{ZZ\gamma} = g_{Z\gamma\gamma} = e$, left out in
the equation, is used.
These couplings are $C$ odd dimensionless
functions of $q_1^2$, $q_2^2$, and $P^2$; i.e. $\hat{s}$.  
In addition, $h^V_1$ and $h^V_2$
are $P$ even, and thus violate $CP$. The other pair, $h^V_3$ and $h^V_4$,
are $CP$ conserving.

In order to avoid violating $S$--matrix unitarity, the couplings should 
asymptotically approach zero (their SM value) 
at high energies\cite{delicate-cancellation,gl22}.  Therefore, the 
$ZV\gamma$ couplings have to be energy-dependent and are thus modified with
form factors $h^V_i(q^2_1,q^2_2,P^2)$ which vanish at high $q^2_1$, 
$q^2_2$, or $P^2$.  However, since $q_1^2\approx m_Z^2$, $q_2^2 \approx 0$ 
and $P^2=\hat{s}$, only the high $\hat{s}$ behavior should be included in the 
form factor for the $q\bar{q}\rightarrow Z\gamma$ diagrams. The 
convention\cite{gl20} is to use a generalized dipole form factor such that:
\begin{eqnarray*}
h_i^V(m_Z^2,0,\hat{s})=\frac{h_{i0}^V}{(1+\hat{s}/\Lambda^2)^n}.
\end{eqnarray*}
The constraints on the $h_{i0}^V$ can be derived from partial wave 
unitarity of the general $f\bar{f}\rightarrow Z\gamma$ process 
\cite{gl23},\cite{gl24}.  Assuming only one coupling is non-zero at a time,
the following unitarity limits can be derived for $\Lambda \gg m_Z$ 
\cite{gl20},\cite{gl25}:
\begin{eqnarray*}
|h^{Z/\gamma}_{10}|,|h^{Z/\gamma}_{30}| < \frac{(\frac{2}{3}n)^n}
              {(\frac{2}{3}n -1)^{n-3/2}} \; \frac{0.126/0.151 {\rm{TeV}}^3}
              {\Lambda^3}    
\end{eqnarray*}
\begin{eqnarray*}
|h^{Z/\gamma}_{20}|,|h^{Z/\gamma}_{40}| < \frac{(\frac{2}{5}n)^n}
              {(\frac{2}{5}n -1)^{n-5/2}} \; \frac{2.1/2.5\cdot 10^{-3}
              {\rm{TeV}}^5}{\Lambda^5}. \\
\end{eqnarray*}
From the above equations, unitarity is satisfied for
$n>3/2$ for $h^V_{1,3}$, and $n>5/2$ for $h^V_{2,4}$.
In this paper
$n=3$ for $h^V_{1,3}$, and $n=4$ for $h^V_{2,4}$ are used.
This choice ensures
the same asymptotic energy behavior for the $h^V_{1,3}$ and $h^V_{2,4}$
couplings.  The dependence of results on the choice of $n$ is discussed in 
\cite{gl20}. 

The anomalous couplings $h^V_i$ are related to the $Z\gamma$ transition
dipole and quadrupole moments.  The $CP$--even combinations of $h^V_3$ and 
$h^V_4$ correspond to the electric dipole and magnetic quadrupole 
transition moments; the $CP$--odd combinations of $h^V_1$ and $h^V_2$
correspond to magnetic dipole and electric quadrupole transition moments.
The relations between the couplings and moments depends on both 
the center of mass energy $\sqrt{\hat{s}}$ and on the momentum of the 
final state photon \cite{gl28}.  They are:
\begin{eqnarray*}
d_{Z_T}=-\frac{e}{\sqrt{2}}\frac{k^2}{M^3_Z} (h_{30}^Z-h_{40}^Z), \\
Q^e_{Z_T}=\frac{e}{M^2_Z}\sqrt{10}(2h_{10}^Z), \\
\mu_{Z_T}=-\frac{e}{\sqrt{2}}\frac{k^2}{M^3_Z} (h_{10}^Z-h_{20}^Z), \\
Q^m_{Z_T}=\frac{e}{M^2_Z}\sqrt{10}(2h_{30}^Z), 
\end{eqnarray*}
where $d_{Z_T}$ $(\mu_{Z_T})$ is the transition electric (magnetic) dipole 
moment, $Q^e_{Z_T}$ $(Q^m_{Z_T})$ is the transition electric (magnetic) 
quadrupole moment, and $k$ is the photon energy.

\section{Appendix 2: Binned Likelihood Fit}
\label{sec-a2}

A binned likelihood fit was applied to the $p_T$ spectra of $\gamma$ and $W$
to set limits on the anomalous coupling parameters.
The observed numbers of events ($N_{i}$) in a particular $p_T$ bin can
be described in terms of the numbers of expected signal events ($n_{i}$)
and background events ($b_{i}$)
using a Poisson distribution function:
\begin{eqnarray*}
P_{i} = \frac{(n_{i} + b_{i})^{N_{i}}}{N_{i}!}e^{-(n_{i} + b_{i})},
\end{eqnarray*}
where $n_{i}$ and in some cases $b_{i}$ are functions of the anomalous
coupling parameters.

The uncertainties in the $n_{i}$ and $b_{i}$ were incorporated by convoluting
with Gaussian distributions:
\begin{eqnarray*}
P_{i}' = \int\limits_{\,\,\,-\infty}^{\,\,\,\,\,\,\,\,\infty}
\int\limits_{\,\,\,-\infty}^{\,\,\,\,\,\,\,\,\infty}\left(
\frac{(f_{n}n_{i} + f_{b}b_{i})^{N_{i}}}{N_{i}!}e^{-(f_{n}n_{i} + f_{b}b_{i})}
\right) \\ \frac{1}{2\pi\sigma_{n}\sigma_{b}}
e^{-\frac{(f_{n}-1)^{2}}{2\sigma_{n}^{2}}}
e^{-\frac{(f_{b}-1)^{2}}{2\sigma_{b}^{2}}}
df_{n}df_{b},
\end{eqnarray*}
where $f_{n}$ and $f_{b}$ are multiplicative factors to $n_{i}$ and $b_{i}$
with mean values of 1.0; $\sigma_{n}$ and $\sigma_{b}$ are the
fractional uncertainties of $n_{i}$ and $b_{i}$.
These uncertainties include the uncertainties in the integrated luminosity
and the theoretical prediction of the signal and background cross sections.
To exploit the prediction that anomalous couplings lead to an excess of
events with high $E_T$ photons or jets (depending on the analysis), 
a bin with no events at high $E_T$ was used in the fit. 
The bin boundary was selected sufficiently above the highest observed 
transverse momentum event in the data sample that the detector 
resolution could not move the last data point across the boundary.
For more detail, see \cite{glthesis}.
The joint probability of all $p_T$ bins is then
\begin{eqnarray*}
P = \prod\limits_{i=1}^{N_{max}} P_{i}^{\prime},
\end{eqnarray*}
where $N_{max}$ is the number of $p_T$ bins.
The log likelihood function of this joint probability is defined as
\begin{eqnarray*}
{\cal L} = - {\rm ln} (P).
\end{eqnarray*}
The limits on the coupling parameters were obtained by maximizing this
quantity. 
The 95\% confidence level limit on the 
parameters of the log-likelihood function (the coupling parameters for
the case here) is the contour where the log-likelihood is 1.92 lower than the 
maximum.

\clearpage

\vspace{1.0in}

\begin{figure}
\epsfysize = 6.8in
\epsfxsize = 5.0in
\centerline{\epsfbox{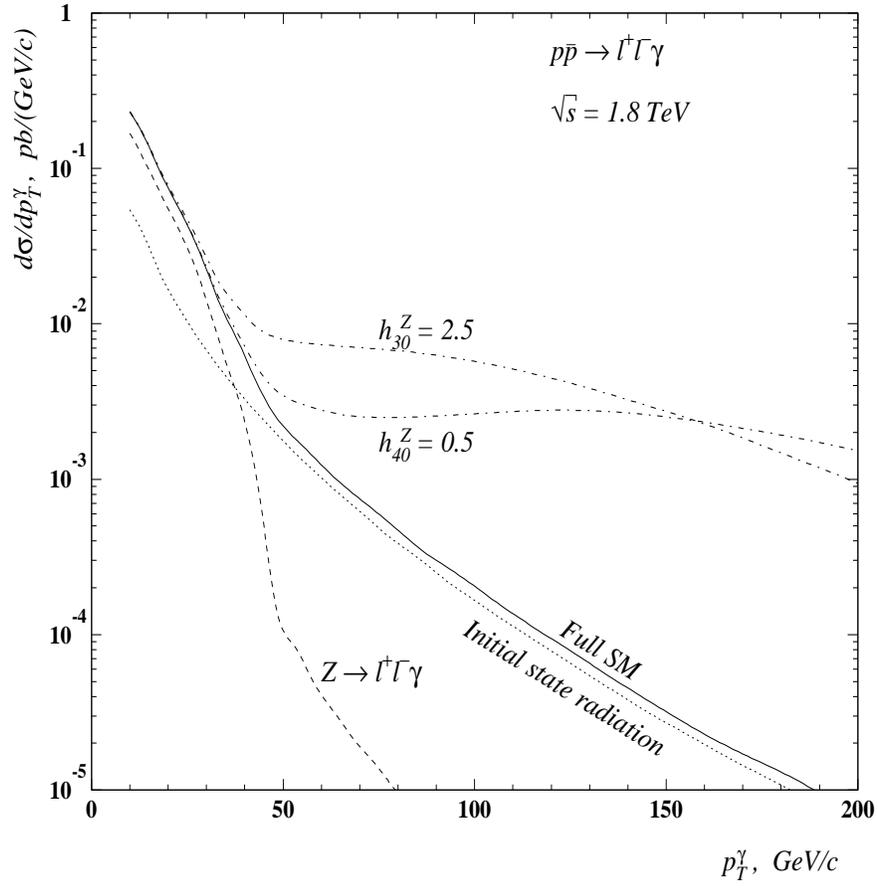}}
\vskip -1.5in
\caption{The $p_T^{\gamma}$ spectrum, $d\sigma/dp^{\gamma}_T$,
 for SM $Z\gamma$ events from final state radiation (dashed 
line), initial state radiation (dotted line), the combination of initial state
and final state radiation (solid line) as well as for two 
anomalous $ZZ\gamma$ couplings. }
\label{fig-zgpt}
\end{figure}

\clearpage

\clearpage

\end{document}